\documentclass[fleqn,usenatbib]{mnras}

\usepackage{newtxtext,newtxmath}

\usepackage[T1]{fontenc}
\usepackage{ae,aecompl}

\usepackage{amsfonts}
\usepackage{commath}
\usepackage{xfrac}
\usepackage{subfigure}
\usepackage{caption}
\usepackage{booktabs}
\usepackage{pdflscape}
\usepackage{adjustbox}
\usepackage{csquotes}
\usepackage{cuted}
\usepackage{multicol}

\usepackage{graphicx}	
\usepackage{amsmath}	

\usepackage{amssymb}	
\usepackage{multirow}
\usepackage{arydshln}
\usepackage{threeparttable}
\usepackage{cleveref}

\title[Hierarchical CMB component separation]{Hierarchical Bayesian CMB Component Separation with the No-U-Turn Sampler}

\author[R.D.P. Grumitt et al.]{
R.D.P. Grumitt,$^{1}$\thanks{E-mail: richard.grumitt@physics.ox.ac.uk}
Luke R.P. Jew,$^{1}$
and C. Dickinson\,$\!^{2,3}$
\\
$^{1}$Sub-department of Astrophysics, University of Oxford, Denys Wilkinson Building, Keble Road, Oxford OX1 3RH, U.K.\\
$^{2}$Jodrell Bank Centre for Astrophysics, Alan Turing Building, Department of Physics and Astronomy, School of Natural Sciences,\\ 
The University of Manchester, Oxford Road, Manchester, M13 9PL, U.K. \\
$^{3}$Cahill Centre for Astronomy and Astrophysics, California Institute of Technology, Pasadena, CA 91125, USA \\
}

\date{Accepted XXX. Received YYY; in original form ZZZ}

\pubyear{2020}

\begin{document}
\label{firstpage}
\pagerange{\pageref{firstpage}--\pageref{lastpage}}
\maketitle

\begin{abstract}
In this paper we present a novel implementation of Bayesian CMB component separation. We sample from the full posterior distribution using the No-U-Turn Sampler (NUTS), a gradient-based sampling algorithm. Alongside this, we introduce new foreground modelling approaches. We use the mean-shift algorithm to define regions on the sky, clustering according to naively estimated foreground spectral parameters. Over these regions we adopt a complete pooling model, where we assume constant spectral parameters, and a hierarchical model, where we model individual pixel spectral parameters as being drawn from underlying hyper-distributions. We validate the algorithm against simulations of the \textit{LiteBIRD} and C-BASS experiments, with an input tensor-to-scalar ratio of $r=5\times 10^{-3}$. Considering multipoles $30\leq\ell<180$, we are able to recover estimates for $r$. With \textit{LiteBIRD}-only observations, and using the complete pooling model, we recover $r=(12.9\pm 1.4)\times 10^{-3}$. For C-BASS and \textit{LiteBIRD} observations we find $r=(9.0\pm 1.1)\times 10^{-3}$ using the complete pooling model, and $r=(5.2\pm 1.0)\times 10^{-3}$ using the hierarchical model. Unlike the complete pooling model, the hierarchical model captures pixel-scale spatial variations in the foreground spectral parameters, and therefore produces cosmological parameter estimates with reduced bias, without inflating the their uncertainties. Measured by the rate of effective sample generation, NUTS offers performance improvements of $\sim10^3$ over using Metropolis-Hastings to fit the complete pooling model. The efficiency of NUTS allows us to fit the more sophisticated hierarchical foreground model, that would likely be intractable with non-gradient based sampling algorithms.  
\end{abstract}




\section{Introduction}\label{sec: intro}

One of the major outstanding goals of CMB cosmology is the detection of primordial $B$-modes in the CMB polarization \citep{2015PhRvL.114j1301B, 2016ARA&A..54..227K}. The challenge of detecting these $B$-modes has become a problem of accurate component separation, that is the extraction of the CMB $B$-mode signal from foreground contaminated observations of the radio and microwave sky \citep{2009A&A...503..691B, doi:10.1063/1.3160888, 2016JCAP...03..052E}. We can parametrize the strength of the CMB $B$-mode signal through the tensor-to-scalar ratio, $r$, which gives the ratio of the amplitude of tensor-to-scalar perturbations in the early universe \citep{2003moco.book.....D}. Given current constraints on $r$, along with the targeted sensitivities of next-generation CMB experiments of $\sigma(r)\sim 10^{-3}$, future CMB experiments must be able to detect a CMB signal that is potentially sub-dominant to foreground emission across all of the sky, at all frequencies \citep{2011JCAP...07..025K, 2016arXiv161002743A, 2018JCAP...04..023R, 2018SPIE10698E..1YS, 2018PhRvL.121v1301B, 2019JCAP...02..056A, 2019BAAS...51c.338S, 2019BAAS...51g.194H}. This presents two primary challenges. First we must be sure to have data of a sufficient sensitivity with enough frequency coverage to be able to model foreground spectral energy distributions (SEDs) with sufficient accuracy and precision. Secondly, our component separation algorithms must be able to extract the CMB signal from our noisy observations with high fidelity, and properly quantify the uncertainty in the extracted signal. 

In this paper we focus on the second challenge, in particular studying Bayesian parametric component separation. Numerous CMB component separation algorithms have been developed, see e.g., \cite{0004-637X-676-1-10, 2008A&A...491..597L, doi:10.1063/1.3160888, 2010MNRAS.408.2319S, 2014A&A...571A..12P, 2018JCAP...04..023R, 2016A&A...594A...9P, 2016A&A...594A..10P, 2003MNRAS.346.1089D, 2003MNRAS.345.1101M, 2011MNRAS.418..467R, 2019A&A...627A..98S}. Each of these algorithms offer different advantages and disadvantages in terms of accuracy, computational efficiency and quantification of uncertainty. One of the main motivations for using Bayesian parametric component separation is the ability to obtain properly motivated probability distributions for our model parameters, and hence a proper quantification of the uncertainty. This does come at the cost of Bayesian inference being computationally expensive, especially when using Markov Chain Monte Carlo (MCMC) techniques. Further, uncertainties in our modelling of foreground emission can lead to biases in our inference. This is potentially highly problematic in the case of $B$-mode science, where the potential biases can be of the same order as the value of $r$ we are attempting to measure. The impact of such modelling errors have received significant previous attention, see e.g., \cite{2016MNRAS.458.2032R,2017MNRAS.472.1195C,2018JCAP...04..023R}. These modelling problems are very closely linked with the available data. If it becomes apparent that more complex modelling is required, more comprehensive data covering a wide range of frequencies will be required to constrain the additional model parameters \citep{2019MNRAS.tmp.2372J}.

Bayesian parametric component separation has notably been applied through the \textsc{Commander} algorithm, which utilizes Gibbs sampling to jointly sample the CMB sky signal, power spectrum, and foreground and instrumental parameters \citep{doi:10.1080/01621459.1990.10476213, 2004ApJS..155..227E, 2008ApJ...676...10E, 2014A&A...571A..12P, 2016A&A...594A...9P, 2016A&A...594A..10P}. Gibbs sampling consists of drawing successive parameter samples from the conditional distributions of your parameters, as opposed to directly sampling from the full joint distribution. In the case of complex, high-dimensional distributions this can offer significant performance improvements over sampling directly from the joint distribution using algorithms such as Metropolis-Hastings, which quickly prove intractable for CMB component separation \citep{hastings70}. However, even Gibbs sampling becomes computationally challenging as we move to higher resolution analyses. This has largely limited the application of Bayesian parametric component separation to studies of the large-scale CMB signal, up to multipoles of $\ell\sim 50$.

In this paper we present a new implementation of Bayesian parametric CMB component separation, using the No-U-Turn Sampler (NUTS) to explore the parameter space. NUTS is a self-tuning variant of Hamiltonian Monte Carlo (HMC), originally presented in \cite{Homan:2014:NSA:2627435.2638586}. Since this original exposition, the algorithm has undergone a number of developments, with state-of-the-art implementations in \textsc{Stan} \citep{stan_development_team_stan_2012, osti_1430202} and \textsc{PyMC3} \citep{pymc3_citation}. HMC algorithms make use of first-order gradient information to generate efficient proposal steps. This allows HMC to avoid the random-walk behaviour of standard Metropolis-Hastings and Gibbs sampling methods, which becomes particularly problematic as the dimension of the parameter space increases.

To validate the performance of our component separation algorithm, we apply it to simulated observations of \textit{LiteBIRD}, a planned next-generation CMB satellite \citep{2018SPIE10698E..1YS}, and the C-Band All-Sky Survey (C-BASS), a $5\,\mathrm{GHz}$ ground-based experiment observing the sky in total intensity and polarization \citep{2018MNRAS.480.3224J}. Previous CMB component separation analyses have considered the extent to which foreground spectral parameters should be allowed to vary, such that a balance can be struck between model realism and simplicity. These studies have considered various approaches to defining regions on the sky, over which foreground spectral parameters are typically assumed to be constant \citep{2009MNRAS.392..216S, 2011PhRvD..84f3005E, 2016PhRvD..94h3526S, 2017PhRvD..95d3504A, 2019A&A...623A..21I, 2019JCAP...02..039K, 2019arXiv190508888T}. In this paper we use the mean-shift algorithm to define regions on the sky, clustering according to naively estimated synchrotron and dust spectral parameters. We initially fit a complete pooling model where we assume foreground spectral parameters to be constant in each region. We then fit a hierarchical foreground model. Hierarchical modelling has recently been employed in the context of blind CMB component separation in \cite{2019arXiv191008077W}. For the hierarchical analysis in this paper, we assume individual pixel spectral parameters are drawn from underlying Gaussian distributions, jointly fitting for the mean and variance of the Gaussian hyper-distributions, and the individual pixel-by-pixel spectral parameters in each region. In doing so, we are able to provide a faithful generative description of the underlying foreground emission, whilst reducing the propensity for fitting noisy outliers when assuming total independence between pixel spectral parameters \citep{doi:10.1198/004017005000000661, GelmanHill:2007}.

Bench-marking against the rate at which the algorithm generates effective CMB amplitude samples, we find that NUTS offers performance improvements of $\sim 10^{3}$ over sampling with the Metropolis-Hastings algorithm for the complete pooling model. Sampling from the posterior distribution of the hierarchical model is particularly challenging. Hierarchical models are known to exhibit geometrical pathologies that make it extremely difficult to achieve convergence using non-gradient based sampling algorithms. In these situations variants of HMC are often the only tractable approaches to sampling from the posterior \citep{hmc_hierarchical2013}.        

The outline of this paper is as follows: In Section \ref{sec: comp models} we describe the diffuse component SED models used in our analysis. In Section \ref{sec: simulations} we describe the \textit{LiteBIRD} and C-BASS simulations used in our algorithm validation. In Section \ref{sec: clustering} we describe the mean-shift clustering algorithm used to define regions on the sky. In Section \ref{sec: comp sep nuts} we describe the component separation algorithm, give a general description of the NUTS algorithm, and discuss the complete pooling and hierarchical foreground models used during validation. In Section \ref{sec: validation} we present the results from our algorithm validation. We conclude in Section \ref{sec: conclusions}.
\section{Diffuse Component Models}\label{sec: comp models}

In this section we describe the diffuse component models employed in the validation of our component separation algorithm. Our focus here is on component separation for CMB polarization studies, and as such we consider only spectral models for synchrotron, thermal dust and CMB emission. Additional contributions can potentially arise from polarized anomalous microwave emission (AME) and free-free emission. However, in both of these cases the level of polarized emission is expected to be very low. For AME, theoretical considerations from spinning dust models suggest that AME should be very weakly polarized, with expected polarization fractions of $\sim 10^{-6}$ \citep{2016ApJ...831...59D}. Existing measurements place upper limits on the AME polarization fraction of $\sim 1$ per cent \citep{2018NewAR..80....1D}. Free-free emission is caused by the scattering of electrons off ions in the interstellar medium. Given the random nature of this scattering, free-free emission is intrinsically un-polarized, with upper limits on the polarization fraction of $\sim 1$ per cent \citep{2011MNRAS.418..888M}. At the edges of bright \ion{H}{ii} regions, higher polarization fractions of $\sim 10$ per cent are possible due to additional Thompson scattering \citep{Rybicki1985,1998ApJ...495..580K}. However, for the purposes of CMB polarization studies, these effects are expected to be largely negligible. 

It is important to note here that the true underlying foreground SEDs are more complicated than those used in this analysis. Mis-modelling the underlying sky emission can result in biases in recovered cosmological parameter estimates. This has received significant attention in $B$-mode forecasting analyses, see e.g., \cite{2016JCAP...03..052E, 2016MNRAS.458.2032R, 2018JCAP...04..023R}. In this paper we present a hierarchical foreground model (described in Section \ref{subsec: hierarchical}), that enables us to model spatial variations in foreground spectral parameters, without inflating parameter uncertainties. Given the simulated datasets considered here, fitting foreground spectral parameters with any parametric method will ultimately be limited by the number of frequency channels available with sufficient sensitivity. For example, in \cite{2019MNRAS.tmp.2372J}, it was found that to constrain models including the synchrotron curvature would require additional low-frequency channels between $5\,\mathrm{GHz}$ and $30\,\mathrm{GHz}$. Given a maximum observed frequency of $\sim 400\,\mathrm{GHz}$, dust spectral parameters in the modified blackbody (MBB) model are also very poorly constrained \citep{2018arXiv180601026R, 2019MNRAS.tmp.2372J}. This problem could be alleviated to some extent by considering model re-parametrizations that yield more tractable posterior geometries. However, at its core this is a problem in the data.

\subsection{Synchrotron}

Synchrotron emission is caused by electrons spiralling in the Galactic Magnetic Field (GMF), and is the dominant diffuse component at low frequencies ($\nu\lesssim 100\,\mathrm{GHz}$ in polarization), contributing to both total intensity and polarized emission. In ordered magnetic fields, the polarization fraction of synchrotron emission can be as high as $70$ per cent, with more typical values of around $40$ per cent at high Galactic latitudes \citep{Rybicki1985, 2015MNRAS.452..656V, 2016A&A...594A..25P}. Over a wide range of frequencies, from $\mathcal{O}(10\,\mathrm{MHz})$ up to $\mathcal{O}(100\,\mathrm{GHz})$, the synchrotron SED can be reasonably approximated as a power-law spectrum \citep{1987MNRAS.225..307L,1988A&AS...74....7R,2003A&A...410..847P,Davies2006,2011A&A...525A.138G}. We may parametrize the power law spectrum as,
\begin{equation}
    S_{\mathrm{s}} = A_{\mathrm{s}}\left(\frac{\nu}{\nu_{0}}\right)^{-\beta_{\mathrm{s}}},
\end{equation}
where $A_{\mathrm{s}}$ is the reference synchrotron amplitude, $\nu$ is the observing frequency, $\nu_{0}$ is some reference frequency and $\beta_{\mathrm{s}}$ is the synchrotron spectral index. 

In reality, the synchrotron spectrum is modified by a combination of intrinsic effects, e.g., spectral ageing, along with pixel and beam averaging effects \citep{1969MNRAS.146..221M,2016MNRAS.458.4443H, 2017MNRAS.472.1195C,2018JCAP...04..023R}. These complications can be modelled through additional spectral curvature terms in the power-law, or through the moment expansion method presented in \cite{2017MNRAS.472.1195C}. However, for the purposes of our validation analysis in this paper it is sufficient to consider a simple power-law model. 

Analysis in \cite{2019MNRAS.tmp.2372J} showed that experiments such as \textit{LiteBIRD} struggle to constrain synchrotron spectral parameters, with additional low-frequency data being necessary to begin to constrain the synchrotron spectral index. To constrain more complex synchrotron curvature models would require additional low-frequency channels. The European Low-Frequency Survey (ELFS) is a proposed experiment, covering frequencies between $6$ and $30\,\mathrm{GHz}$, that would allow us to constrain more complex synchrotron SEDs \citep{jaz2019a, 2019BAAS...51g.111G}.

\subsection{Thermal Dust}

Thermal dust emission is caused by thermal emission from interstellar dust grains. In general, dust grains are not spherically symmetric and emit preferentially along their longer axis \citep{2015A&A...576A.104P,2018arXiv180104945P}. These interstellar dust grains align with local magnetic fields, resulting in emission in both total intensity and polarization. Thermal dust polarization fractions can be as high as $20$ per cent, with a median value across the sky of approximately $8$ per cent \citep{2015A&A...576A.104P}. Thermal dust is the dominant component at high frequencies ($\nu\gtrsim 100\,\mathrm{GHz}$).

The thermal dust spectrum can be approximated by an MBB model, given by,
\begin{equation}
    S_{\mathrm{d}} = A_{\mathrm{d}}\left(\frac{\nu}{\nu_{0}}\right)^{\beta_{\mathrm{d}}+1}\frac{\exp{(\gamma\nu_{0})} -1}{\exp{(\gamma\nu)}-1},
\end{equation}
where $A_{\mathrm{d}}$ is the reference dust amplitude, $\nu_{0}$ is some reference frequency, $\gamma = h/k_{B}T_{\mathrm{d}}$, $T_{\mathrm{d}}$ is the dust temperature, and $\beta_{\mathrm{d}}$ is the dust spectral index. This model is a simplification, in reality multiple dust populations will exist along the line of sight. More complex models have previously been considered in \cite{2018ApJ...853..127H}. However, these more complex models encounter the same issues as for synchrotron emission. The additional parameters introduced require additional data at frequencies $\nu\gtrsim 400\,\mathrm{GHz}$ to constrain them. Given the simulated frequency coverage we consider in this paper, we already struggle to place constraints on the spectral parameters of the single MBB model.

\subsection{CMB}

The CMB follows a blackbody spectrum given by,
\begin{equation}
    S_{\mathrm{cmb}} = A_{\mathrm{cmb}}\frac{x^{2}\exp{(x)}}{(\exp{(x)} - 1)^2},
\end{equation}
where $A_{\mathrm{cmb}}$ is the CMB amplitude, $x=h\nu/k_{B}T_{\mathrm{cmb}}$ and $T_{\mathrm{cmb}}=2.7255\,\mathrm{K}$ is the mean CMB temperature \citep{Fixsen2009}. Throughout this paper we work in units of Rayleigh-Jeans brightness temperature unless otherwise stated. 
\section{Simulations}\label{sec: simulations}

To validate the performance of our component separation algorithm we generate a set of simulated Stokes $Q$ and $U$ maps, corresponding to the frequencies and sensitivities of the C-BASS and \textit{LiteBIRD} experiments \citep{2018MNRAS.480.3224J,2018SPIE10698E..1YS}. The frequencies and polarization sensitivities of the C-BASS and \textit{LiteBIRD} simulations are given in Table \ref{tab: cbass litebird details}.

\textit{LiteBIRD} is a planned next-generation CMB satellite, aiming to measure the tensor-to-scalar ratio with a sensitivity of $\sigma(r)\sim 10^{-3}$. To accomplish this task \textit{LiteBIRD} will target large angular scales up to $\ell\sim 200$, covering the reionization peak at $\ell\sim 10$ and the recombination peak at $\ell\sim 80$ in the primordial $B$-mode power spectrum. The experiment is proposed to cover frequencies from $40\,\mathrm{GHz}$ to $402\,\mathrm{GHz}$, with the lowest resolution $40\,\mathrm{GHz}$ channel having a resolution of approximately $70\,\mathrm{arcmin}$. We smooth all of our simulated maps to this $70\,\mathrm{arcmin}$ resolution, which is sufficient for our validation analysis targeting angular scales $\ell\lesssim 200$.

C-BASS is a $5\,\mathrm{GHz}$ experiment, observing the whole sky in intensity and polarization at a native resolution of $45\,\mathrm{arcmin}$, at a sensitivity of $0.1\,\mathrm{mK/beam}$ ($4320\,\mathrm{\mu\mbox{K-arcmin}}$). The primary purpose of the survey is to provide improved constraints on polarized synchrotron emission, to aid in CMB component separation analyses \citep{2018MNRAS.480.3224J}.

We simulate maps of polarized emission, containing contributions from synchrotron, thermal dust and the CMB, using \textsc{PySM} \citep{2017MNRAS.469.2821T}. We adopt the SEDs for our sky components described in Section \ref{sec: comp models}. For synchrotron emission we use the \textsc{PySM} s1 model. This used the 9-year \textit{WMAP} $23\,\mathrm{GHz}$ maps \citep{2013ApJS..208...20B}, smoothed to $3^{\circ}$, as synchrotron $Q/U$ templates. Small scales are added to these templates by extrapolating the map power spectra to high $\ell$ and obtaining Gaussian realizations of the power spectra. Details of the implementation can be found in \cite{2017MNRAS.469.2821T}. These synchrotron templates are then extrapolated to higher frequencies using a spatially varying spectral index map taken from \cite{2008A&A...490.1093M}. Across the whole sky, the synchrotron spectral index map has a mean of $\langle\beta_s\rangle\approx-3.0$ and a standard deviation of $\sigma(\beta_s)\approx 0.06$. Analysis of the synchrotron angular power spectrum in \cite{2018A&A...618A.166K} found that the spectral index map used in \textsc{PySM} lacks power on all angular scales. However, for the purposes of our validation analysis it is sufficient.  

For thermal dust we use the \textsc{PySM} d1 model, which uses the \textit{Planck} $353\,\mathrm{GHz}$ maps as $Q/U$ thermal dust templates. The templates are scaled using the spatially varying dust temperature and spectral index maps obtained from the \textit{Planck} \textsc{Commander} analysis \citep{2016A&A...594A..10P}. Across the whole sky the mean of the dust temperature map is $\langle T_{\mathrm{d}}\rangle\approx 20.9\,\mathrm{K}$ and the standard deviation is $\sigma(T_d)\approx 2.2\,\mathrm{K}$. For the dust spectral index the mean value is $\langle\beta_{\mathrm{d}}\rangle\approx 1.54$ and the standard deviation is $\sigma(\beta_{\mathrm{d}})\approx 0.04$. The dust templates are smoothed to $2.6^{\circ}$, with small scales being added using the same prescription as for synchrotron emission. 

\begin{figure*}
\centering
\subfigure[$\mathbfit{A}_{\mathrm{cmb}}^Q$]{\includegraphics[width=0.33\textwidth]{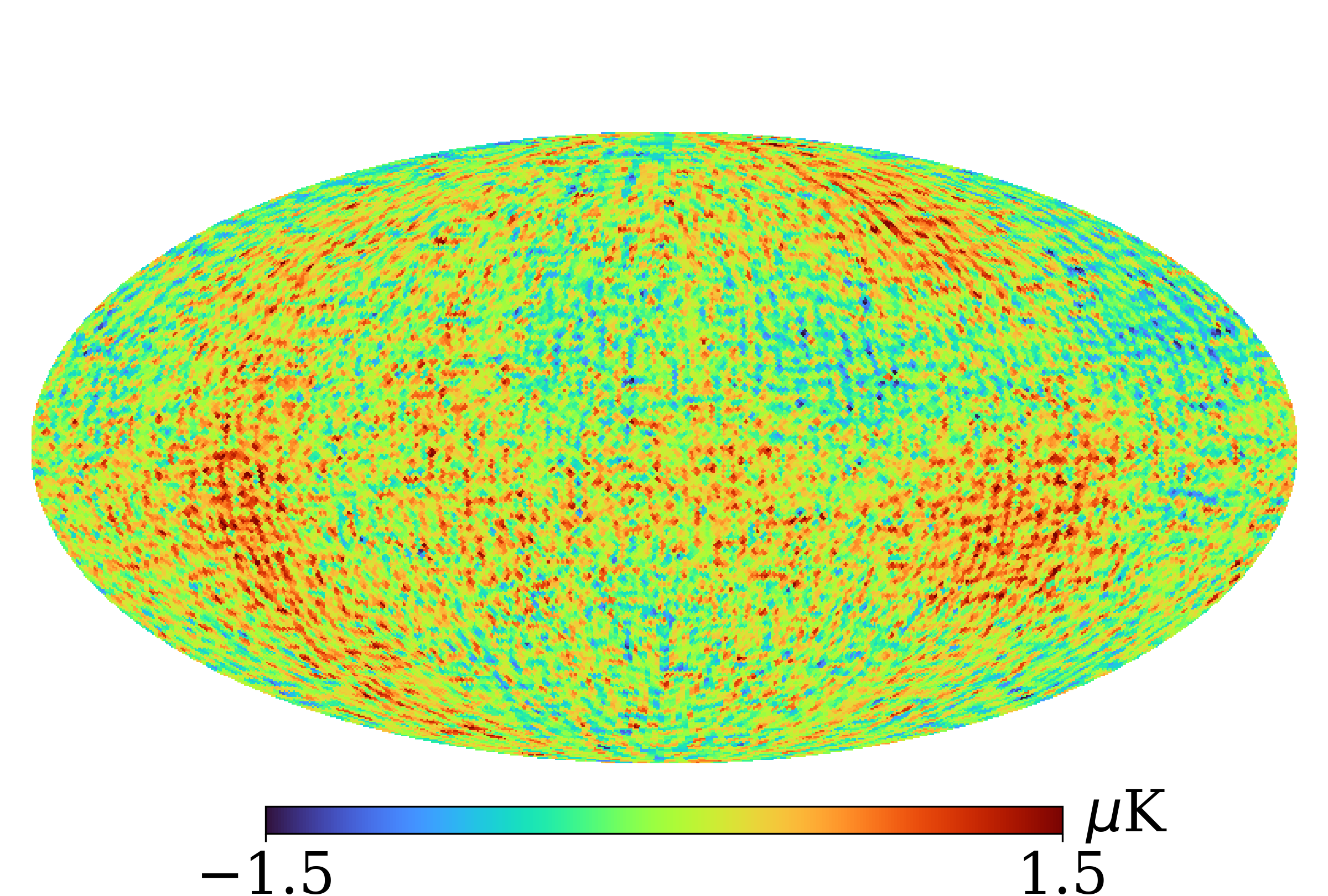}}
\subfigure[$\mathbfit{A}_{\mathrm{d}}^{Q}$ @ $402\,\mathrm{GHz}$]{\includegraphics[width=0.33\textwidth]{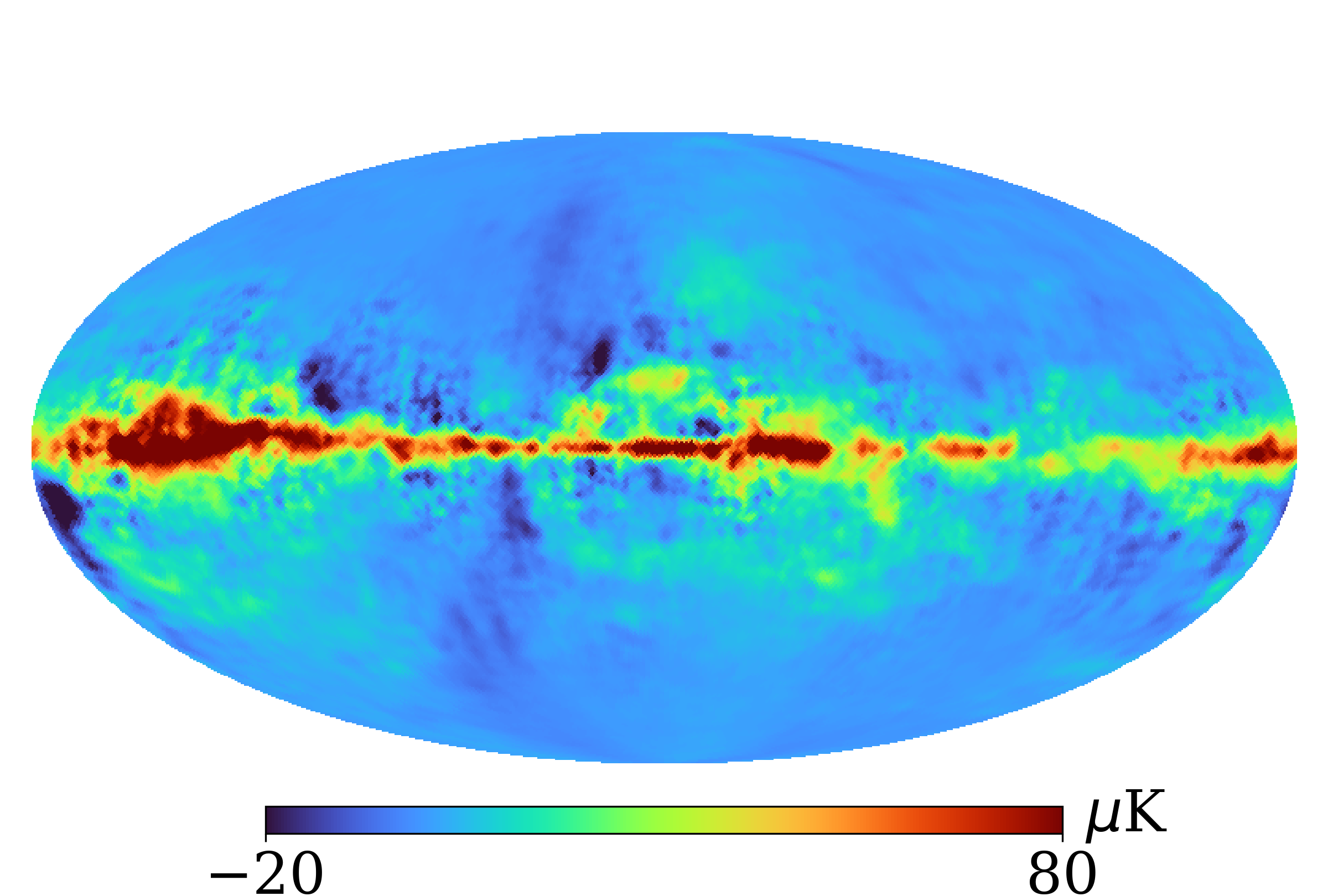}}
\subfigure[$\mathbfit{A}_{\mathrm{s}}^Q$ @ $40\,\mathrm{GHz}$]{\includegraphics[width=0.33\textwidth]{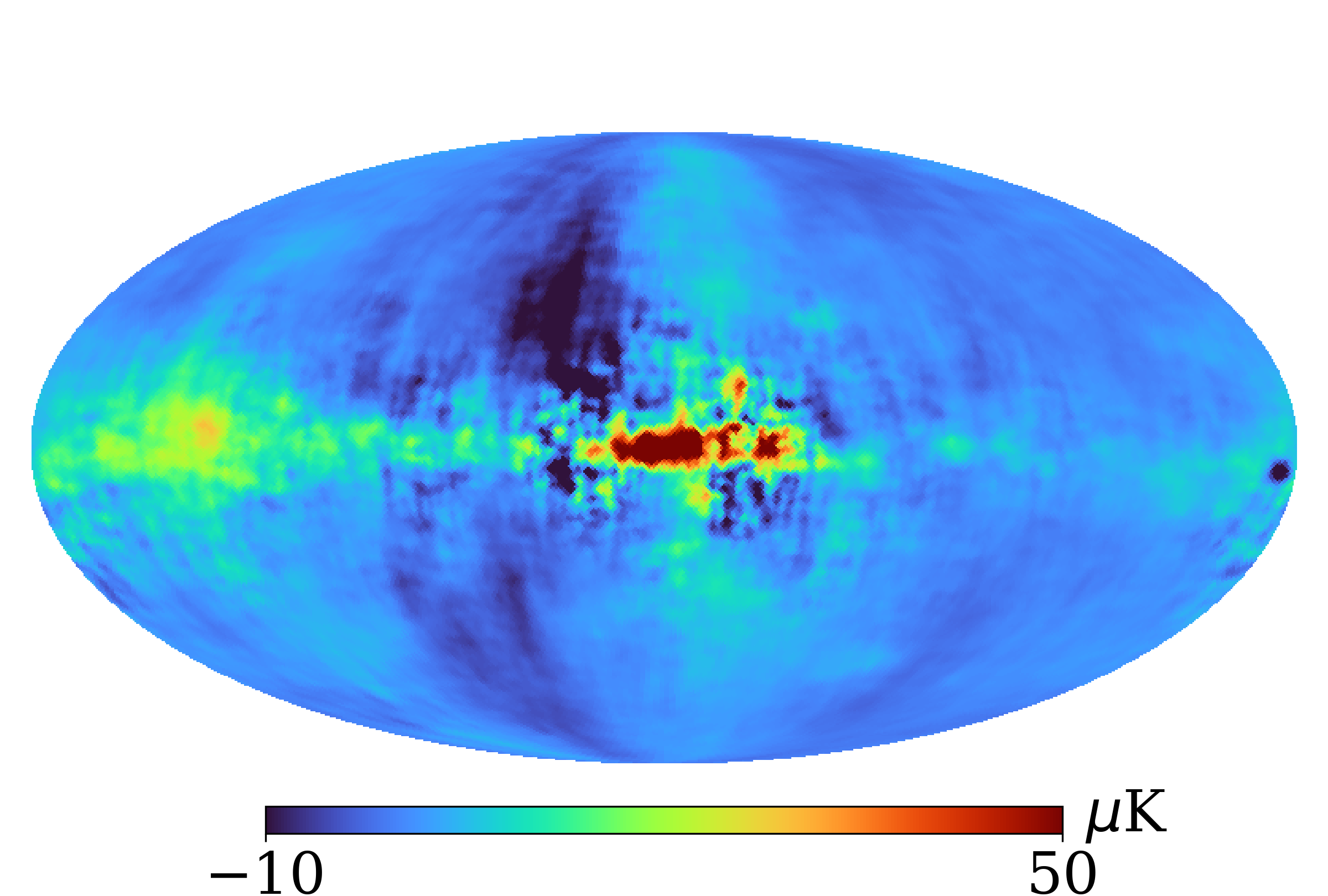}}
\subfigure[$\mathbfit{A}_{\mathrm{cmb}}^U$]{\includegraphics[width=0.33\textwidth]{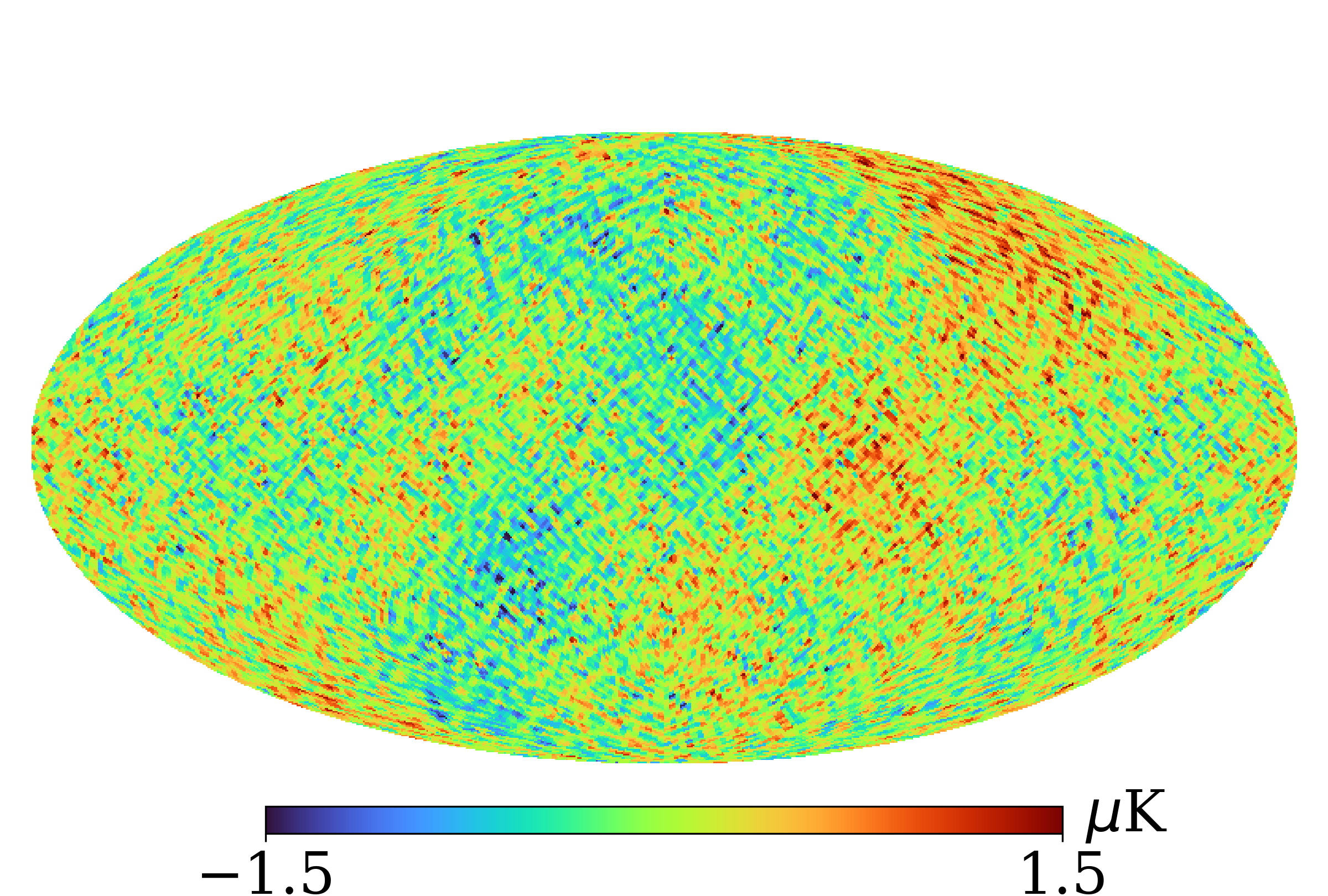}}
\subfigure[$\mathbfit{A}_{\mathrm{d}}^U$ @ $402\,\mathrm{GHz}$]{\includegraphics[width=0.33\textwidth]{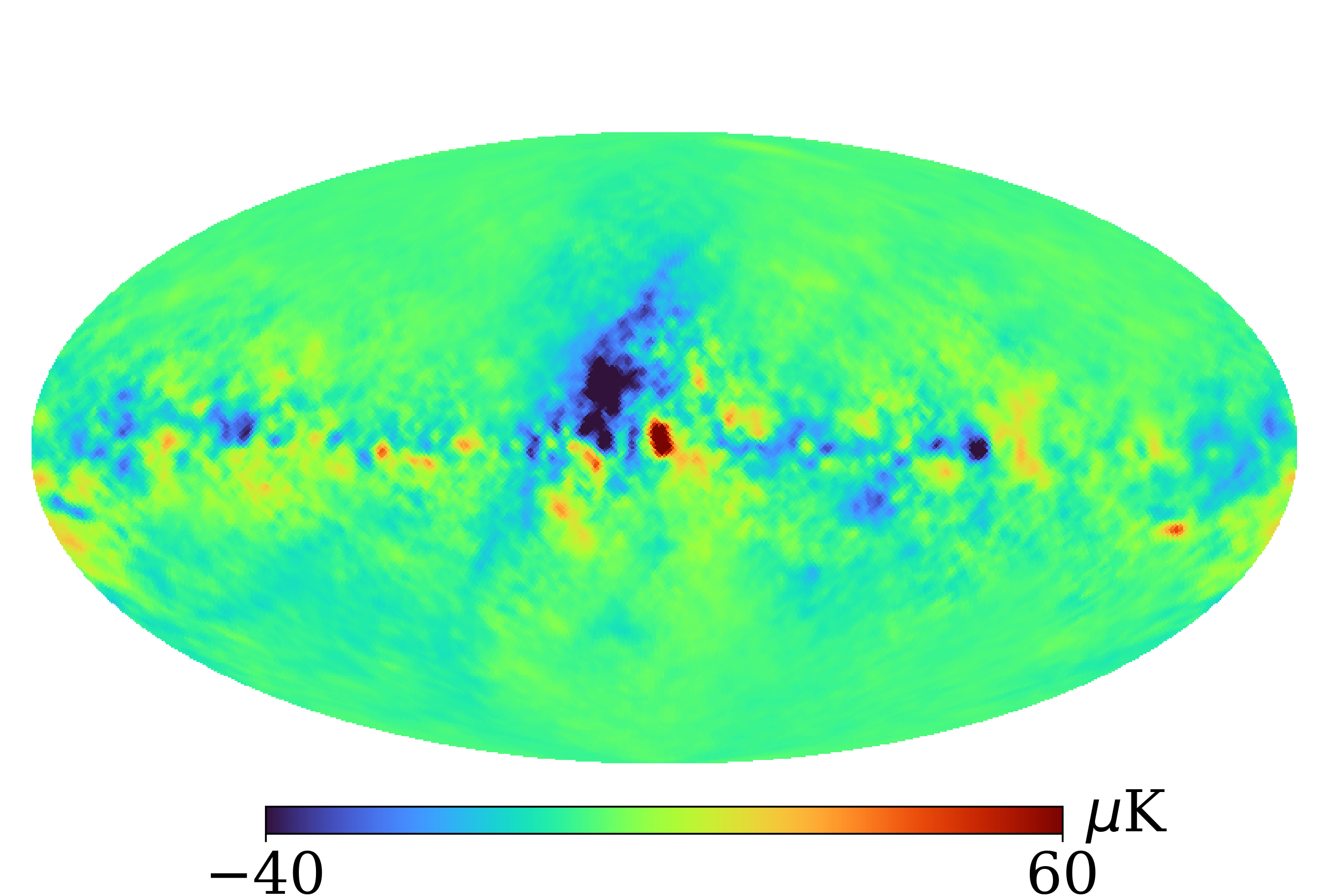}}
\subfigure[$\mathbfit{A}_{\mathrm{s}}^U$ @ $40\,\mathrm{GHz}$]{\includegraphics[width=0.33\textwidth]{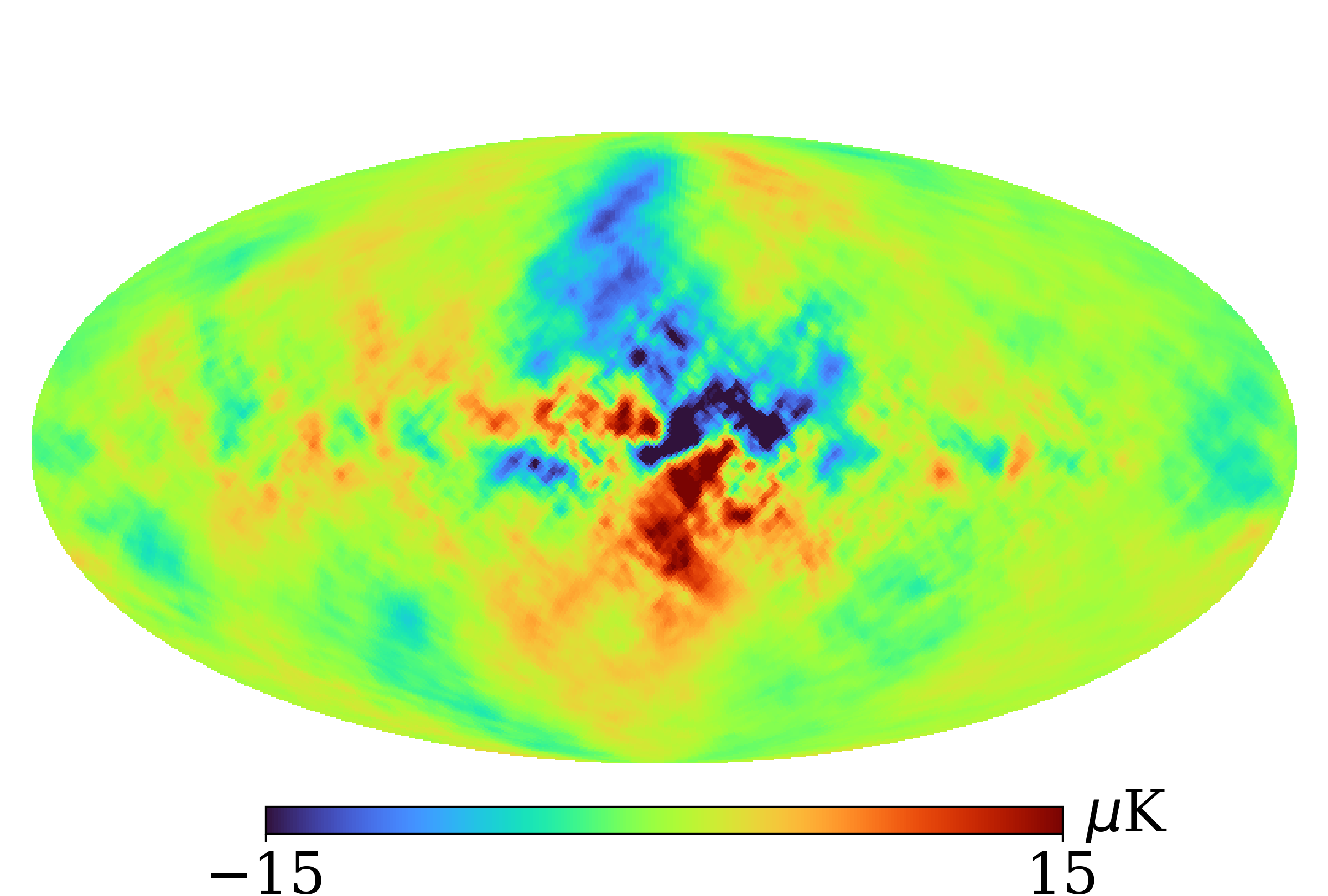}}
\subfigure[$\beta_{\mathrm{d}}$]{\includegraphics[width=0.33\textwidth]{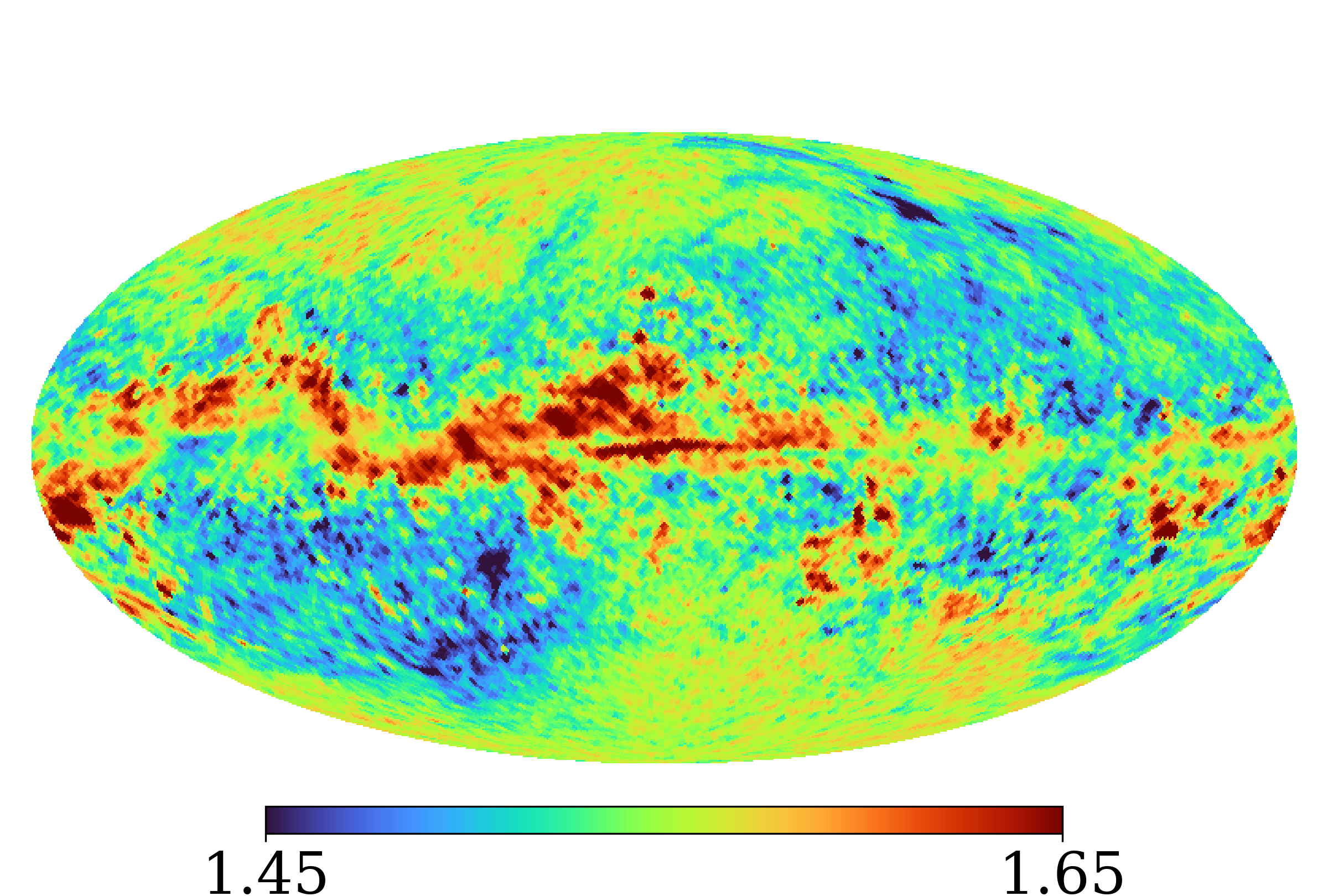}}
\subfigure[$T_{\mathrm{d}}$]{\includegraphics[width=0.33\textwidth]{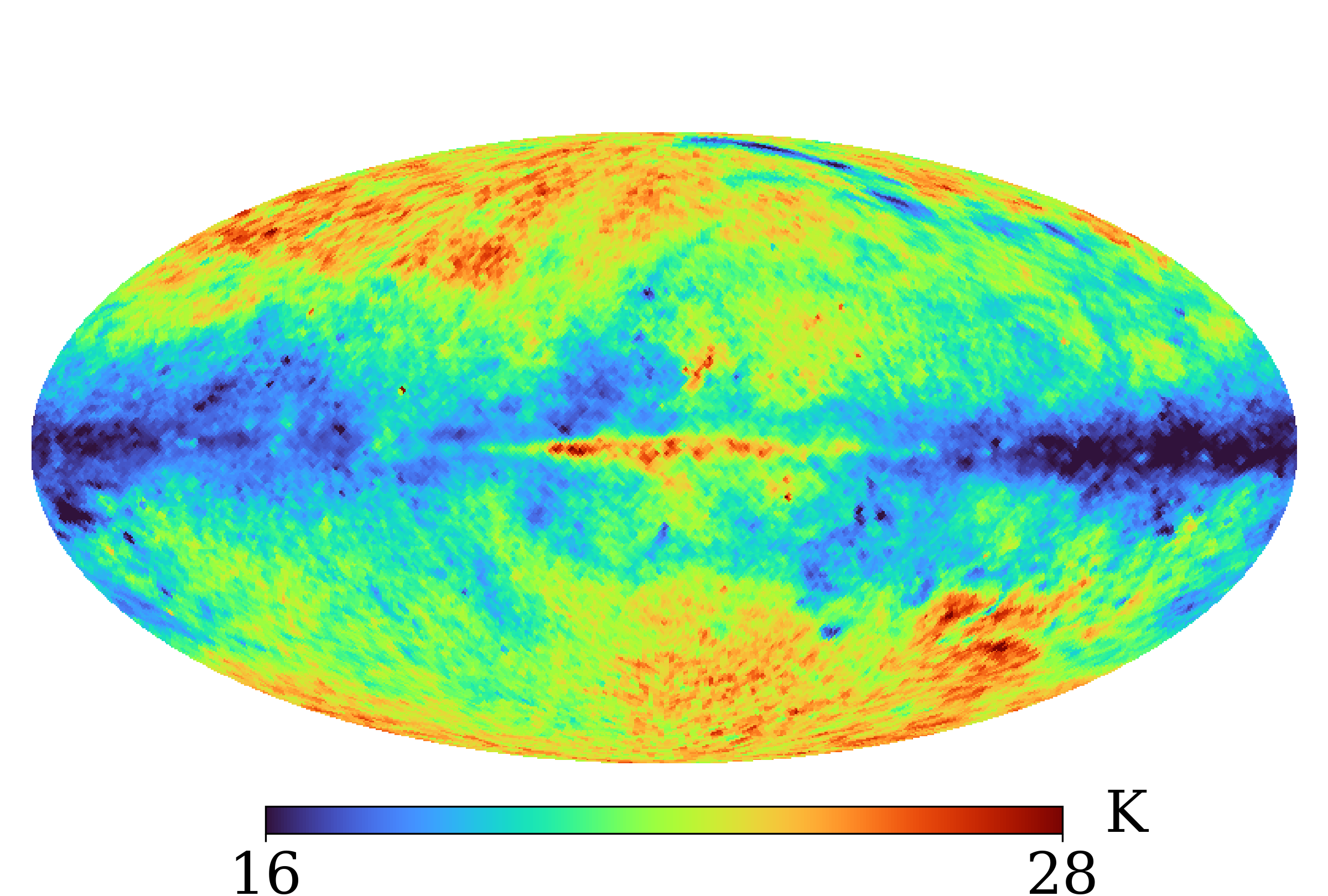}}
\subfigure[$\beta_{\mathrm{s}}$]{\includegraphics[width=0.33\textwidth]{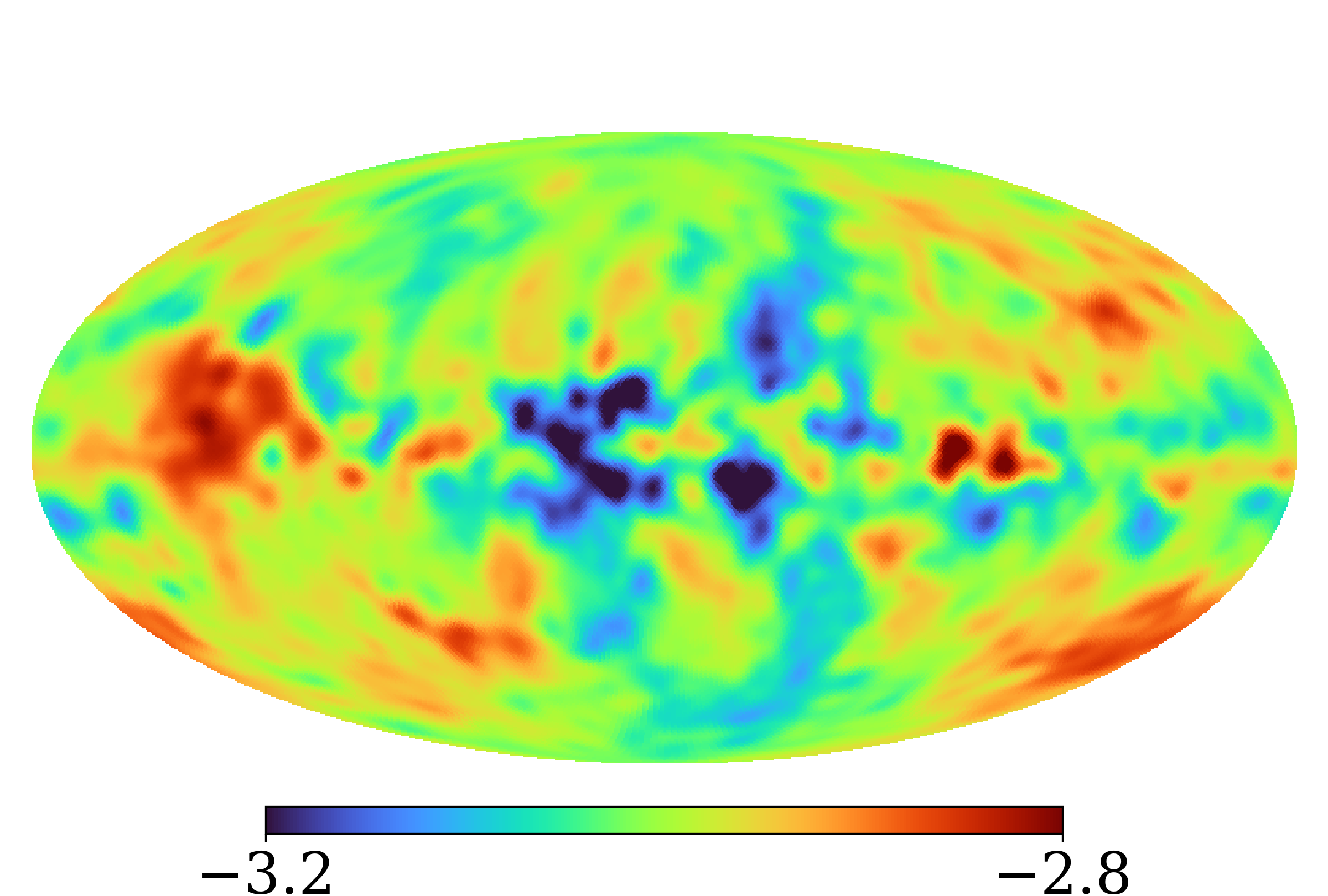}}
\caption{Input component parameter maps used in our simulations. The synchrotron and dust amplitude maps are shown here at reference frequencies of $40\,\mathrm{GHz}$ and $402\,\mathrm{GHz}$ respectively. When performing the component separation with C-BASS and \textit{LiteBIRD} simulations, we set the synchrotron reference frequency to $\nu_{0}=5\,\mathrm{GHz}$.}
\label{fig: input maps}
\end{figure*}

The CMB map was generated as a realization of lensed CMB power spectra, $\mathbf{C}_{\ell} = (C_{\ell}^{TT}, C_{\ell}^{EE}, C_{\ell}^{BB}, C_{\ell}^{TE})$, calculated using \textsc{CAMB} \citep{2000ApJ...538..473L}. For the $B$-mode simulation we set $r=5\times 10^{-3}$, and the lensing amplitude $A_{L}=1$. In this validation study we do not consider the detailed impact of delensing on recovered $B$-mode estimates. In future work it would be valuable to combine the foreground analysis presented here, with a Bayesian delensing scheme such as in \cite{2020arXiv200200965M}, to form a complete forward model for our cosmological observations. The component amplitude templates and spectral parameter maps used in our simulations are shown in Fig. \ref{fig: input maps}.

\begin{table}
    \centering
    \caption{Frequencies and polarization sensitivities for the C-BASS and \textit{LiteBIRD} experiments, used in our simulations for algorithm validation. Sensitivities are given in CMB thermodynamic temperature units. Simulated maps were smoothed to a common resolution of $70\,\mathrm{arcmin}$, corresponding to the lowest resolution, $40\,\mathrm{GHz}$ \textit{LiteBIRD} channel. This resolution is sufficient for targeting multipoles $\ell\lesssim 200$, i.e., angular scales corresponding to the reionization and recombination peaks of the primordial $B$-mode power spectrum.}
    \label{tab: cbass litebird details}
    \begin{threeparttable}
    \begin{tabular}{lcc}
        \hline
        \hline
         Experiment & Frequency  & Polarization\\
         Name & [GHz] & Sensitivity \\
         & & $[\mu\mbox{K$_{\mathrm{cmb}}$-arcmin}]$ \\
         \hline
         \hline
         C-BASS$^{a}$ & 5 & 4320 \\
         \hline
         \multirow{15}{*}{\textit{LiteBIRD}$^{b}$} & 40 & 27.9 \\
          & 50 & 19.6 \\
          & 60 & 15.6 \\
          & 68 & 12.3 \\
          & 78 & 10.0 \\
          & 89 & 9.4 \\
          & 100 & 7.6 \\
          & 119 & 6.4 \\
          & 140 & 5.1 \\
          & 166 & 7.0 \\
          & 195 & 5.8 \\
          & 235 & 8.0 \\
          & 280 & 9.1 \\
          & 337 & 11.4 \\
          & 402 & 19.6 \\
         \hline
         \hline
    \end{tabular}
    \begin{tablenotes}
    \item $^{a}$\cite{2018MNRAS.480.3224J}
    \item $^{b}$\cite{2018SPIE10698E..1YS} 
    \end{tablenotes}
    \end{threeparttable}
\end{table}

\section{Mean-shift clustering of sky regions}\label{sec: clustering}

The modelling of spectral parameters in CMB component separation presents a number of challenges. In the face of limited data it can prove difficult to properly constrain spectral parameters. Attempting to allow full pixel-by-pixel variations in the spectral parameters in this situation can result in significant increases in post component separation noise and in the prior dominating the posterior. Given this, one may seek to reduce the number of degrees of freedom in the sky model by fitting for global spectral parameters. However, this approach will inevitably lead to modelling errors that have the potential to bias cosmological measurements made with the derived CMB map \citep{2019arXiv190508888T}. These challenges have motivated modelling approaches where spatially uniform spectral parameters are assumed over a set of defined sky regions. These have included regions defined as super-pixels on low \textsc{Nside} \textsc{Healpix} maps, and regions defined according to similarities in their spectral properties \citep{2005ApJ...622..759G, 2009MNRAS.392..216S, 2011PhRvD..84f3005E, 2016PhRvD..94h3526S, 2017PhRvD..95d3504A, 2019A&A...623A..21I, 2019JCAP...02..039K, 2019arXiv190508888T}. A detailed discussion of the modelling approaches that can be adopted for spectral parameters in these sky regions is given in Sections \ref{subsec: const spectral} and \ref{subsec: hierarchical}. In this section we describe the mean-shift clustering algorithm, implemented in \textsc{Scikit-Learn} \citep{Comaniciu2002, Pedregosa:2011:SML:1953048.2078195}, that we have used to define regions on the sky for our component separation analyses. The mean-shift clustering algorithm has previously been used in \cite{2020MNRAS.495..578J} to identify pixels with good detections of the synchrotron spectral index.

We use the mean-shift clustering algorithm to construct sky regions according to their location on the sky and their spectral properties. Specifically, we cluster according to the Cartesian coordinates of pixel centres on the unit sphere, $(x^{1}, x^{2}, x^{3})$, and the naive spectral indices between the C-BASS $5\,\mathrm{GHz}$ and \textit{LiteBIRD} $40\,\mathrm{GHz}$ channels, and the \textit{LiteBIRD} $337\,\mathrm{GHz}$ and $402\,\mathrm{GHz}$ channels in polarized intensity. The two sets of frequency maps are used as synchrotron and thermal dust tracers respectively, with the naive spectral indices in a pixel, $p$, being given by,
\begin{equation}
    \beta^{i,j}_{p} = \frac{\ln(m_{i}(p)/m_{j}(p))}{\ln{(\nu_{i}/\nu_{j})}},
\end{equation}
where $m_{i}(p)$ and $m_{j}(p)$ are the map values in the pixel $p$, at the frequencies $\nu_{i}$ and $\nu_{j}$ respectively. Given these parameters we may form the feature vector,
\begin{equation}
    \mathbfit{z}_p = (x^{1}_{p}/\zeta, x^{2}_{p}/\zeta, x^{3}_{p}/\zeta, \beta^{5,40}_{p}, \beta^{337,402}_{p}),
\end{equation}
where $\zeta$ is a spatial vector scaling factor. Setting the value of $\zeta$ to be less than $1$ allows us to preferentially weight proximity on the sky as being favourable over proximity in spectral index space. The function of the mean-shift algorithm is then to cluster points in this five dimensional feature space.

The mean-shift algorithm proceeds by assigning a walker to each pixel, giving us the starting vectors, $\mathbfit{z}^{0}_{p}=\mathbfit{z}_{p}$. The walkers then step through the feature space towards regions of higher density, with the $t^{\mathrm{th}}$ update being calculated as,
\begin{equation}
    \mathbfit{z}^{t}_{p} = \mathbfit{z}^{t-1}_{p} + \mathbfit{s}(\mathbfit{z}^{t-1}_{p}),
\end{equation}
where $\mathbfit{s}$ is the mean-shift vector, given by,
\begin{equation}
    \mathbfit{s}(\mathbfit{z}^{t}_{p}) = \frac{\sum_{q}K(\mathbfit{z}^{t}_{q} - \mathbfit{z}^{t}_{p})\mathbfit{z}^{t}_{q}}{\sum_{q}K(\mathbfit{z}^{t}_{q} - \mathbfit{z}^{t}_{p})}.
\end{equation}
For our purposes we choose the kernel, $K$, to be a top-hat defined by,
\begin{equation}
    K(\mathbf{\Delta}) = 
    \begin{cases}
    1,& \vert \mathbf{\Delta} \vert\leq \omega, \\
    0,& \vert \mathbf{\Delta} \vert>\omega,
    \end{cases}
\end{equation}
where $\omega$ is the bandwidth parameter. Walkers take steps until they converge i.e., walker positions in feature space no longer change with new updates (up to some threshold). Regions are then defined as a set of pixels whose walkers have converged on the same position in the 5-dimensional feature space. Any regions containing fewer pixels than some arbitrary minimum are reassigned to the nearest region in feature space containing a sufficient number of pixels.

\begin{figure*}
    \centering
    \includegraphics[width=0.9\textwidth]{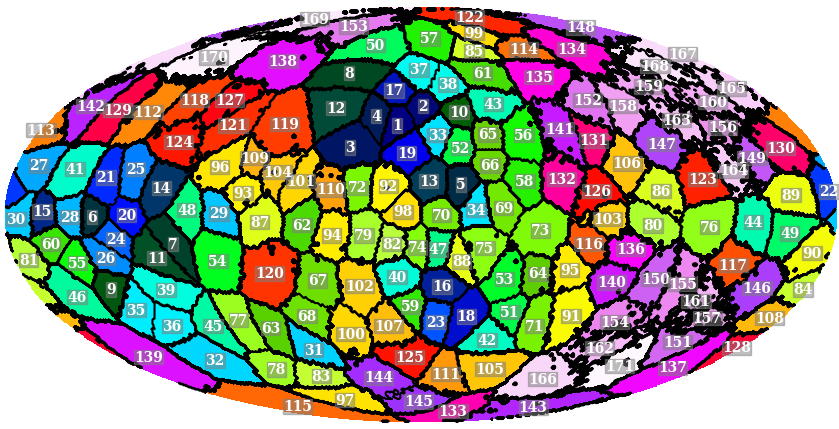}
    \caption{Regions defined using the mean-shift clustering algorithm, produced using \textsc{Nside}=64 maps. We set $\zeta=0.5$, $\omega=0.3$ and the minimum number of pixels in a region to 10. Given our set-up, we obtain 171 regions. The smallest region contains 138 pixels, and the largest region contains 645 pixels. The mean number of pixels is $\langle N_{\mathrm{pix}}\rangle\approx 290$, and the standard deviation of the number of pixels is $\sigma(N_{\mathrm{pix}})\approx 95$. In regions of low SNR we can see that the region borders become less smooth, due to the naive spectral indices used to cluster pixels on the sky becoming contaminated by noise.}
    \label{fig: regions}
\end{figure*}

In Fig. \ref{fig: regions} we show the regions obtained using the frequency channels outlined above, which are used as the region definitions for the component separation analyses in this paper. We used the simulated maps at a \textsc{Healpix} \textsc{Nside} of 64, setting $\zeta=0.5$, $\omega=0.3$ and the minimum number of pixels in a region to 10 \citep{2005ApJ...622..759G}. Using the foreground tracers and parameters described here, we obtain 171 regions. The smallest region on the sky contains 138 pixels, and the largest region contains 645 pixels. The mean number of pixels in a region is $\langle N_{\mathrm{pix}}\rangle\approx 290$, and the standard deviation in the number of pixels is $\sigma(N_{\mathrm{pix}})\approx 95$. In regions of low signal-to-noise ratio (SNR) the borders of regions become less smooth. This can be mitigated to some extent by defining regions on lower \textsc{Nside} maps, at the cost of the larger pixel size meaning regions become more coarse. Using lower \textsc{Nside} maps can lead to a significant degradation in computational performance when assuming foreground spectral parameters in a given region are related in some manner.

A range of possible, non-trivial, extensions to the clustering algorithm exist that could help to alleviate some of the issues surrounding region definition in areas of the sky with low SNR. Instead of using naive spectral indices as a tracer of the spectral properties of diffuse emission over the sky, a more sophisticated estimation of the spectral indices could be performed, accounting for noise properties across the sky. The spectral index estimation could also be improved with better tracers of synchrotron and thermal dust emission. It is worth noting here that the use of the C-BASS map as a synchrotron template is particularly important. If instead we had only used low-frequency \textit{LiteBIRD} channels as our synchrotron tracers the naive synchrotron spectral index estimates would have been heavily noise dominated. Using the methods outlined here, we found it would only possible to define very coarse regions on \textsc{Nside}=8 maps.

\section{Parametric Bayesian CMB component separation}\label{sec: comp sep nuts}

We have developed a new implementation of Bayesian pixel-by-pixel CMB component separation, using the NUTS algorithm to explore our parameter space \citep{Homan:2014:NSA:2627435.2638586}. The primary benefit in using NUTS to sample from the target distribution is in its avoidance of the random walk behaviour that slows more conventional sampling algorithms such as Metropolis-Hastings and Gibbs sampling. The component separation code is written in the \textsc{Python} programming language, with the NUTS algorithm being implemented through the \textsc{PyMC3} library \citep{pymc3_citation}. In its current form, the whole-sky component separation is parallelized over the sky regions defined using the mean-shift clustering algorithm. For our validation purposes in this paper, we do not consider monopole and dipole corrections, or instrumental factors such as colour corrections in our modelling. The extension of our modelling to include such complications is left to future work. 

Bench-marking the algorithm performance against the rate of effective sample generation, NUTS offers potential performance improvements of $\sim 10^{3}$ compared to sampling with Metropolis-Hastings\footnote{These simple bench-marking tests were performed on a single Intel Xeon CPU, running at $\sim 2.6\,\mathrm{GHz}$.}. Close to the Galactic plane sampling becomes more difficult, with the CMB completely sub-dominant to foregrounds. In this situation computational performance can be degraded such that the sampler exhibits undesirable random walk behaviour. By parallelizing over sky regions, and masking the most contaminated sky regions close to the Galactic plane, it should be possible to achieve rapid convergence. We also note that these bench-marking tests have been performed without extensive optimization of the component separation code. By exploring re-parametrizations, model prior choice, optimizing sampling parameters etc. it is likely that we would be able to achieve further performance improvements. It would also be worthwhile considering the potential for GPU acceleration with \textsc{PyMC3}. For the hierarchical model, the posterior exhibits geometrical pathologies that make sampling with non-gradient based algorithms essentially intractable \citep{hmc_hierarchical2013}. However, using NUTS we are able to achieve comparable computational performance to the complete pooling model. 

The outline of this section is as follows: In Section \ref{subsec: likelihood} we describe the general sky model and likelihood used in our component separation analysis. In Section \ref{subsec: nuts algo} we give an overview of the NUTS algorithm. In Section \ref{subsec: const spectral} we discuss the complete pooling model, where we fit for constant spectral parameters over sky regions. In Section \ref{subsec: hierarchical} we discuss our hierarchical modelling approach, where we directly fit for the underlying hyper-distributions of the spectral parameters. In Section \ref{subsec: convergence} we describe the convergence checks and diagnostics performed during sampling.  

\subsection{Sky model and likelihood}\label{subsec: likelihood}

In attempting to observe the CMB we actually observe multiple sky components. It is the goal of our component separation to extract the CMB from these additional confusing components. For a given sky pixel, $p$, we may write the observed value in that pixel as,
\begin{equation}
    d_{p,\lambda}(\nu) = s_{p,\lambda}(\nu) + n_{p,\lambda}(\nu),
\end{equation}
where $\nu$ is the observing frequency, $\lambda=\left\{I,Q,U\right\}$ represents one of the Stokes parameters, $s_{p,\lambda}(\nu)$ is the true sky signal and $n_{p,\lambda}(\nu)$ is the noise term. For our analysis in this paper we restrict ourselves to $\lambda=\left\{Q,U\right\}$.

The true sky signal includes contributions from diffuse emission, compact sources, line emission etc.  For our current modelling purposes we focus on diffuse emission, as discussed in Section \ref{sec: comp models}. Additional contributions from compact sources, line emission etc. are important considerations when performing component separation on real experimental data. A variety of strategies exist for mitigating their contribution, e.g. through masking point sources, and direct modelling during the fitting process. However, this sits beyond the scope of our algorithm validation.  

The noise term consists largely of contributions from instrumental white noise and $1/f$ noise, which acts to introduce large-scale correlated noise in the sky maps. For ground-based experiments mitigating atmospheric noise is a significant challenge, largely limiting accessible angular scales to $\ell\gtrsim 30$. The ability of such experiments to accurately recover low multipoles will be vital for future $B$-mode experiments \citep{2017PhRvD..95d3504A}. 

Given a set of sky maps at frequencies $\mathbf{\nu}=\left[\nu_{1},\ldots,\nu_{N}\right]$, the Gaussian likelihood for a pixel, $p$, is given by,
\begin{multline}
    -\ln{\mathcal{L}_{p,\lambda}(\mathbf{\Theta}_{p,\lambda})} = \sum_{i}\frac{1}{2}\ln{\left(2\pi\sigma_{p,\lambda}^{2}(\nu_{i})\right)} + \\ \frac{\left(d_{p,\lambda}(\nu_{i}) - s_{p,\lambda}(\nu_{i}, \mathbf{\Theta}_{p,\lambda})\right)^{2}}{2\sigma_{p,\lambda}^{2}(\nu_{i})},
    \label{eq: likelihood}
\end{multline}
where $\mathbf{\Theta}_{p,\lambda}$ are the model parameters, and $\sigma_{p,\lambda}$ is the pixel noise. We assume for simplicity that the noise is independent between pixels and frequency channels. This is not necessarily optimal when we consider the complications described above. However, the approximation is sufficient for the purposes of algorithm validation.

\subsection{The No-U-Turn Sampler}\label{subsec: nuts algo}

The key measure of the efficiency of a sampling algorithm is in its ability to produce effective/independent samples, or equivalently reduce the correlation between samples. Indeed, this is where NUTS significantly out-performs standard Metropolis-Hastings and Gibbs sampling algorithms. Even though an individual step in these simpler sampling algorithms is less computationally expensive, their random-walk behaviour results in highly correlated samples and hence very inefficient generation of effective samples. At its core, the NUTS algorithm is an extension of HMC, which was originally developed for performing calculations in lattice field theory \citep{DUANE1987216}. In this section we give an overview of the HMC algorithm, and the additional tuning procedures NUTS implements to avoid the need for hand-tuned HMC implementations. For a detailed discussion of NUTS and HMC see \cite{Homan:2014:NSA:2627435.2638586,doi:10.1111/2041-210X.12681,2017arXiv170102434B,betancourt2017}.

HMC essentially proceeds by generating physical trajectories through parameter space, akin to simulating particle trajectories through a potential. A simple implementation of HMC may proceed as follows:
\begin{enumerate}
	\item Given a set of parameters, $\mathbf{\Theta}=\left[\Theta_{1},\ldots,\Theta_{\mathrm{d}}\right]$, with corresponding joint density, $p\left(\mathbf{\Theta}\right)$, we introduce a set of auxiliary momentum variables, $\mathbfit{r}=\left[r_{1},\ldots,r_{\mathrm{d}}\right]$. We take the distribution over the momenta to be a Gaussian distribution centered on zero, i.e., $\mathbfit{r}\sim\mathcal{N}(\mathbf{0},\mathbfss{M})$, where $\mathbfss{M}$ is the mass matrix. This defines a kinetic energy term,
	\begin{equation}
	    K(\mathbf{\Theta},\mathbfit{r}) = -\ln p(\mathbfit{r}|\mathbf{\Theta}) = \frac{1}{2}\mathbfit{r}^{\top}\mathbfss{M}^{-1}\mathbfit{r} + \ln{\left|\mathbfss{M}\right|} + \mathrm{const.}
	\end{equation}
	We may then define the Hamiltonian of our system as,
	\begin{equation}
	    \mathcal{H}(\mathbfit{r},\mathbf{\Theta}) = -\ln p(\mathbfit{r}|\mathbf{\Theta}) - \ln p\left(\mathbf{\Theta}\right) = K(\mathbf{\Theta},\mathbfit{r}) + V\left(\mathbf{\Theta}\right),
	\end{equation}
	where we define the potential energy term, $V\left(\mathbf{\Theta}\right) = - \ln p\left(\mathbf{\Theta}\right)$.
    \item We then evolve our position in parameter space by integrating Hamilton's equations,
    \begin{align}
        \frac{\mathrm{d}\mathbf{\Theta}}{\mathrm{d}t} &= \frac{\partial\mathcal{H}}{\partial\mathbfit{r}},\\
        \frac{\mathrm{d}\mathbfit{r}}{\mathrm{d}t} &= -\frac{\partial\mathcal{H}}{\partial\mathbf{\Theta}}.
    \end{align}
    Practically this is done through a leapfrog algorithm. To generate a new sample in our Markov chain we draw $\mathbfit{r}$ from $\mathcal{N}(\mathbf{0}, \mathbfss{M})$. The leapfrog steps then proceed as,
    \begin{align}
	&\mathbfit{r}^{t+\epsilon/2} = \mathbfit{r}^{t} - \frac{\epsilon}{2}\nabla_{\mathbf{\Theta}}V(\mathbf{\Theta}^{t}),\\
	&\mathbf{\Theta}^{t+\epsilon} = \mathbf{\Theta}^{t} + \epsilon\mathbfss{M}\mathbfit{r}^{t+\epsilon/2},\\
	&\mathbfit{r}^{t+\epsilon} = \mathbfit{r}^{t+\epsilon/2}-\frac{\epsilon}{2}\nabla_{\mathbf{\Theta}}V(\mathbf{\Theta}^{t+\epsilon}),
	\end{align}
	where $\epsilon$ is the leapfrog step-size. The leapfrog steps used to update our position have the convenient property of being a symplectic integrator. That is, the numerical trajectory generated by the leapfrog steps preserve the volume of phase space, as is the case for the Hamiltonian trajectories they approximate. A more detailed discussion of the numerical integration of Hamilton's equations can be found in \cite{Leimkuhler:835066}. These leapfrog steps are performed $L$ times to generate a new proposal position, $\left(\mathbf{r}^*,\mathbf{\Theta}^*\right)$.
	\item The new proposal position is then accepted with a probability of acceptance given by,
	\begin{equation}
	    p_{\alpha} = \mathrm{min}\left\{1, \exp\left(\mathcal{H}(\mathbfit{r},\mathbf{\Theta}) - \mathcal{H}(\mathbfit{r}^*,\mathbf{\Theta}^*)\right)\right\}.
	\end{equation}
	We note that what we have done here is essentially generate a Metropolis-Hastings proposal step with a very high chance of being accepted.
	\item By repeating this sampling procedure $N$ times we may generate the parameter samples for our Markov chain.
\end{enumerate}

The mass matrix used to define the distribution over the momenta acts to rotate and re-scale parameter space. Choosing $\mathbfss{M}^{-1}$ to be the covariance of the target distribution will help to de-correlate the target distribution, which can lead to significant performance improvements when dealing with highly correlated parameters. For practical implementations the mass matrix can be estimated during a tuning phase. Starting with the identity matrix we can generate an initial sample set, from which we update our estimate of the mass matrix using the sample covariance. We may then iterate over this tuning process to generate an accurate estimate of the covariance of the target distribution \citep{2017arXiv170102434B}. Whilst estimating off-diagonal elements of the mass matrix does help in de-correlating the target distribution, using the off-diagonal elements does not necessarily scale well to high-dimensional problems given the need to invert the mass matrix at the end of tuning, and perform matrix multiplications during leapfrog steps. For our analysis in this paper we only tune diagonal elements of the mass matrix. However, this can significantly improve sampling efficiency for single pixel analyses, or analyses assuming complete independence between pixels. In this case, one can employ the tuning steps in \cite{exoplanet:exoplanet}, using the default tuning schedule described in \cite{stan_development_team_stan_2012}.

The efficiency of HMC as described above critically depends on the choice of $\epsilon$ and $L$ used in the leapfrog steps. If $\epsilon$ is chosen to be too small, the sampler will waste computation taking very small steps along the Hamiltonian trajectories. In contrast, if $\epsilon$ is chosen to be too large, the simulation of the Hamiltonian trajectory will become inaccurate and the sampler will produce proposal steps with low acceptance probabilities. If $L$ is chosen to be too small, the sampler will generate samples close to one another, resulting in undesirable random walk behaviour. If $L$ is chosen to be too large on the other hand, the sampler will generate paths through parameter space that loop back on themselves. This results in proposal steps close to the starting value, with the additional waste of generating the extended trajectory. In even more severe scenarios, a poor choice of $L$ that results in the sampler jumping from one side of parameter space to another at each iteration can result in a non-ergodic chain i.e. a chain that is not guaranteed to converge on the target distribution \citep{neal_mcmc_hmc2012}. 

The need to finely tune $\epsilon$ and $L$ means that standard implementations of HMC typically require costly tuning runs. This can significantly reduce the utility and general applicability of HMC in realistic problems. The NUTS algorithm overcomes these problems by automatically tuning these sampling parameters. The value of $\epsilon$ is tuned during an initial tuning phase to meet some target acceptance probability. The target acceptance probability can be adjusted depending on the degree of curvature in the posterior, with a higher acceptance rate (or equivalently, smaller step-size) being needed for highly curved distributions. The value of $L$ is modified during sampling to meet a No-U-Turn criterion. That is, the leapfrog integrator is iterated over until the simulated trajectory begins to turn back on itself, or some maximum number of leapfrog simulations are performed. In doing so, the sampler is able to maximize the distance between the proposal step and the initial position, before looping back on itself and wasting computation. Details on these tuning procedures can be found in \cite{Homan:2014:NSA:2627435.2638586,stan_development_team_stan_2012,pymc3_citation}.

\subsection{Complete pooling of spectral parameters}\label{subsec: const spectral}

As discussed in Section \ref{sec: clustering}, in the face of limited data and low SNR, allowing spectral parameters to vary completely from pixel-to-pixel is sub-optimal, resulting in increased levels of post component separation noise and the posterior potentially becoming prior dominated. As a first alternative to allowing full pixel-by-pixel variations we may instead assume spectral parameters to be constant over the regions defined as in Section \ref{sec: clustering} i.e., we assume a complete pooling of the spectral parameters.

The priors used in the complete pooling model are given in Table \ref{tab: complete pooling priors}. We assign informative Normal priors to the spectral parameters. The standard deviations on the priors for $\beta_{\mathrm{s}}$ and $\beta_{\mathrm{d}}$ are chosen to be $0.3$, corresponding to the bandwidth used in clustering sky regions and encompassing most of the range over which these parameters have been measured \citep{2016A&A...594A..10P, 2018A&A...618A.166K}. For the dust temperature we set a prior based on constraints on the dust temperature found in the \textit{Planck} \textsc{Commander} analysis \citep{2016A&A...594A...9P, 2016A&A...594A..10P}. These help to down-weight the more extreme regions of parameter space, offering significant computational performance improvements, and helping to regularize the posterior by reducing the biasing effect of the probability mass associated with extreme parameter values. Detailed discussion of prior choice, in particular around the use of weakly informative priors, can be found in \cite{gelman2006,gelman2008,evans2011,polson2012,doi:10.1111/rssa.12276,simpson2017,2017Entrp..19..555G}. In addition to the Normal priors on the spectral parameters we multiply these by the associated Jeffreys priors as in the \textit{Planck} \textsc{Commander} analysis \citep{1946RSPSA.186..453J, Jeffreys61, 0004-637X-676-1-10, 2014A&A...571A..12P, 2016A&A...594A...9P, 2016A&A...594A..10P}.

\begin{table}
    \centering
    \caption{Priors for parameters in the complete pooling model. We assign informative Normal priors to the spectral parameters, and flat priors to the amplitude parameters. In addition to the Normal priors for the spectral parameters, we multiply these by their associated Jeffreys priors as in the \textit{Planck} \textsc{Commander} analysis. Spectral parameters are assumed to be constant over a given region, whilst amplitude parameters are allowed to vary within each pixel, $p$. We restrict our analysis to polarization, so that $\lambda=\{Q,U\}$.}
    \begin{tabular}{ll}
    \hline
    \hline
        $\Theta$ & $p(\Theta)$ \\
    \hline
    \hline
         $\beta_{\mathrm{s}}$ & $\mathcal{N}(\mu=-3, \sigma=0.3)$\\
         $\beta_{\mathrm{d}}$ & $\mathcal{N}(\mu=1.6, \sigma=0.3)$\\
         $T_{\mathrm{d}}$ & $\mathcal{N}(\mu=21.0, \sigma=2.0)$\\
         $A_{\mathrm{s}}^{p, \lambda}$  & $\mathrm{Unif}(-\infty, \infty)$\\
         $A_{\mathrm{d}}^{p, \lambda}$  & $\mathrm{Unif}(-\infty, \infty)$\\
         $A_{\mathrm{cmb}}^{p, \lambda}$  & $\mathrm{Unif}(-\infty, \infty)$\\
    \hline
    \hline
    \end{tabular}
    \label{tab: complete pooling priors}
\end{table}

Our emission model for a pixel, $p$ in some sky region is given by,
\begin{equation}
    s_{p,\lambda}(\nu) = A_{\mathrm{s}}^{p, \lambda}f_{\mathrm{s}}(\nu,\beta_{\mathrm{s}})+A_{\mathrm{d}}^{p, \lambda}f_{\mathrm{d}}(\nu,\beta_{\mathrm{d}},T_{\mathrm{d}})+A_{\mathrm{cmb}}^{p, \lambda}f_{\mathrm{cmb}}(\nu).
    \label{eq: emission model}
\end{equation}
The functions, $f_{\mathrm{s}}(\nu,\beta_{\mathrm{s}})$, $f_{\mathrm{d}}(\nu,\beta_{\mathrm{d}},T_{\mathrm{d}})$ and $f_{\mathrm{cmb}}(\nu)$ are the spectral forms of the synchrotron, dust and CMB components, as defined in Section \ref{sec: comp models}. Note that we assume the spectral parameters to be identical for $\lambda=\{Q,U\}$. We model our data as being Normally distributed i.e., we assume the Gaussian likelihood in Equation \ref{eq: likelihood}. 

Complete pooling offers a potentially effective approach to account for the spatial variation in spectral parameters whilst avoiding the generation of excessive post component separation noise. However, with additional data points at low and/or high frequencies it is possible to adopt a more sophisticated, hierarchical model of the spectral parameters in these regions. It is worth noting that for the complete pooling model one can analytically marginalize over the amplitude parameters as in \cite{2017PhRvD..95d3504A}. This greatly reduces the dimension of parameter space and hence improves the sampling efficiency. We have not implemented sampling of this marginal distribution for our analysis here, where we study the computational performance of NUTS in sampling the full posterior. Indeed, the intrinsic efficiency of NUTS makes this unnecessary.

\subsection{Hierarchical modelling of spectral parameters}\label{subsec: hierarchical}

In statistical modelling, we often encounter scenarios where our model contains a set of latent variables that are related in some way. In such a scenario it is neither ideal to treat the latent variables as being entirely independent, or to simply fit for a single, global variable. Instead we can take a hierarchical approach. In a hierarchical Bayesian model we introduce a set of population-level hyper-parameters, which define the distribution from which our individual latent variables are drawn \citep{doi:10.1198/004017005000000661, GelmanHill:2007}. 

In the context of CMB component separation, we may model the pixel spectral parameters within our regions as being drawn from some underlying hyper-distributions. In our particular case we model the spectral parameters as being drawn from underlying Normal distributions, parametrized by the hyper-parameters, $\{(\mu_{\beta_{\mathrm{s}}},\sigma_{\beta_{\mathrm{s}}}), (\mu_{\beta_{\mathrm{d}}},\sigma_{\beta_{\mathrm{d}}}), (\mu_{T_{\mathrm{d}}},\sigma_{T_{\mathrm{d}}})\}$. Each pair corresponds to the mean and standard deviation of the underlying Normal hyper-distribution for the synchrotron spectral index, the dust spectral index and the dust temperature respectively. During component separation we jointly fit for the population-level hyper-parameters and the associated pixel-level spectral parameters. The hierarchical approach allows us to model the pixel-level variations in the spectral parameters, with the hyper-distributions reducing the propensity of the model to overreact to noise, as would be the case if we assumed total independence between pixel-level spectral parameters \citep{KATAHIRA201637}.  

Our emission model takes the same form as in Equation \ref{eq: emission model}, and we again assume the Gaussian likelihood in Equation \ref{eq: likelihood}. The priors for our hierarchical model are listed in Table \ref{tab: hierarchical priors}. Analogously to the complete pooling model, we set informative priors on the spectral hyper-parameters. For the means of the hyper-distributions we set the same Normal priors as for the spectral parameters in the complete pooling model. For the standard deviations of the hyper-distributions we set Half-Normal priors, with scale parameters set to correspond to the standard deviations of the mean priors. The Half-Normal prior constrains the standard deviations to be positive, with the scale parameters chosen to encapsulate the likely degree of variation of spectral parameters in a given region. Setting informative priors on the hyper-parameters in a hierarchical model can be particularly important in ensuring the robust computational performance of the sampling algorithm. Hyper-parameters are highly correlated with the associated pixel-level parameters, and small changes in the values of the hyper-parameters can induce large changes in the target distribution. This can result in funnel-like geometries in the posterior when the data is limited i.e., a region of high density but low volume below a region of low density but high volume. The funnel regions are highly curved, which can lead to major computational difficulties during sampling, in the worst case leading to a failure in geometric ergodicity. This problem can be partly mitigated through setting informative priors that down-weight more extreme parameter values as we have done here \citep{gelman2006, hmc_hierarchical2013}. 

\begin{table}
    \centering
    \caption{Priors for parameters in the hierarchical model. For the means of the spectral hyper-distributions we assign Normal priors, and for the standard deviations of the hyper-distributions we assign Half-Normal priors. The priors for the pixel-level spectral parameters are then defined through their conditional dependence on the hyper-parameters. As with the complete pooling model, we assign flat priors to the amplitude parameters and allow them to vary from pixel to pixel. We restrict our analysis to polarization.}
    \begin{tabular}{ll}
    \hline
    \hline
        $\Theta$ & $p(\Theta)$ \\
    \hline
    \hline
         $\mu_{\beta_{\mathrm{s}}}$ & $\mathcal{N}(\mu=-3, \sigma=0.3)$\\
         $\sigma_{\beta_{\mathrm{s}}}$ & $\mbox{Half-Normal}(\sigma=0.35)$\\
         $\mu_{\beta_{\mathrm{d}}}$ & $\mathcal{N}(\mu=1.6, \sigma=0.3)$\\
         $\sigma_{\beta_{\mathrm{d}}}$ & $\mbox{Half-Normal}(\sigma=0.35)$\\
         $\mu_{T_{\mathrm{d}}}$ & $\mathcal{N}(\mu=21, \sigma=2)$\\
         $\sigma_{T_{\mathrm{d}}}$ & $\mbox{Half-Normal}(\sigma=2.5)$\\
         $\beta_{\mathrm{s}}^{p}$ & $\mathcal{N}(\mu=\mu_{\beta_{\mathrm{s}}}, \sigma=\sigma_{\beta_{\mathrm{s}}})$\\
         $\beta_{\mathrm{d}}^{p}$ & $\mathcal{N}(\mu=\mu_{\beta_{\mathrm{d}}}, \sigma=\sigma_{\beta_{\mathrm{d}}})$\\
         $T_{\mathrm{d}}^{p}$ & $\mathcal{N}(\mu=\mu_{T_{\mathrm{d}}}, \sigma=\sigma_{T_{\mathrm{d}}})$\\
         $A_{\mathrm{s}}^{p, \lambda}$  & $\mathrm{Unif}(-\infty, \infty)$\\
         $A_{\mathrm{d}}^{p, \lambda}$  & $\mathrm{Unif}(-\infty, \infty)$\\
         $A_{\mathrm{cmb}}^{p, \lambda}$  & $\mathrm{Unif}(-\infty, \infty)$\\
    \hline
    \hline
    \end{tabular}
    \label{tab: hierarchical priors}
\end{table}

In addition to our choice of informative priors, we re-parametrize our spectral parameters by introducing the auxiliary variables,
\begin{equation}
    \Gamma_{c}^{p} \sim \mathcal{N}(0, 1), c=\{\mathrm{s},\mathrm{d},\mathrm{cmb}\}. 
\end{equation}
In the case of the synchrotron spectral index we may re-express $\beta_{\mathrm{s}}$ as,
\begin{equation}
    \beta_{\mathrm{s}}^{p} = \mu_{\beta_{\mathrm{s}}} + \Gamma_{\mathrm{s}}^{p}\sigma_{\beta_{\mathrm{s}}},
\end{equation}
with analogous expressions for $\beta_{\mathrm{d}}^{p}$ and $T_{\mathrm{d}}^{p}$. Thus, instead of directly sampling the $\{\beta_{\mathrm{s}}^{p}, \beta_{\mathrm{d}}^{p}, T_{\mathrm{d}}^{p}\}$, we instead sample a set of Gaussian latent variables and obtain the pixel-by-pixel spectral parameters through a translation and scaling with the hyper-parameters. This is known as the non-centred paramterization and has the convenient effect of reducing correlations between the hyper-parameters and the pixel-level spectral parameters. A detailed discussion of the geometrical pathologies of hierarchical models and practical approaches to their mitigation can be found in \cite{hmc_hierarchical2013}. 

\subsection{Convergence Checks}\label{subsec: convergence}

Given an infinite number of samples it can be shown the the NUTS algorithm will converge on the target distribution. However, it remains important to perform a number of checks to reassure ourselves of convergence after a finite number of samples. To this end, we output a number of convergence diagnostics that we describe below.

The first covergence diagnostic we output is the Gelman-Rubin statistic \citep{1992StaSc...7..457G,doi:10.1080/10618600.1998.10474787}. This compares the variance between multiple, independently initialized chains with the variance within each chain, and is defined as,
\begin{equation}
	\hat{R} = \frac{\hat{V}}{\hat{W}},
\end{equation}
where $\hat{V}$ is the between-chain variance and $\hat{W}$ is the within chain variance. If convergence has been achieved the between-chain and within-chain variance will be equal. In reality we apply the threshold, $\hat{R}\leq 1.1$ to reassure ourselves that our chains satisfy the necessary geometric ergodicity conditions.

We also output the number of effective samples in each chain, $n_{\mathrm{eff}}$ \citep{geyer1992,brooks2011handbook}. When sampling from a target distribution using some MCMC algorithm, we may draw a total of $N$ samples, but these samples are not totally independent. The effective sample size provides a measure of the number of independent samples in a chain, defined as,
\begin{equation}
	n_{\mathrm{eff}} \equiv \frac{N}{\sum_{t=-\infty}^{\infty}\rho_{t}}=\frac{N}{1+2\sum_{t=1}^{\infty}\rho_{t}},
\end{equation}
where $\rho_{t}$ is the auto-correlation within a chain at a lag $t$. Details on the estimation of the auto-correlation can be found in \cite{stan_development_team_stan_2012}\footnote{It is worth noting that, if a chain is estimated to have a negative auto-correlation between samples, one can obtain $n_{\mathrm{eff}}>N$.}. The appropriate number of effective samples to be able to properly capture the target distribution is to some extent a question of judgement. However, in \cite{kruschke2011doing} a threshold of $\sim 1000$ effective samples is proposed to be confident in expectations calculated with parameter chains. As such, we adopt this as a confidence threshold for our sampling output. 

Finally, we also output warnings when divergences occur during sampling. A divergence takes place when the sampler encounters a region of the target distribution where the curvature is too high to be resolved given the tuned step-size. In practice, divergences are detected when the value of the Hamiltonian diverges from its initial value when simulating trajectories through parameter space. This is significant in that divergences can mean that the conditions for geometric ergodicity are not met, and therefore using the resultant chains to construct statistical estimators can lead to biased inferences \citep{hmc_hierarchical2013,2016arXiv160400695B,2017arXiv170102434B}.

\section{Algorithm Validation}\label{sec: validation}

We validate the component separation algorithm's performance against the simulated experimental observations described in Section \ref{sec: simulations}. For ease of discussion, we assign the following labels to our three validation sets (i.e., simulation and modelling runs):
\begin{itemize}
    \item CP(L): The \textit{LiteBIRD} only analysis, fitting the complete pooling model described in Section \ref{subsec: const spectral}.
    \item CP(LC): The C-BASS and \textit{LiteBIRD} analysis, fitting the complete pooling model.
    \item H(LC): The C-BASS and \textit{LiteBIRD} analysis, fitting the hierarchical model described in Section \ref{subsec: hierarchical}.
\end{itemize}
In all three validation sets we use regions defined on \textsc{Nside}=64 maps, as described in Section \ref{sec: clustering}, for our spectral modelling. We note here that we do not fit the hierarchical model to the simulation set consisting of just \textit{LiteBIRD} observations. It was found to be very challenging to control for the occurrence of divergences when fitting the hierarchical model to \textit{LiteBIRD}-only observations, leaving the convergence properties of the resulting MCMC chains suspect. These problems around controlling divergences can be understood when we consider the lack of low-frequency channels in \textit{LiteBIRD}. Given limited available information to constrain synchrotron spectral parameters, the posterior geometry for the hierarchical model becomes extremely difficult to sample. Applying the hierarchical model to \textit{LiteBIRD}-only observations likely requires a careful study of prior choice for model hyper-parameters and extended tuning phases to help mitigate the occurrence of divergences.

For each validation set, we evaluate the normalized deviations of the recovered parameters. For some parameter $\mathbf{\Theta}$, the normalized deviations are defined as,
\begin{equation}
    \eta_{\lambda} = \frac{\mathbf{\Theta}^{\mathrm{in},\lambda} - \mathbf{\Theta}^{\mathrm{out},\lambda}}{\sigma_{\Theta}^{\lambda}}, \quad \lambda=\{Q,U\},
\end{equation}
where $\mathbf{\Theta}^{\mathrm{out},Q/U}$ is the output parameter $Q/U$ map, $\mathbf{\Theta}^{\mathrm{in},Q/U}$ is the input parameter $Q/U$ map, and $\sigma_{\Theta}^{Q/U}$ is the corresponding $Q/U$ standard deviation map. We note that for spectral parameters the $Q$ and $U$ normalized deviations will be identical. If our observed $\mathbf{\Theta}^{\mathrm{out},Q/U}$ are drawn from a Gaussian distribution with mean given by $\mathbf{\Theta}^{\mathrm{in},Q/U}$ and standard deviation given by $\sigma_{\Theta}^{Q/U}$, the normalized deviations should be distributed as a standard Gaussian, $\mathcal{N}(0,1)$.

\begin{figure*}
    \centering
    \subfigure[$\eta$: CMB amplitude]{\includegraphics[width=0.85\columnwidth]{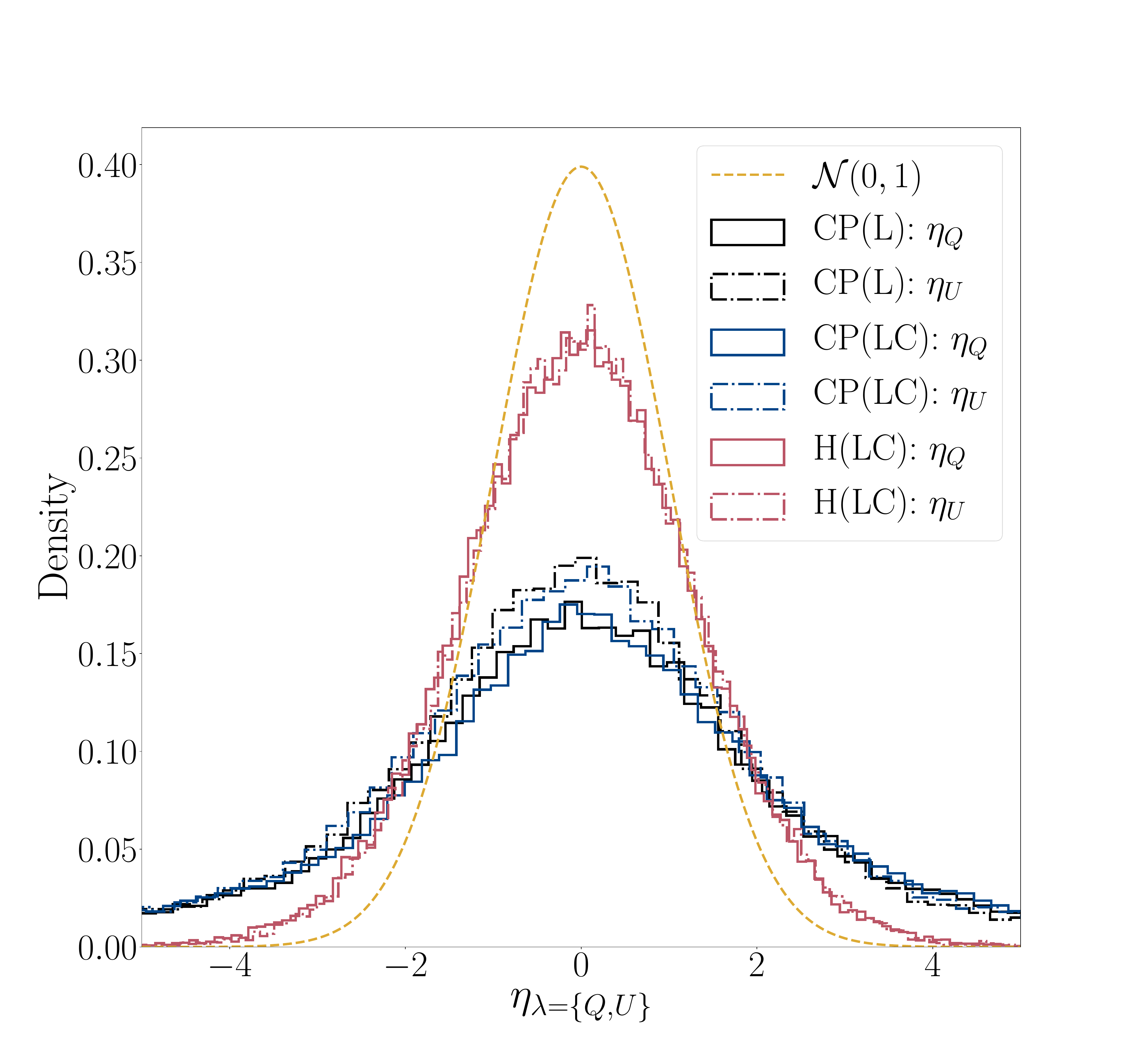} \label{fig: cmb norm dev}}
    \subfigure[$\eta$: Synchrotron amplitude]{\includegraphics[width=0.85\columnwidth]{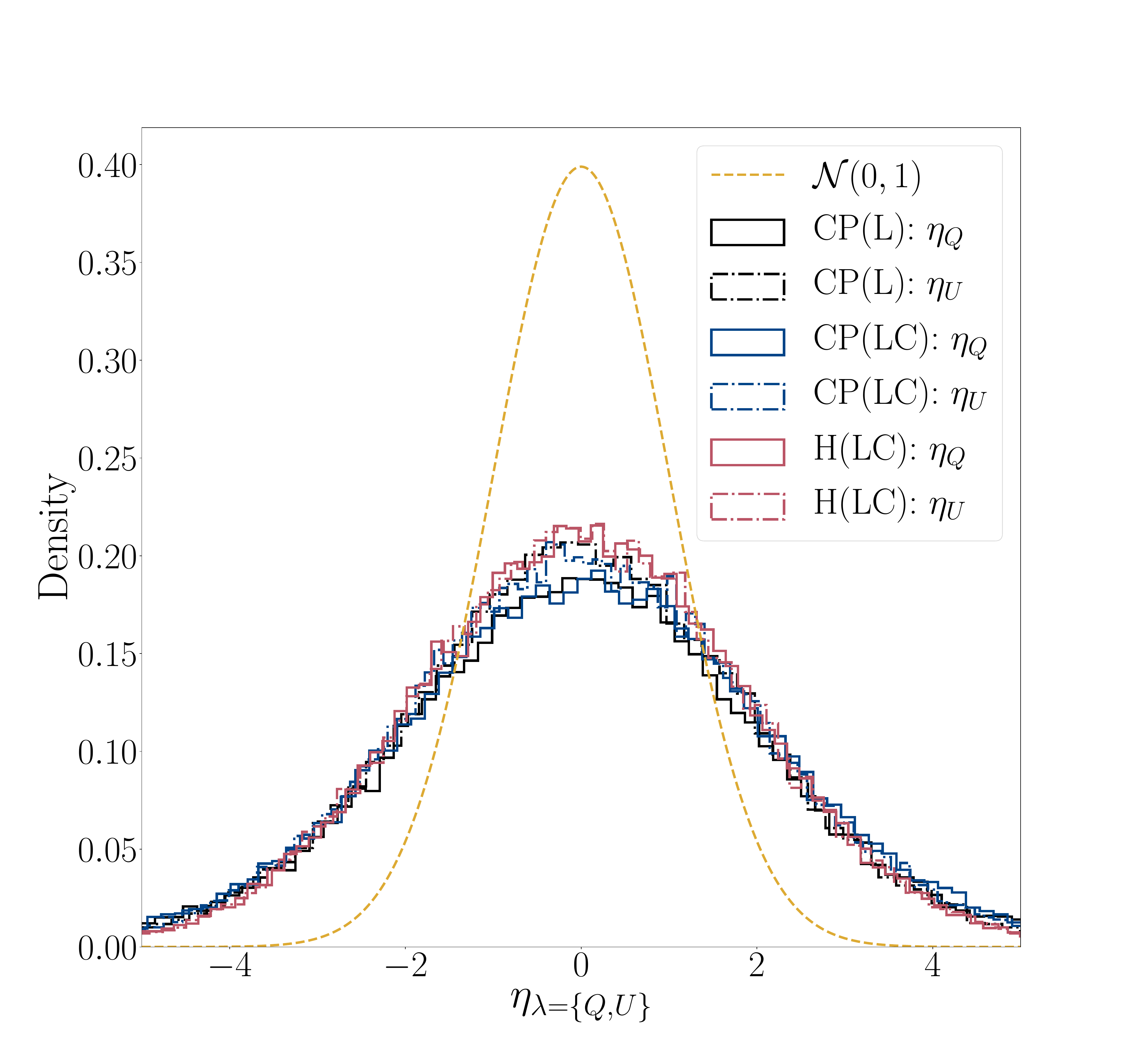} \label{fig: synch amp norm dev}}
    \subfigure[$\eta$: Dust amplitude]{\includegraphics[width=0.85\columnwidth]{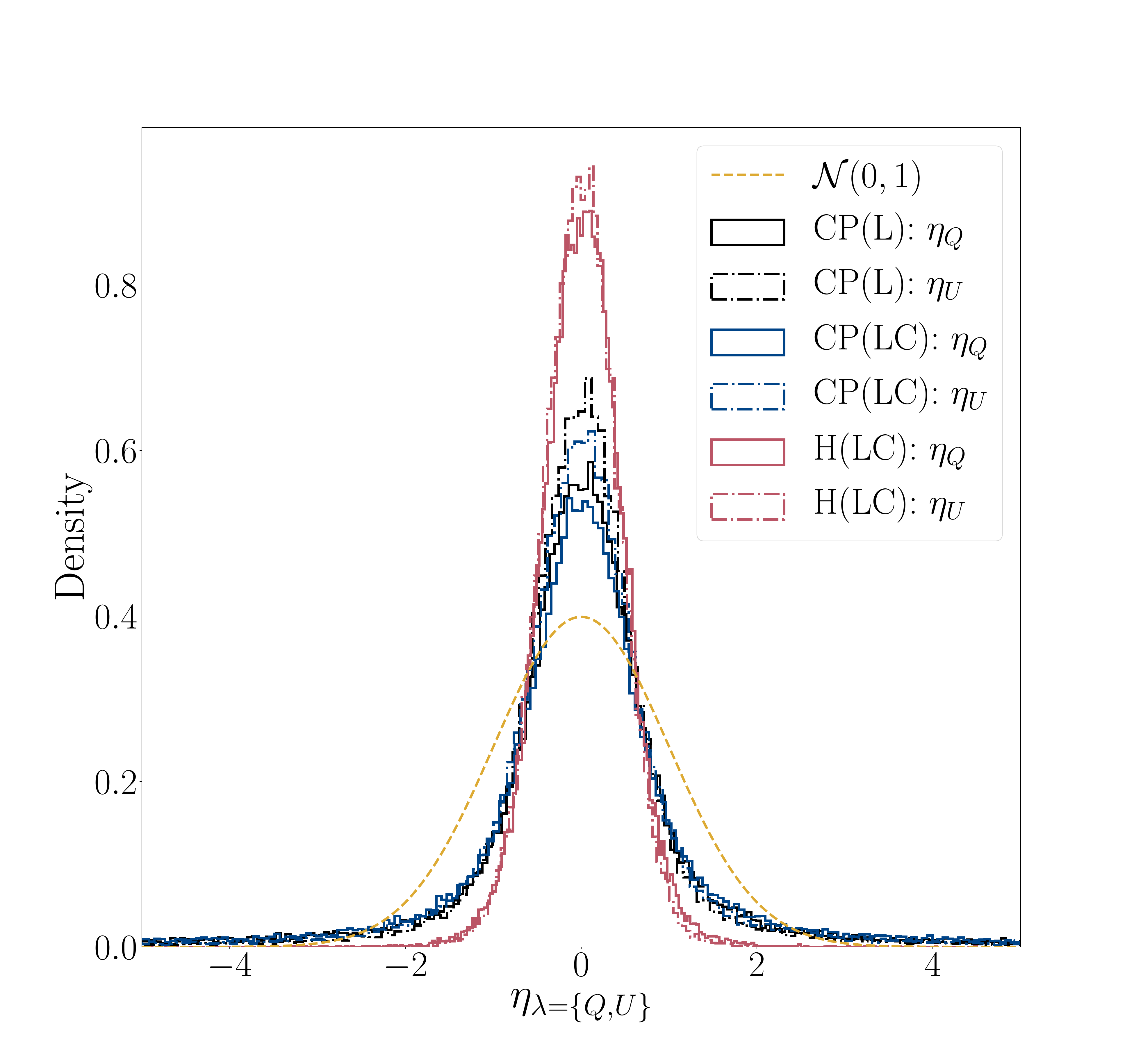} \label{fig: dust amp norm dev}}
    \subfigure[$\eta$: Synchrotron spectral index]{\includegraphics[width=0.85\columnwidth]{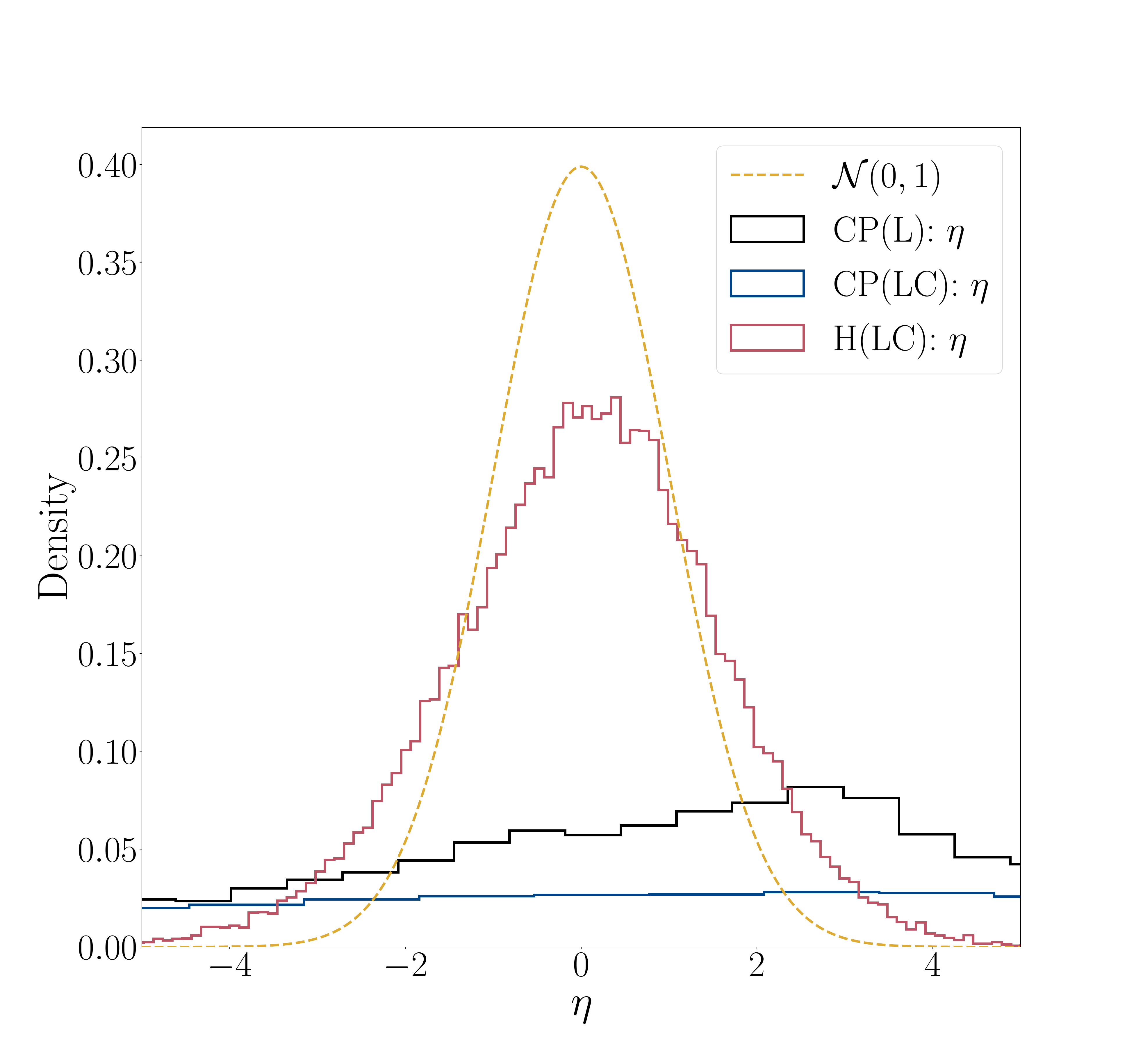} \label{fig: synch beta norm dev}}
    \subfigure[$\eta$: Dust spectral index]{\includegraphics[width=0.85\columnwidth]{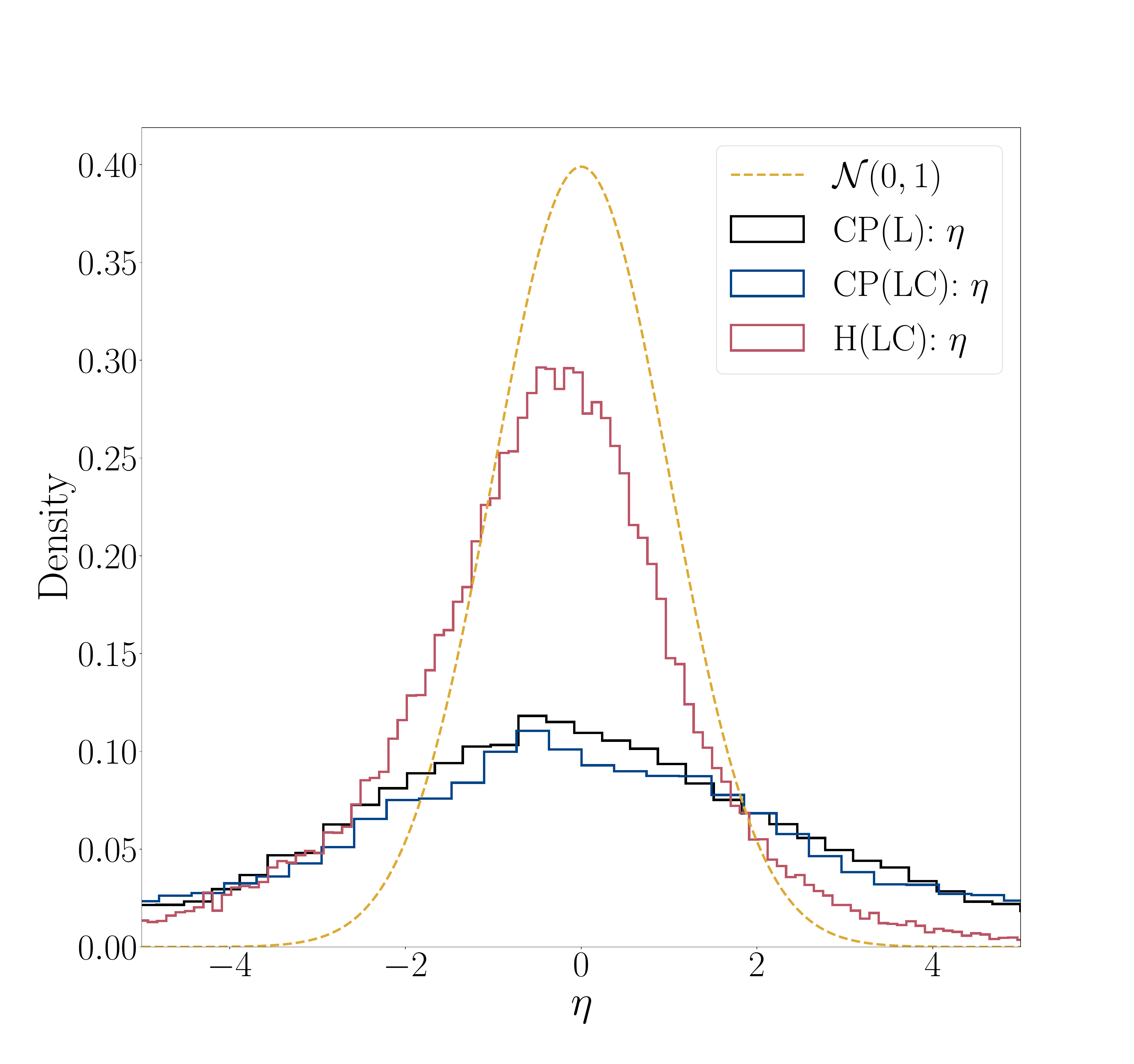} \label{fig: dust beta norm dev}}
    \subfigure[$\eta$: Dust temperature]{\includegraphics[width=0.85\columnwidth]{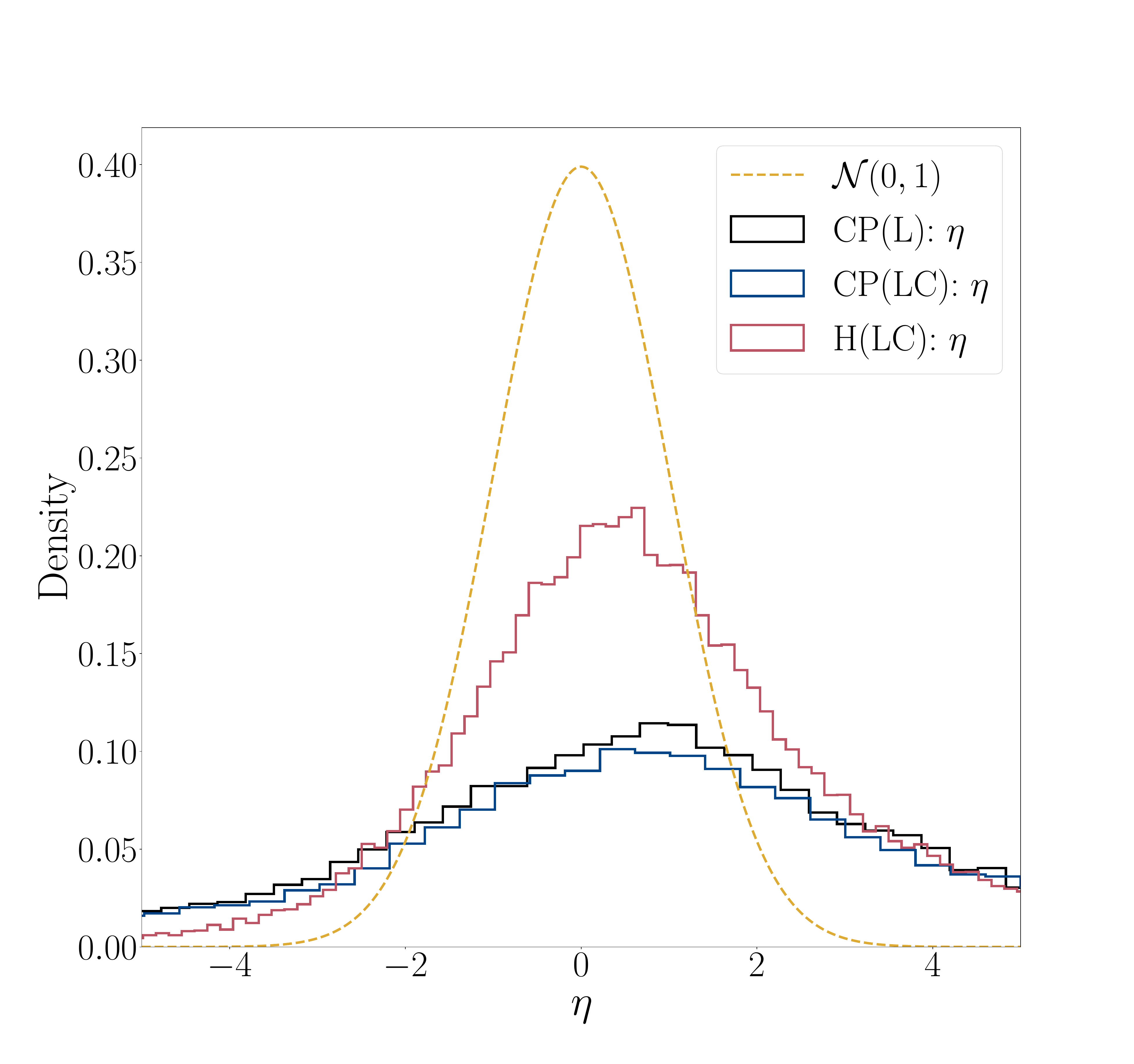} \label{fig: dust temp norm dev}}
    \caption{Histograms of the normalized deviations for our recovered parameter maps, obtained for the three validation sets. Alongside these histograms we plot the standard Gaussian, $\mathcal{N}(0,1)$. Mis-match between the normalized deviation histograms and $\mathcal{N}(0,1)$ indicate departures from Gaussianity in the marginal parameter posteriors. In itself, this is not surprising, given the posterior distributions include contributions from non-Gaussian priors, and in the case of the hierarchical model contains complex correlations between hyper-parameters and pixel-level parameters.}
    \label{fig: norm dev}
\end{figure*}

Histograms of the normalized deviations for the model parameters recovered from each validation set are shown in Fig. \ref{fig: norm dev}. In Table \ref{tab: norm dev} we state the median of the normalized deviations for each parameter and validation set, along with the corresponding median absolute deviation (MAD) values. The distributions of the normalized deviations for each parameter are discussed in the relevant subsections outlined below.

The outline of the remainder of this section is as follows: In Section \ref{subsec: cmb amp val} we discuss the CMB amplitude output, in Section \ref{subsec: fg amp val} we present the synchrotron and dust amplitude constraints, in Section \ref{subsec: synch spectral val} we discuss constraints on the synchrotron spectral index, and in Section \ref{subsec: dust spectral val} we show results for the dust spectral parameters. In Fig. \ref{fig: cmb amp} we show maps of the recovered CMB amplitudes and effective sample size for each validation set. In Fig. \ref{fig: synch dust amp} we show the recovered synchrotron and dust amplitude maps and in Fig. \ref{fig: spectral params} we show the recovered synchrotron and dust spectral parameter maps. In addition to the results presented here, we provide figures showing residual maps for our model parameters as supplementary online material.

\begin{table}
    \centering
    \begin{tabular}{lcccc}
    \hline
    \hline
        Validation & $\mathrm{Med}(\eta_{Q})$ & $\mathrm{Med}(\eta_{U})$ & $\mathrm{MAD}(\eta_{Q})$ & $\mathrm{MAD}(\eta_{U})$ \\
        Set &  &  &  &  \\
    \hline
    \hline
        $A_{\mathrm{cmb}}$: CP(L) & 0.21 & -0.10 & 2.60 & 2.18\\
        $A_{\mathrm{cmb}}$: CP(LC) & 0.12 & -0.02 & 2.66 & 2.25\\
        $A_{\mathrm{cmb}}$: H(LC) & -0.03 & $2\times10^{-3}$ & 1.30 & 1.29\\
        \hline
        $A_{\mathrm{s}}$: CP(L) & $2\times10^{-3}$ & -0.02 & 2.17 & 1.98\\
        $A_{\mathrm{s}}$: CP(LC) & 0.08 & -$2\times10^{-3}$ & 2.14 & 2.00\\
        $A_{\mathrm{s}}$: H(LC) & 0.01 & $1\times10^{-3}$ & 1.89 & 1.88\\
        \hline
        $A_{\mathrm{d}}$: CP(L) & 0.01 & 0.01 & 0.74 & 0.62\\
        $A_{\mathrm{d}}$: CP(LC) & 0.02 & 0.01 & 0.81 & 0.68\\
        $A_{\mathrm{d}}$: H(LC) & 0.03 & $2\times10^{-3}$ & 0.45 & 0.43\\
        \hline
        $\beta_{\mathrm{s}}$: CP(L) & -0.44 & \ldots & 7.05 & \ldots \\
        $\beta_{\mathrm{s}}$: CP(LC) & -1.03 & \ldots & 16.8 & \ldots\\
        $\beta_{\mathrm{s}}$: H(LC) & 0.09 & \ldots & 1.46 & \ldots \\
        \hline
        $\beta_{\mathrm{d}}$: CP(L) & 0.09 & \ldots & 4.18 & \ldots \\
        $\beta_{\mathrm{d}}$: CP(LC) & -0.35 & \ldots & 4.96 & \ldots \\
        $\beta_{\mathrm{d}}$: H(LC) & -0.35 & \ldots & 1.42 & \ldots \\
        \hline
        $T_{\mathrm{d}}$: CP(L) & 0.61 & \ldots & 4.33 & \ldots\\
        $T_{\mathrm{d}}$: CP(LC) & 1.23 & \ldots & 5.11 & \ldots\\
        $T_{\mathrm{d}}$: H(LC) & 0.63 & \ldots & 1.96 & \ldots\\
    \hline
    \hline
    \end{tabular}
    \caption{In columns 2 and 3 we state the medians of the normalized deviations for the recovered model parameters in each validation set, and in columns 4 and 5 we state the corresponding MAD values. For spectral parameters the normalized deviations in $Q$ and $U$ are identical. As such, we only state results for spectral parameters under their corresponding $Q$ columns.}
    \label{tab: norm dev}
\end{table}

\subsection{CMB amplitude}\label{subsec: cmb amp val}

\begin{figure*}
\centering
\subfigure[CP(L): $\mathbfit{A}_{\mathrm{cmb}}^Q$]{\includegraphics[width=0.325\textwidth]{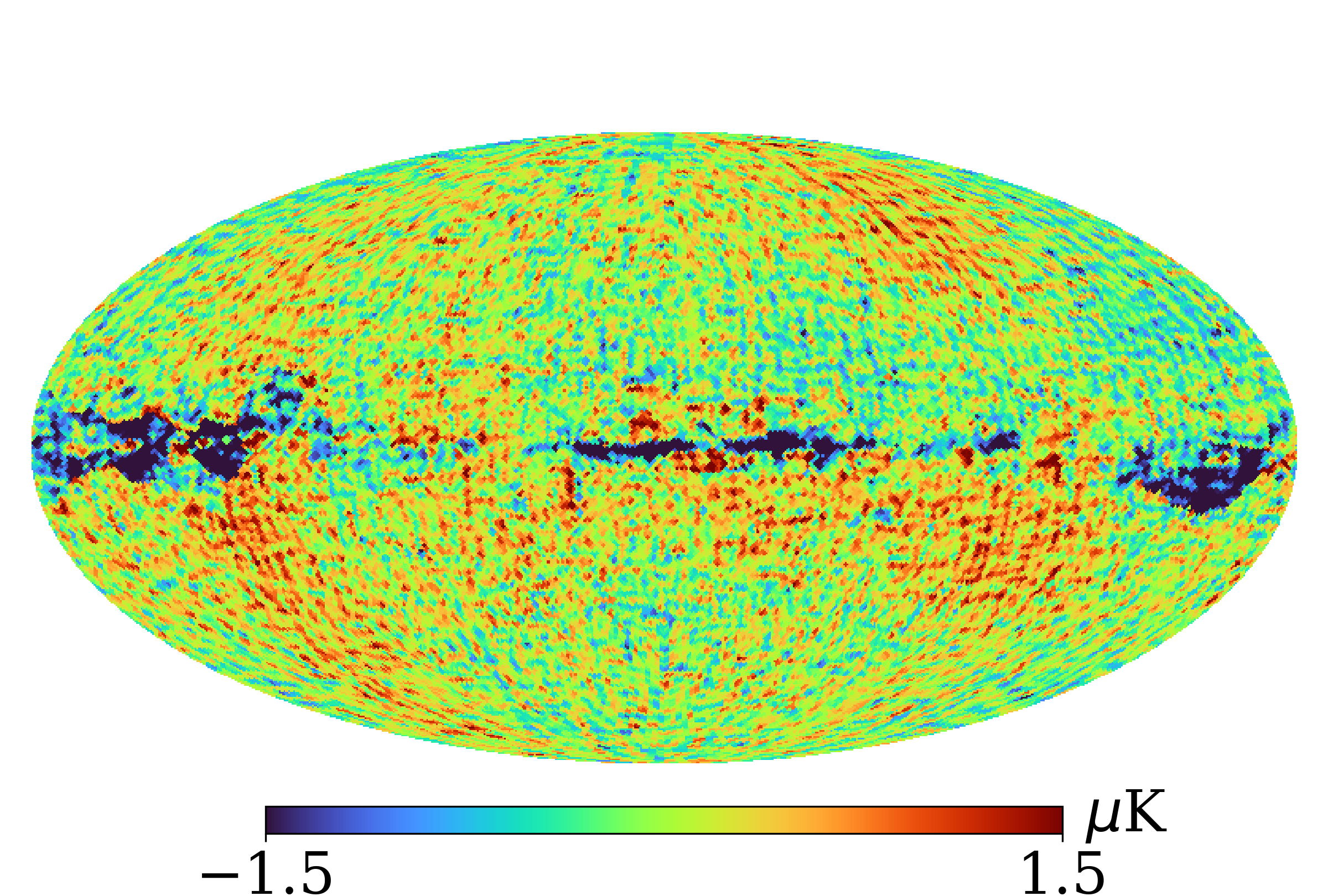} \label{fig: cmb Q CPL}}
\subfigure[CP(LC): $\mathbfit{A}_{\mathrm{cmb}}^Q$]{\includegraphics[width=0.325\textwidth]{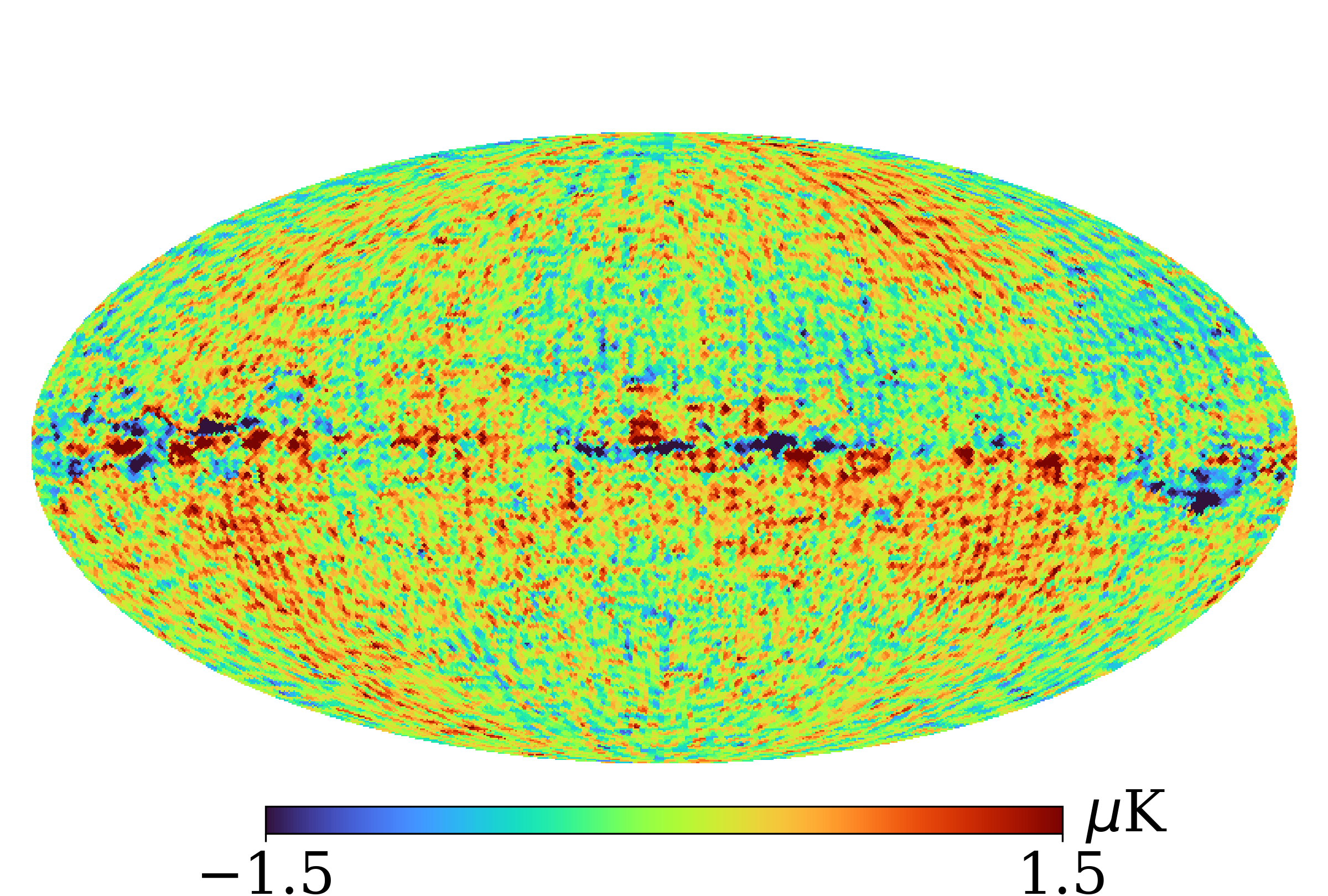} \label{fig: cmb Q CPLC}}
\subfigure[H(LC): $\mathbfit{A}_{\mathrm{cmb}}^Q$]{\includegraphics[width=0.325\textwidth]{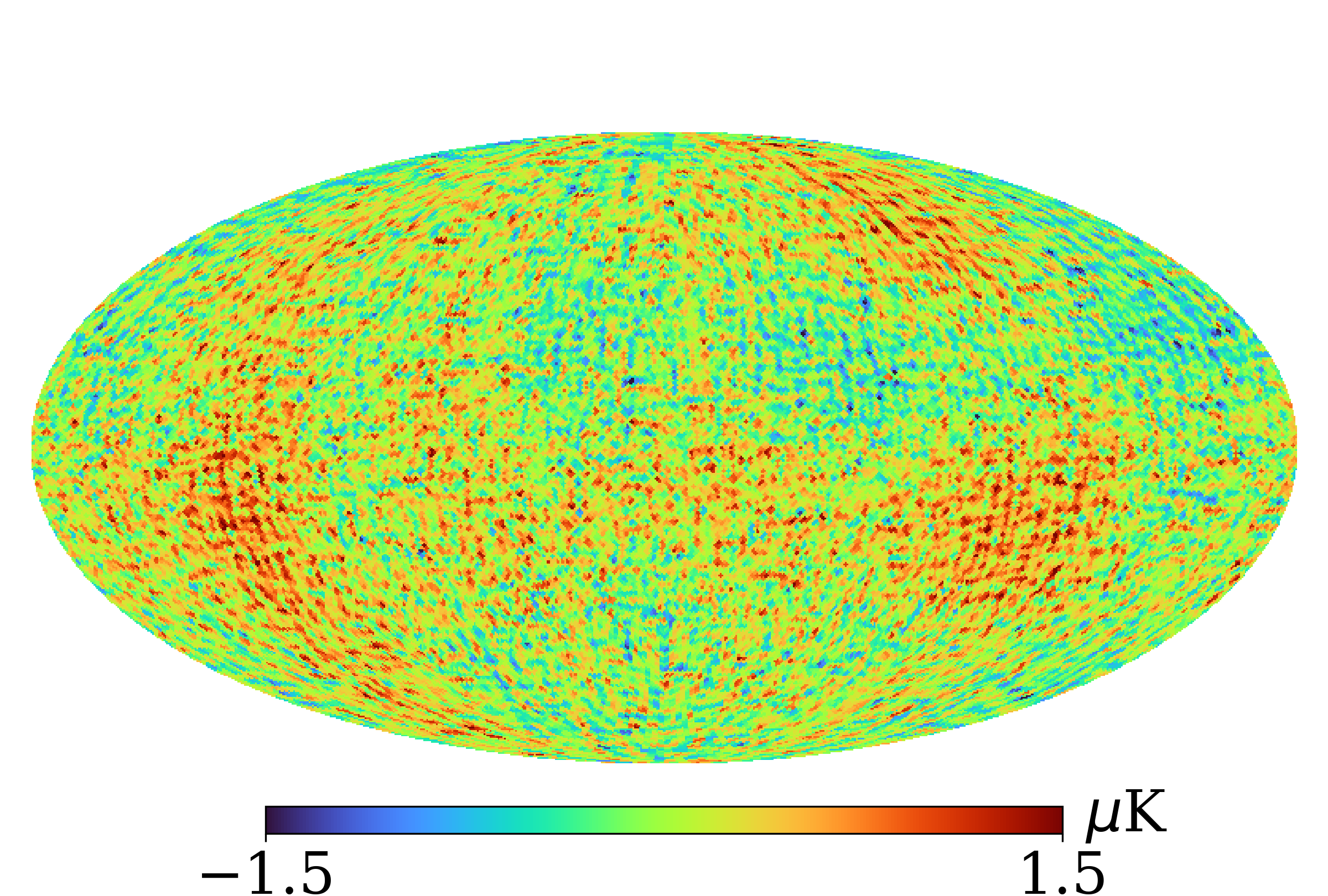} \label{fig: cmb Q HLC}}
\subfigure[CP(L): $\mathbfit{A}_{\mathrm{cmb}}^U$]{\includegraphics[width=0.325\textwidth]{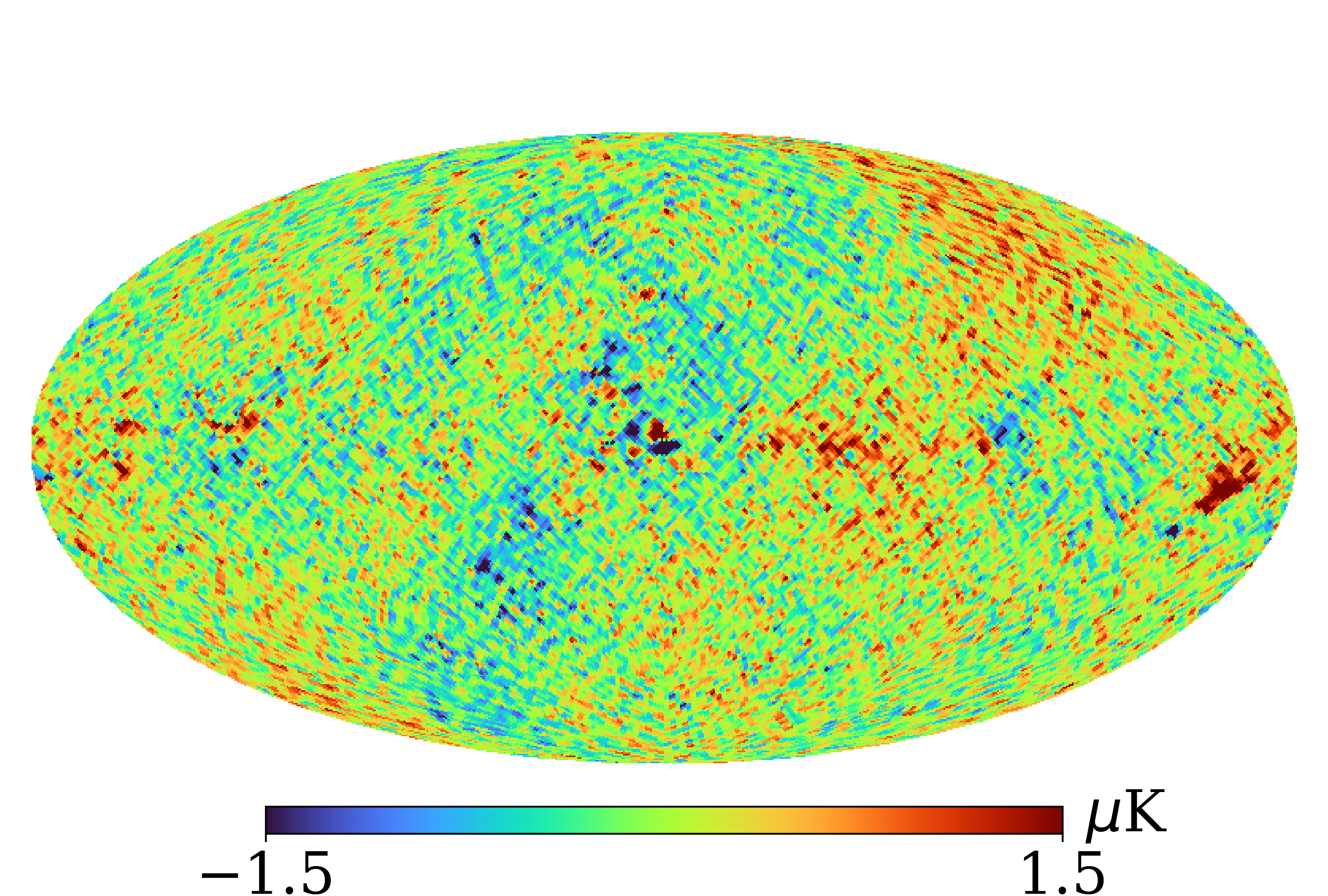} \label{fig: cmb U CPL}}
\subfigure[CP(LC): $\mathbfit{A}_{\mathrm{cmb}}^U$]{\includegraphics[width=0.325\textwidth]{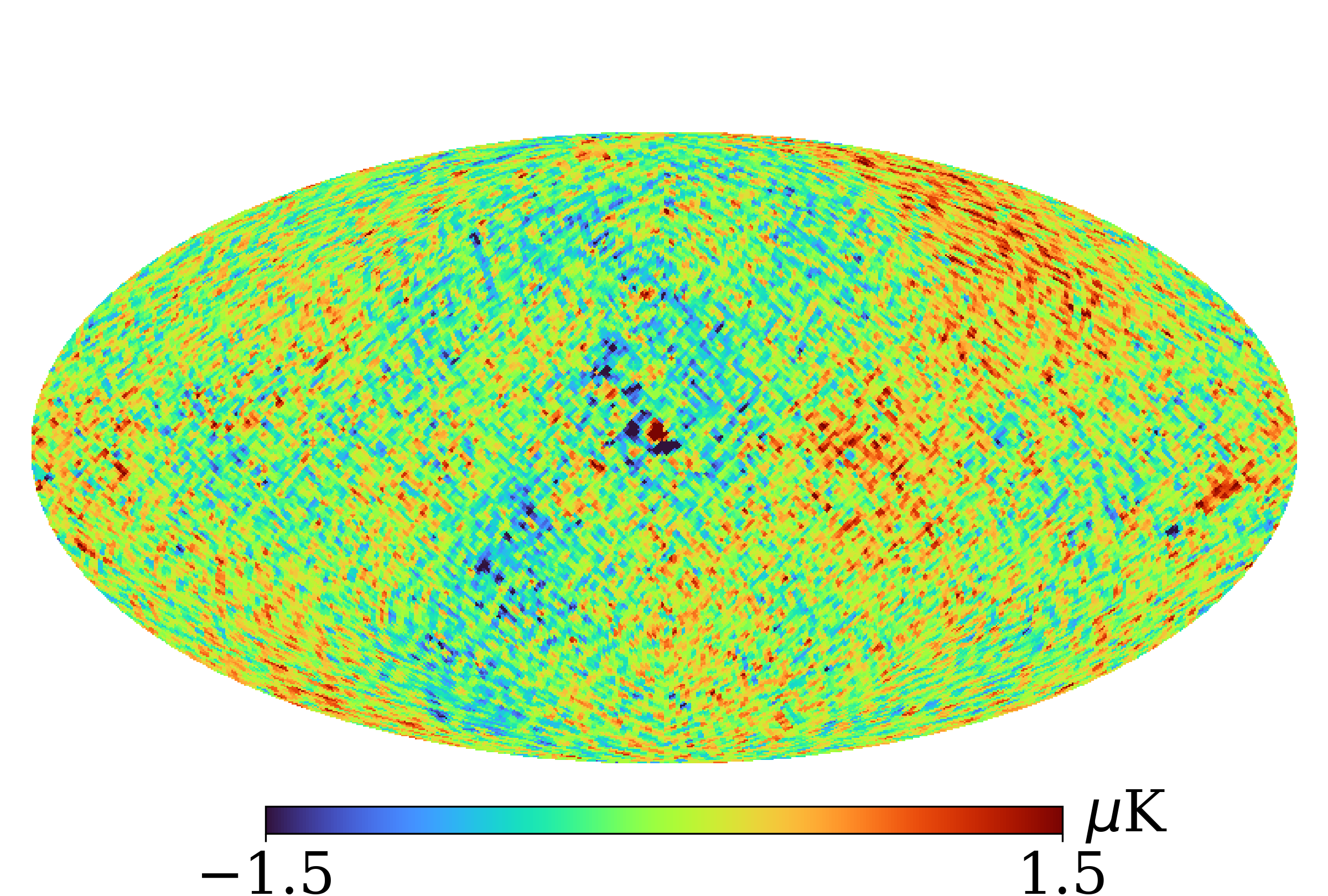} \label{fig: cmb U CPLC}}
\subfigure[H(LC): $\mathbfit{A}_{\mathrm{cmb}}^U$]{\includegraphics[width=0.325\textwidth]{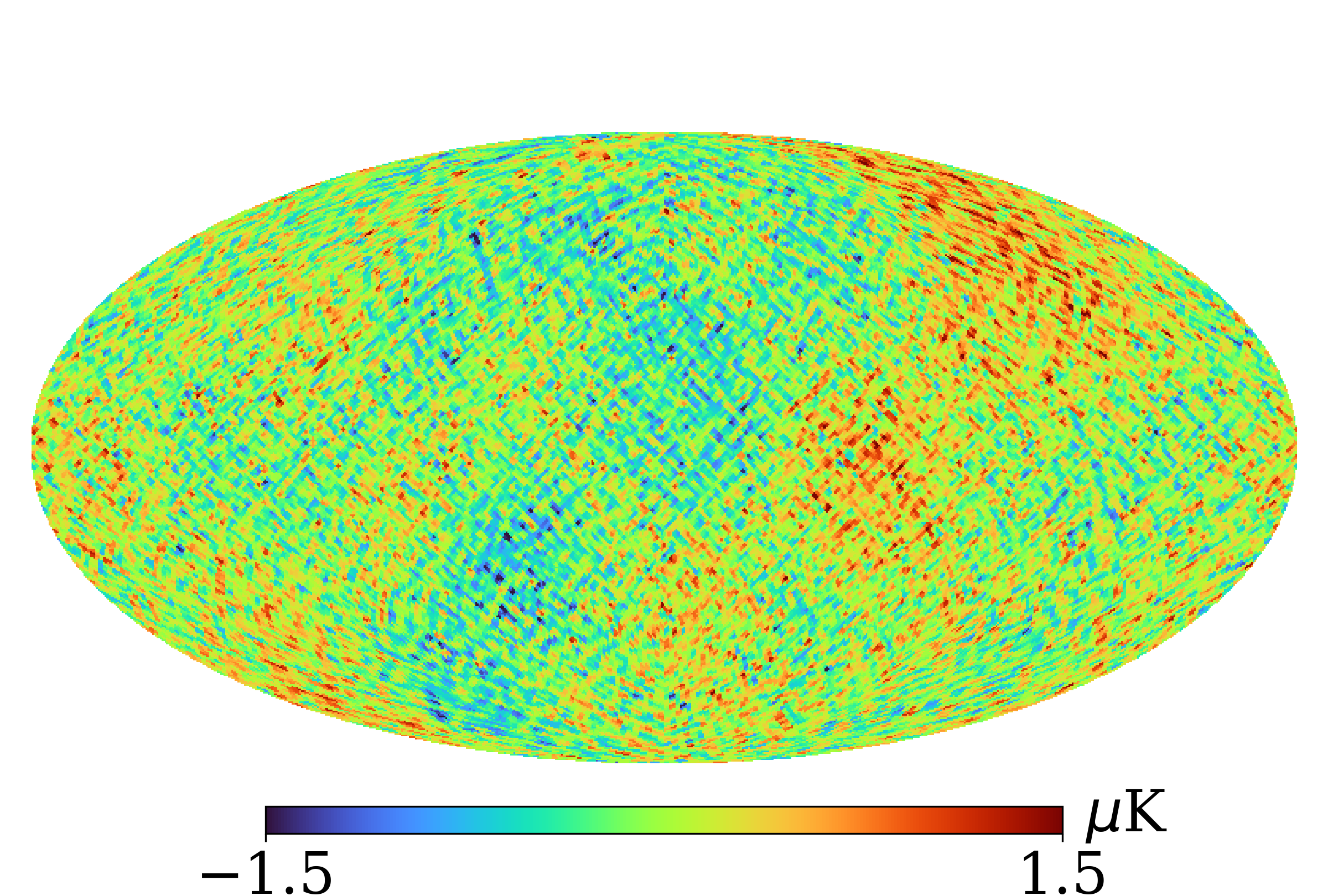} \label{fig: cmb U HLC}}
\subfigure[CP(L): $n_{\mathrm{eff}}^Q$]{\includegraphics[width=0.325\textwidth]{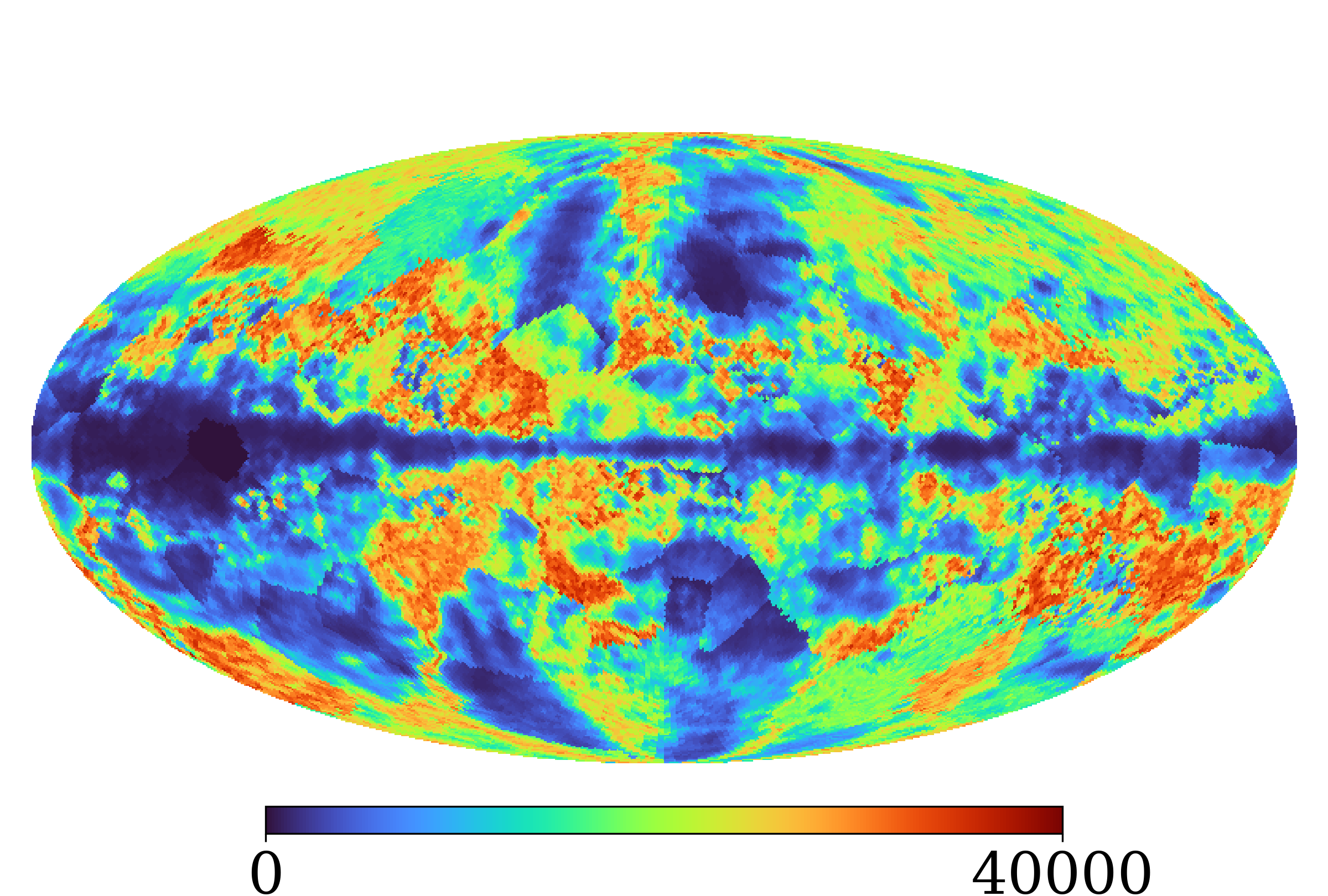} \label{fig: cmb neff Q CPL}}
\subfigure[CP(LC): $n_{\mathrm{eff}}^Q$]{\includegraphics[width=0.325\textwidth]{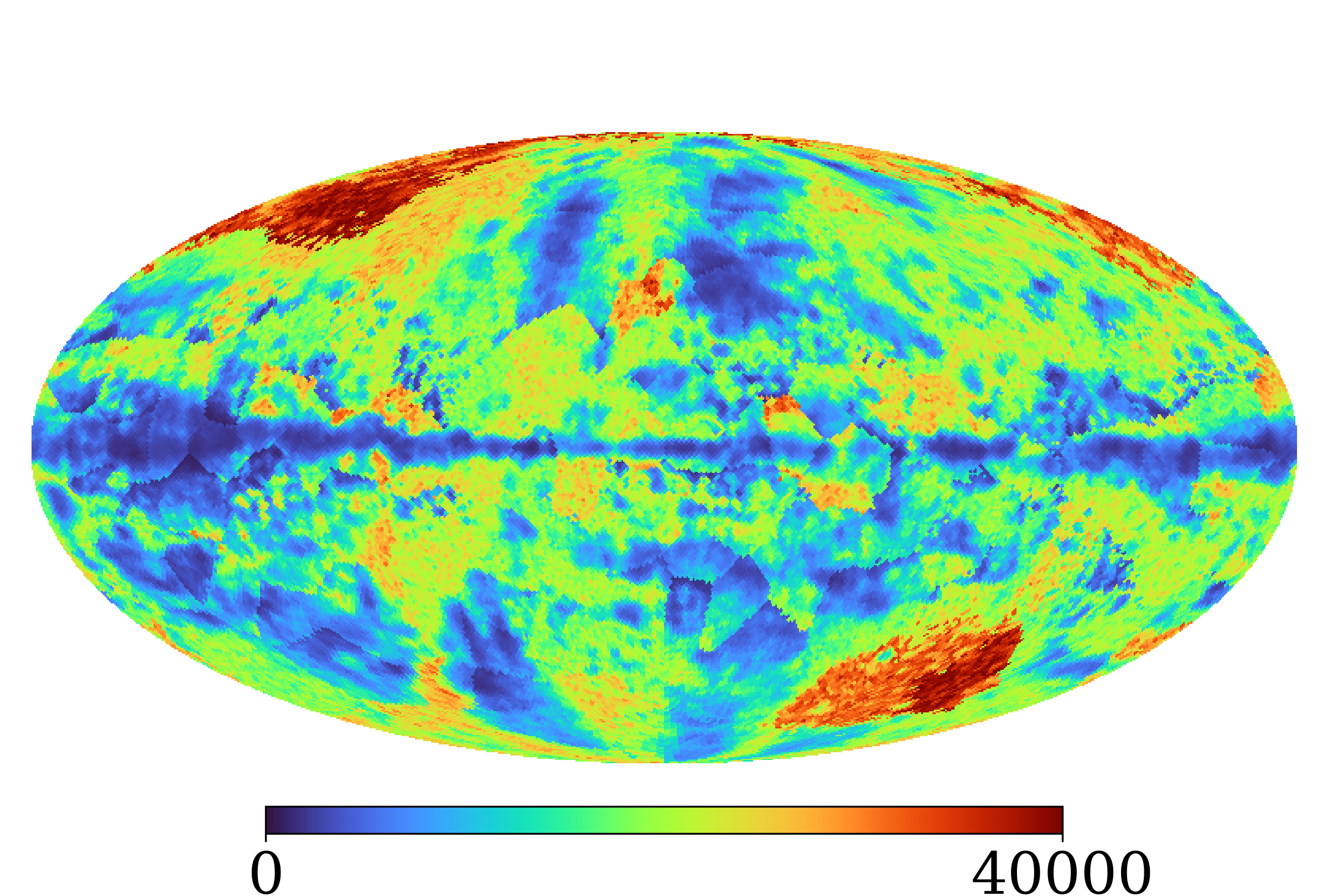} \label{fig: cmb neff Q CPLC}}
\subfigure[H(LC): $n_{\mathrm{eff}}^Q$]{\includegraphics[width=0.325\textwidth]{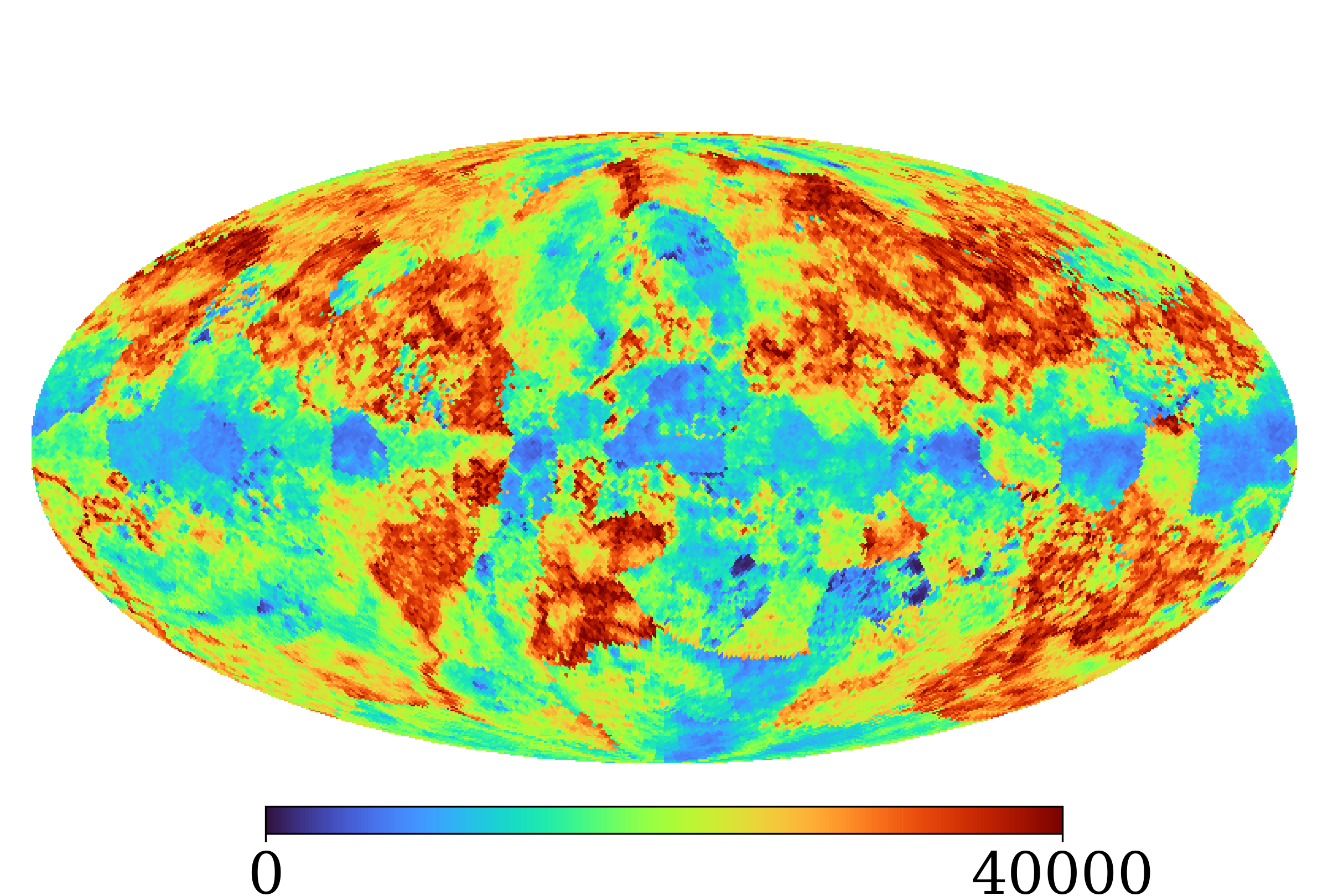} \label{fig: cmb neff Q HLC}}
\subfigure[CP(L): $n_{\mathrm{eff}}^U$]{\includegraphics[width=0.325\textwidth]{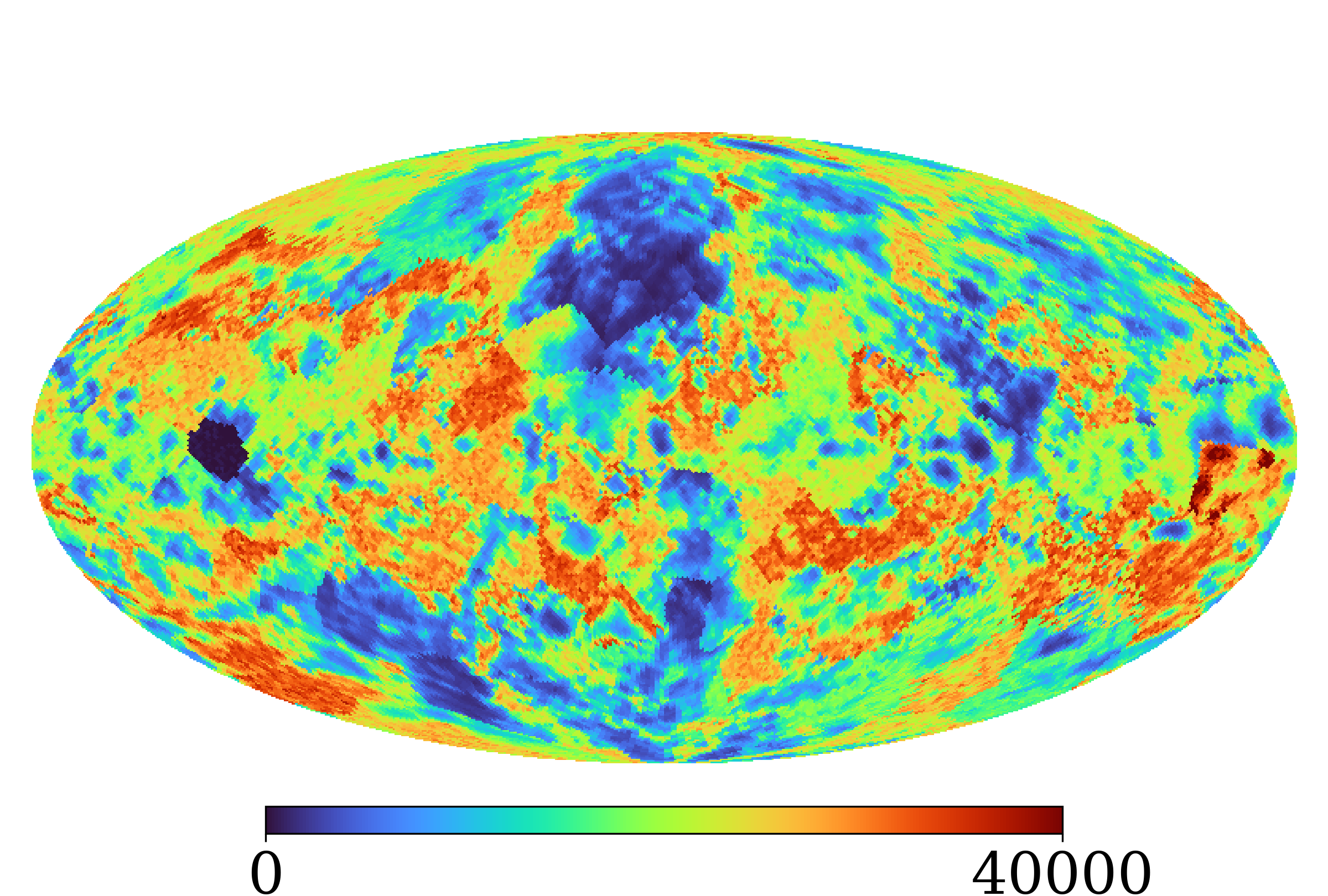} \label{fig: cmb neff U CPL}}
\subfigure[CP(LC): $n_{\mathrm{eff}}^U$]{\includegraphics[width=0.325\textwidth]{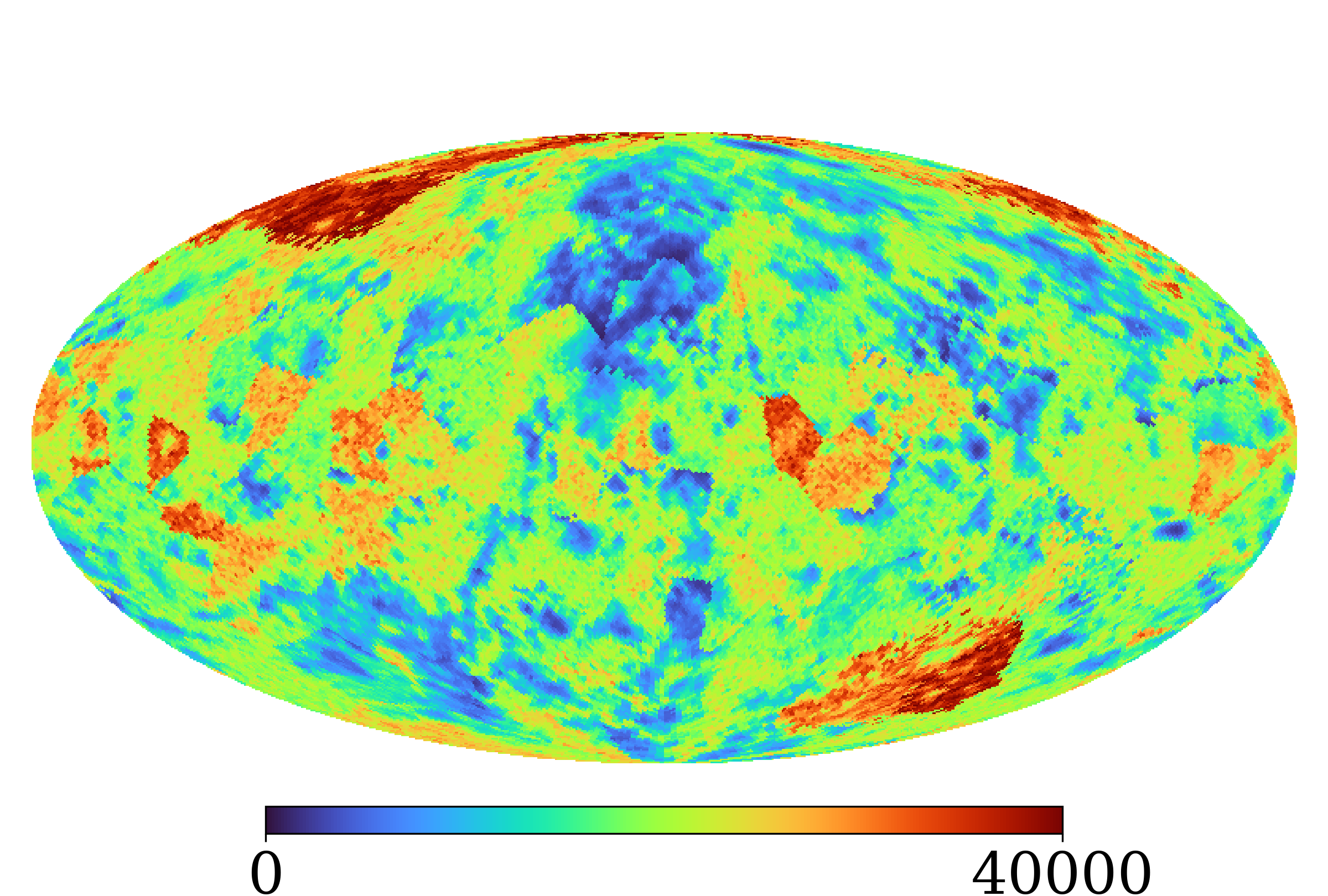} \label{fig: cmb neff U CPLC}}
\subfigure[H(LC): $n_{\mathrm{eff}}^U$]{\includegraphics[width=0.325\textwidth]{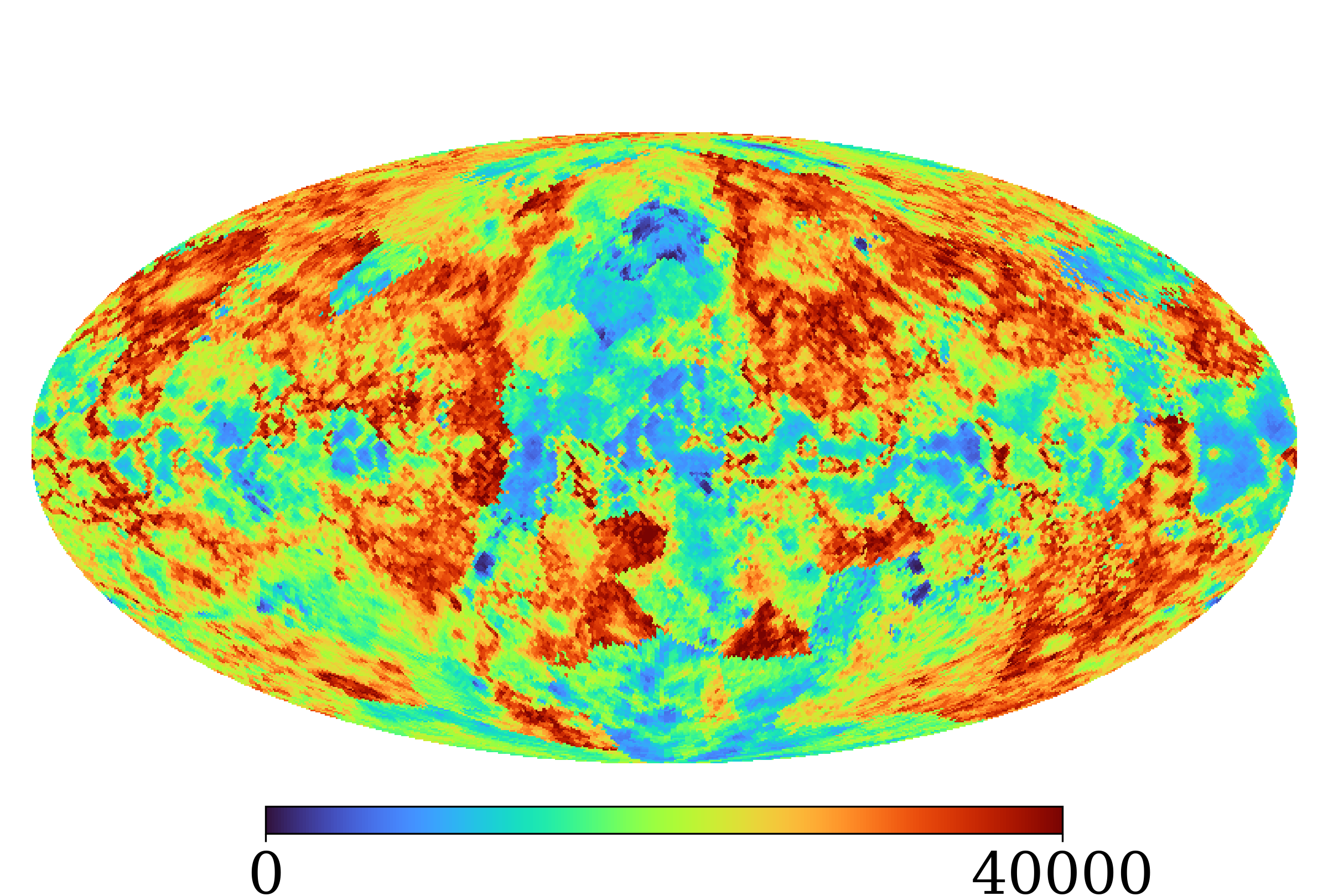} \label{fig: cmb neff U HLC}}
\caption{Output CMB amplitude maps obtained with the NUTS component separation algorithm. In panels (a) to (f) we show the output $\mathbfit{A}_{\mathrm{cmb}}^{Q/U}$ maps for the various validation sets, and in panels (g) to (l) we show the corresponding maps of the effective sample size. Obvious artefacts can be seen in the output $\mathbfit{A}_{\mathrm{cmb}}^{Q/U}$ maps when using a complete pooling model. This is the result of extremely bright foreground emission dominating over the CMB signal, making the extraction of the CMB signal very challenging. These artefacts are well traced by the effective sample size, being most apparent in bright regions close to the Galactic plane and North Polar Spur. By using a hierarchical model we are able to remove these obvious artefacts from our recovered CMB maps.}
\label{fig: cmb amp}
\end{figure*}

The primary output from the NUTS component separation are CMB amplitude maps in $Q$ and $U$. This consists of a set of maps corresponding to individual posterior samples, along with the summary maps of the mean and standard deviation of the amplitude maps. In Fig. \ref{fig: cmb amp} we show the mean CMB $Q$ and $U$ amplitude maps obtained for our three validation sets, along with the associated maps of the effective sample size. When using a complete pooling model obvious artefacts can be seen in the recovered CMB amplitude maps near the Galactic plane. This is to be expected, given the bright diffuse emission in these regions makes the extraction of weak CMB signals extremely challenging. When using a hierarchical model these artefacts are no longer present. By allowing the model to account for the real variation in spectral parameters in our regions, whilst constraining this variation through the fitted hyper-distributions, we are able to achieve a more accurate foreground removal and thereby remove the biases apparent from assuming constant spectral parameters. Regions of the CMB amplitude maps containing component separation artefacts are well traced by the effective sample size. In regions of low effective sample size ($n_{\mathrm{eff}}\lesssim 1000$), the parameter chains exhibit a high degree of auto-correlation, indicative of the sampler struggling to draw independent posterior samples. Thresholding maps of the effective sample size can be used to construct confidence masks for the CMB amplitude maps, which can be used in combination with standard Galactic emission masks.

In Fig. \ref{fig: cmb norm dev} we show histograms of the normalized deviations of the $\mathbfit{A}_{\mathrm{cmb}}^{Q/U}$ maps. For the validation sets studied here the distributions of the normalized deviations are slightly wider than the standard Gaussian. This means the uncertainties reported by the CMB amplitude standard deviation maps underestimate the errors on the recovered CMB amplitudes, and the CMB amplitude posterior shows slight departures from Gaussianity. The complete pooling model results in small biases away from zero in the median of the normalized deviations, which are not present with the hierarchical model.

We perform power spectrum estimation using the \textsc{NaMaster} library \citep{2019MNRAS.484.4127A}. \textsc{NaMaster} is a code for performing pseduo-$C_{\ell}$ estimation, accounting for effects from sky masking, as well as performing full $E$ and $B$-mode purification. This is particularly important when the $B$-mode signal is much weaker than the $E$-mode signal, as is the case for CMB $B$-mode studies. In this situation $E$-to-$B$ leakage when performing power spectrum estimation on a cut sky can lead to the variance of the $B$-mode power spectrum estimators being dominated by the variance of the leaked $E$-modes. Details on the construction of unbiased pseudo-$C_{\ell}$ estimators, and $E$ and $B$-mode purification can be found in \cite{2002ApJ...567....2H, 2017MNRAS.465.1847E, 2019MNRAS.484.4127A}. 

We generate the Galactic emission mask following the procedure in \cite{2018JCAP...04..023R}, with a common mask being used to enable direct comparison between the recovered power spectra. We extrapolated $10^{\circ}$ smoothed $5\,\mathrm{GHz}$ and $402\,\mathrm{GHz}$ polarized intensity maps to $70\,\mathrm{GHz}$, and applied a $5\sigma$ threshold against the standard deviation of the $10^{\circ}$ smoothed CMB polarized intensity map at $70\,\mathrm{GHz}$. The $5\,\mathrm{GHz}$ map was extrapolated using a constant spectral index of $\beta_{\mathrm{s}}=-3$ and the $402\,\mathrm{GHz}$ map was extrapolated using an MBB SED, setting $\beta_{\mathrm{d}}=1.6$ and $T_{\mathrm{d}}=19.4\,\mathrm{K}$. We further mask all pixels with $n_{\mathrm{eff}}\lesssim 1000$ in either the $Q$ or $U$ maps, although most of these pixels are already contained within the Galactic emission masks. The Galactic emission mask produced by masking all pixels below the $5\sigma$ threshold excludes approximately 60\% of the sky.    

\begin{figure*}
    \centering
    \includegraphics[width=0.9\textwidth]{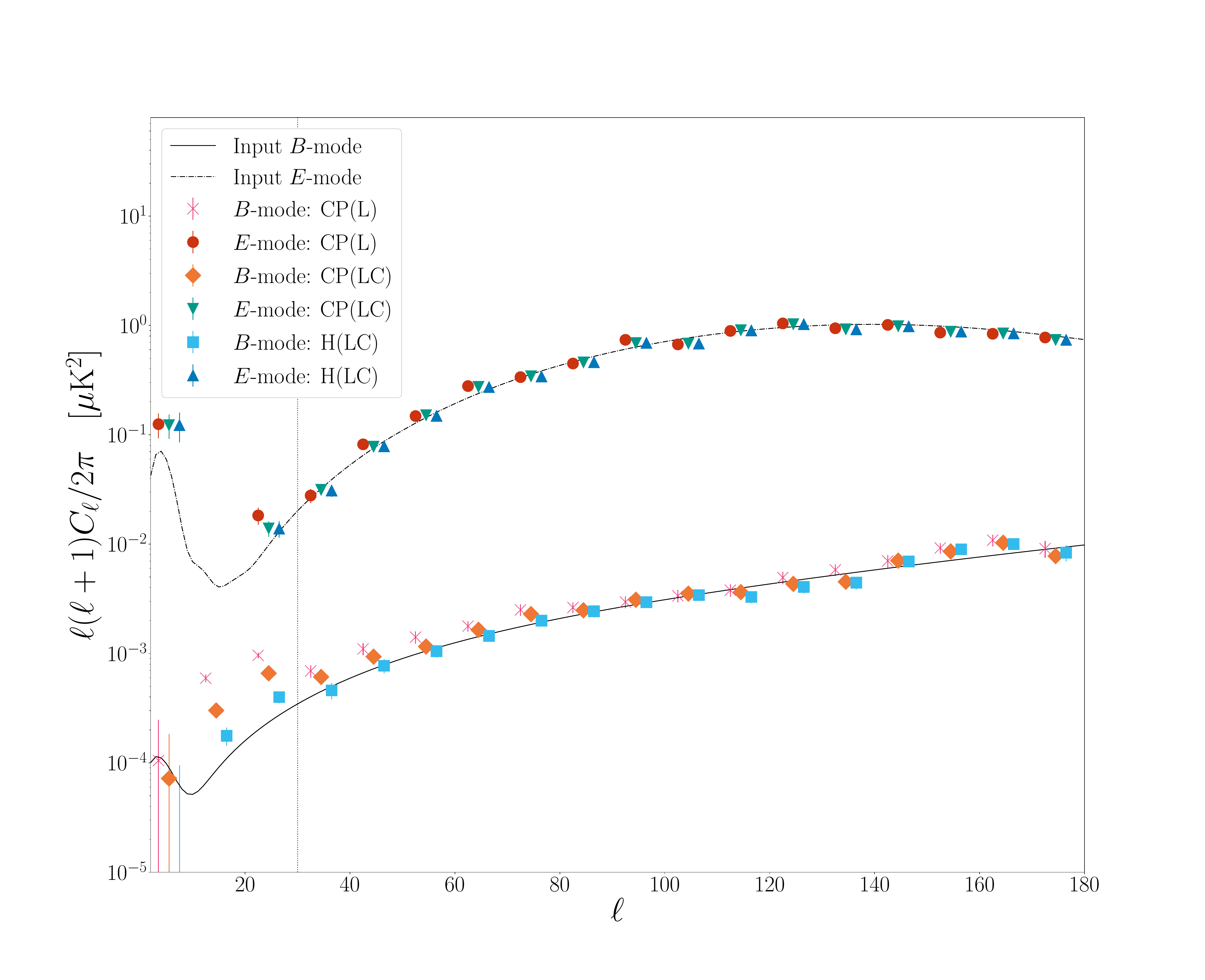}
    \caption{CMB $E$-mode and $B$-mode power spectra for our three validation sets, covering the multipole range, $2\leq\ell< 180$. Power spectra are binned, using a bin width of $\Delta\ell=10$ (excluding the first bin, which contains multipoles $2\leq\ell<10$). We plot the power spectra for the three validation sets together, slightly offset from one another for display purposes. The central point within each triple is at the effective multipole value for that bin. We can see that the $E$-mode power spectrum is recovered with high accuracy for $\ell>30$, demonstrating the overall fidelity of the algorithm. For the $B$-mode power spectrum we can see the results obtained using a complete pooling model contain significant large scale foreground residuals, that are not present when using the hierarchical model. For $\ell\lesssim 30$ the sky mask results in the pseudo-$C_\ell$ estimator becoming sub-optimal, which is particularly apparent in the $10\leq\ell<20$ bin, where the point estimates for the $E$-mode bandpowers are negative. We plot a dotted vertical line at $\ell=30$ to separate these angular scales. For $\ell\gtrsim 150$ the beam correction results in higher variance pseudo-$C_\ell$ estimates. This beam correction, along with the limiting pixel resolution at \textsc{NSIDE}=64, results in bandpower estimates becoming unreliable for $\ell\gtrsim 180$.}
    \label{fig: power spectra}
\end{figure*}

In Fig. \ref{fig: power spectra} we show the $E$ and $B$-mode power spectra derived for the three validation sets. For the purposes of power spectrum estimation, we perform component separation on two splits of the input data with differing noise realisations. We then evaluate the cross-spectra of the recovered CMB maps from the two data-splits, allowing us to avoid complications from noise bias. We show power spectra, covering multipoles $2\leq\ell< 180$, using a bin width of $\Delta\ell=10$ (excluding the first bin, which covers multipoles $2\leq\ell<10$). 

For multipoles $\ell\lesssim 30$ it was found that the purified pseudo-$C_{\ell}$ estimator no longer recovered accurate point estimates for the bandpowers. At these large angular scales, the sky mask results in the pseudo-$C_\ell$ estimator becoming sub-optimal. For these low multipoles, one can obtain better power spectrum estimates by directly sampling the $C_{\ell}$ from the joint distribution, $P(C_{\ell},\mathbfit{A}_{\mathrm{cmb}}|\mathbf{d})$, as is performed in the \textsc{Commander} component separation code \citep{2004PhRvD..70h3511W, 2004ApJS..155..227E, 2008ApJ...676...10E, 2008MNRAS.389.1284T}. We leave the implementation of this additional power spectrum estimation technique to future work. 

For multipoles $\ell\gtrsim 150$, the $70\,\mathrm{arcmin}$ beam correction resulted in higher variance pseudo-$C_{\ell}$ estimates. At these small scales the beam correction begins to inflate the noise present in the CMB amplitude maps. These effects, in combination with the pixel resolution at \textsc{NSIDE}=64, result in bandpower estimates becoming unreliable for $\ell\gtrsim 180$.

To estimate the bandpower covariances, noise realizations were obtained by taking the difference between individual CMB amplitude posterior samples and the mean CMB amplitude map for the respective data-splits, i.e.,
\begin{equation}
    \delta\mathbfit{A}_{\mathrm{cmb}}^{k,j,\lambda} = \mathbfit{A}_{\mathrm{cmb}}^{k,j,\lambda} - \langle\mathbfit{A}_{\mathrm{cmb}}^{j,\lambda}\rangle,\quad\lambda=\{Q,U\},
\end{equation}
where $k$ denotes the $k^{\mathrm{th}}$ posterior sample and $j\in\{1,2\}$ denotes the relevant data-split. The covariance matrix was then estimated by calculating the power spectra of $1000$ signal plus noise simulations. For the signal simulations we generated $1000$ realizations of the theoretical input CMB power spectrum to account for cosmic variance. 

The $E$-mode power spectrum is obtained with high accuracy for $\ell\geq 30$, confirming the overall fidelity of the component separation algorithm. The recovery of the $B$-mode power spectrum is more challenging, given this signal is significantly weaker than the $E$-mode signal and potentially sub-dominant to foregrounds over most of the sky, at all frequencies. It can be seen that when using a complete pooling model the recovered $B$-mode power spectrum contains large-scale foreground residuals, biasing the power spectrum high. In real experimental applications one could attempt to mitigate this by applying even more aggressive Galactic emission masks, although this comes at the cost of increasing the uncertainty in the recovered power spectra. For multipoles $\ell<30$ we can see that the apparent residuals in the $B$-mode power spectrum are reduced when using the hierarchical model. However, as discussed above, the pseudo-$C_\ell$ estimates at these multipoles become sub-optimal, demonstrated clearly here by the negative point estimates obtained for the $E$-mode bandpowers in the $10\leq\ell<20$ bin.     

To quantify the impact of foreground residuals in the $B$-mode power spectrum, we can study the tensor-to-scalar ratio constraints that would be obtained from these power spectra for $\ell\geq30$. To do this we approximate the likelihood for the CMB power spectra as a multivariate Gaussian,
\begin{equation}
    -2\ln\mathcal{L} = \mathrm{const.} + \sum_{\ell\ell'}\left(\tilde{C}_{\ell} - \tilde{C}_{\ell}^{\mathrm{th}}\right)(\Sigma^{-1})_{\ell\ell'}\left(\tilde{C}_{\ell'} - \tilde{C}_{\ell'}^{\mathrm{th}}\right).
\end{equation}
The $\tilde{C}_{\ell}$ are the binned power spectra or bandpowers, with $\tilde{C}_{\ell}^{\mathrm{th}}$ being the corresponding theoretical bandpowers, and $\Sigma$ is the bandpower covariance matrix. The sum here runs over the effective $\ell$ values for each bin. It is worth noting that, in general, the CMB likelihood is non-Gaussian. However, for the higher multipoles we consider here the power spectrum estimates are formed by averaging over the individual $a_{\ell m}$'s corresponding to a given multipole, justifying the use of the Gaussian approximation through the central limit theorem.

We parametrize the theoretical power spectrum as,
\begin{equation}
    C_{\ell}^{\mathrm{th}} = \frac{r}{0.01}C_{\ell}^{BB}(r=0.01) + A_{L}C_{\ell}^{\mathrm{lens}},
\end{equation}
where $A_{L}$ is the lensing amplitude, $C_{\ell}^{BB}(r=0.01)$ is a fiducial primordial $B$-mode power spectrum corresponding to $r=0.01$, and $C_{\ell}^{\mathrm{lens}}$ is the fiducial lensing $B$-mode power spectrum. In a general $B$-mode analysis we would fit jointly for $r$ and $A_{L}$. However, it is challenging to constrain $A_{L}$ solely through the $B$-mode power spectrum here. In a realistic experiment, tight constraints can be put on the lensing $B$-mode through analysis of the $E$-mode, $TE$ correlations and the lensing potential power spectrum. Combined with constraints from external data sets and delensing, one can expect to be able to place a tight prior on $A_{L}$. For the sake of simplicity here, we fix $A_{L}$ to the input value of $1$.

\begin{figure}
    \centering
    \includegraphics[width=\columnwidth]{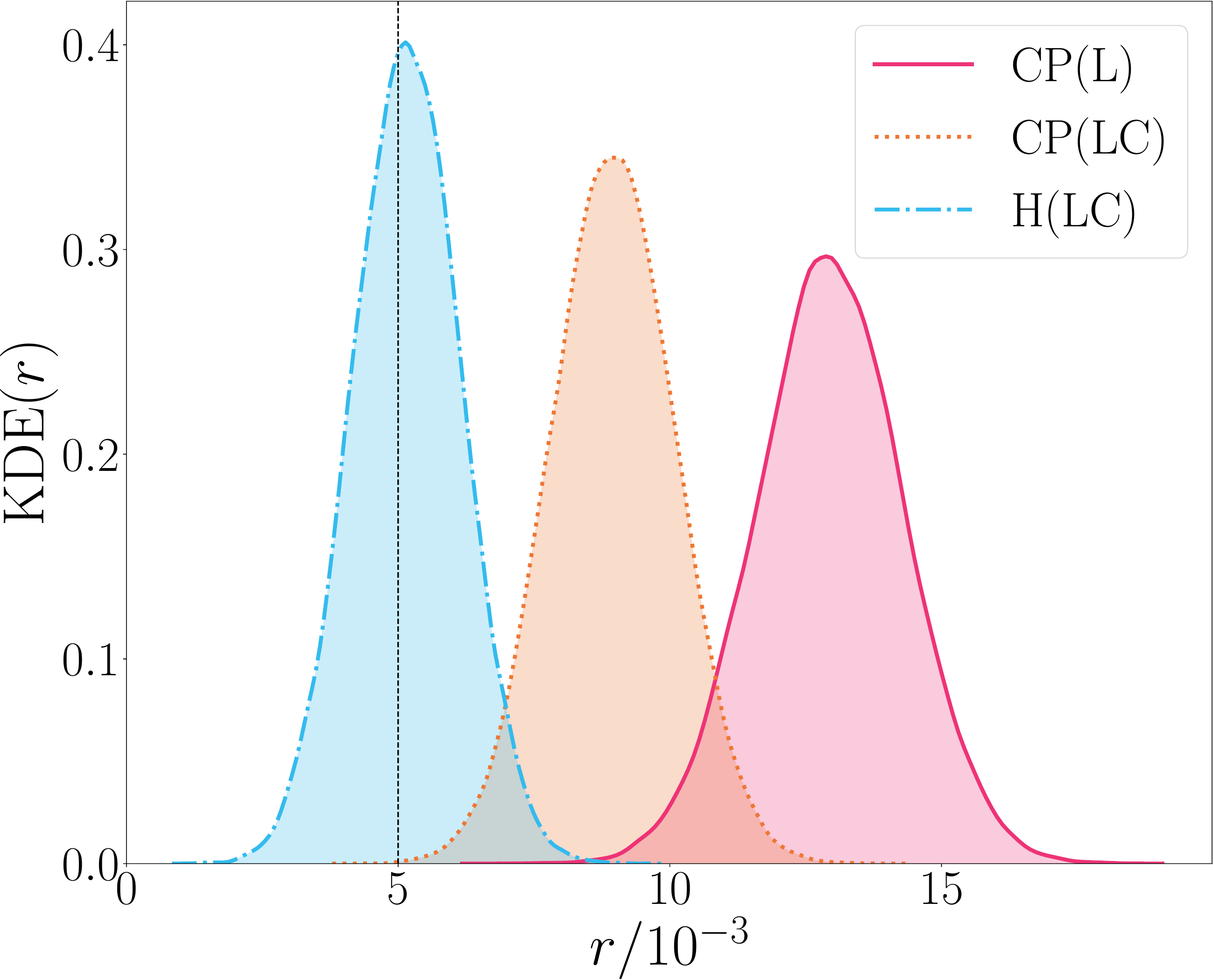}
    \caption{Tensor-to-scalar ratio constraints obtained for the three validation sets, using derived power spectra over the range $30\leq\ell< 180$ and assuming a multivariate Gaussian likelihood. For the CP(L) set we find $r=(12.9\pm 1.4)\times 10^{-3}$, for the CP(LC) set we find $r=(9.0\pm 1.1)\times 10^{-3}$, and for the H(LC) set we find $r=(5.2\pm1.0)\times10^{-3}$. With the addition of C-BASS we see a reduction in the bias on $r$ with the complete pooling model, although a $\sim 3.6\sigma$ bias still remains for the CP(LC) set. This bias is removed when using a hierarchical model. The uncertainty on the measured tensor-to-scalar ratio is smaller for the hierarchical model. The hyper-distributions over foreground spectral parameters allow us to model their spatial variations without increasing parameter uncertainties. Further, foreground residuals that result from the complete pooling model mean the power spectrum model is misspecified, resulting in increased biases and uncertainties on cosmological parameters.}
    \label{fig: tensor-to-scalar}
\end{figure}

Sampling from this likelihood using \textsc{PyMC3}, we obtain the constraints on $r$ shown in Fig. \ref{fig: tensor-to-scalar}. Using the complete pooling model we recover biased estimates of the tensor-to-scalar ratio, obtaining $r=(12.9\pm 1.4)\times 10^{-3}$ for the CP(L) set and $r=(9.0\pm 1.1)\times 10^{-3}$ for the CP(LC) set. This can be understood when we consider the large residuals and artefacts present in the recovered CMB when using a complete pooling model. By comparison, the bias is effectively removed for the H(LC) set, obtaining $r=(5.2\pm 1.0)\times10^{-3}$. It is worth noting that, despite the increased degrees of freedom in using a hierarchical model, the uncertainties obtained for the H(LC) set are smaller than those for the CP(LC) set. By adopting a multi-level structure for the spectral parameters in each region we prevent the model from overreacting to noise, whilst still capturing the spatial variation in spectral parameters. This in turn removes many of the foreground residuals present in the CMB amplitude maps obtained with the complete pooling model. These residuals result in increased uncertainties and biases on $r$, caused by the misspecified power spectrum model in the presence of foreground residuals.

As an aside, we do caution against interpreting these results as direct forecasts for the proposed \textit{LiteBIRD} experiment (and/or some combination with additional experiments). \textit{LiteBIRD} will also obtain constraints on lower multipoles around the reionization peak in the $B$-mode power spectrum, which we have not considered here due to the sub-optimal pseudo-$C_\ell$ estimator used, increasing the sensitivity of any tensor-to-scalar ratio measurement. Further, for simplicity in this validation analysis we smoothed all channels to the $70\,\mathrm{arcmin}$ resolution of the lowest frequency \textit{LiteBIRD} channel. This is likely a somewhat pessimistic approach. However, it is beyond the scope of this work to study the impact of the low resolution of the low-frequency \textit{LiteBIRD} channels on the ability to recover the CMB power spectra at higher multipoles. We have also not considered the effect of mis-modelling foreground SEDs or the impact of experimental systematics, both of which would significantly complicate any $B$-mode measurements. 

As discussed previously, given the frequency coverage considered here we already struggle to constrain spatial variations in foreground spectral parameters for these simple models. Fitting more complex models will require additional low and high-frequency data. We leave the impact of mis-modelling complex foreground SEDs to more detailed forecasting analyses, focusing here on validating the ability of Bayesian hierarchical modelling to reduce biases in recovered CMB estimates without inflating parameter uncertainties. It is worth noting that we consider the performance of the hierarchical model here for an experimental frequency coverage much broader than that of next-generation ground-based experiments such as the Simons Observatory \citep{2019JCAP...02..056A} and CMB-S4 \citep{2016arXiv161002743A}. Given the reduced frequency coverage of these experiments, fitting a full hierarchical model will be significantly more challenging, and will likely require a careful study of the prior choice for model hyper-parameters, and an exploration of model re-parametrizations.

\subsection{Synchrotron and dust amplitudes} \label{subsec: fg amp val}

\begin{figure*}
\centering
\subfigure[CP(L): $\mathbfit{A}_{\mathrm{s}}^Q$ @ $40\,\mathrm{GHz}$]{\includegraphics[width=0.325\textwidth]{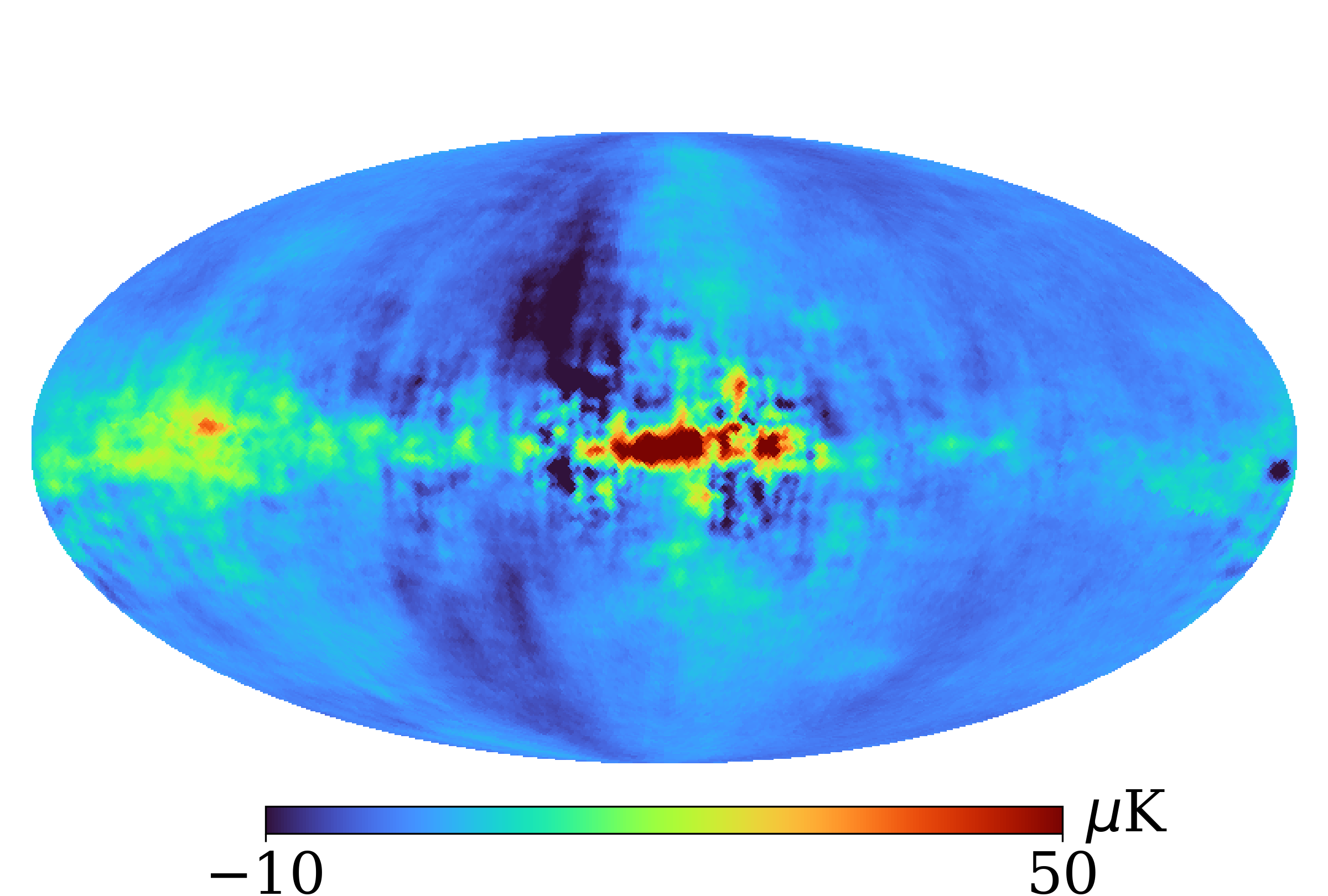} \label{fig: synch Q CPL}}
\subfigure[CP(LC): $\mathbfit{A}_{\mathrm{s}}^Q$ @ $5\,\mathrm{GHz}$]{\includegraphics[width=0.325\textwidth]{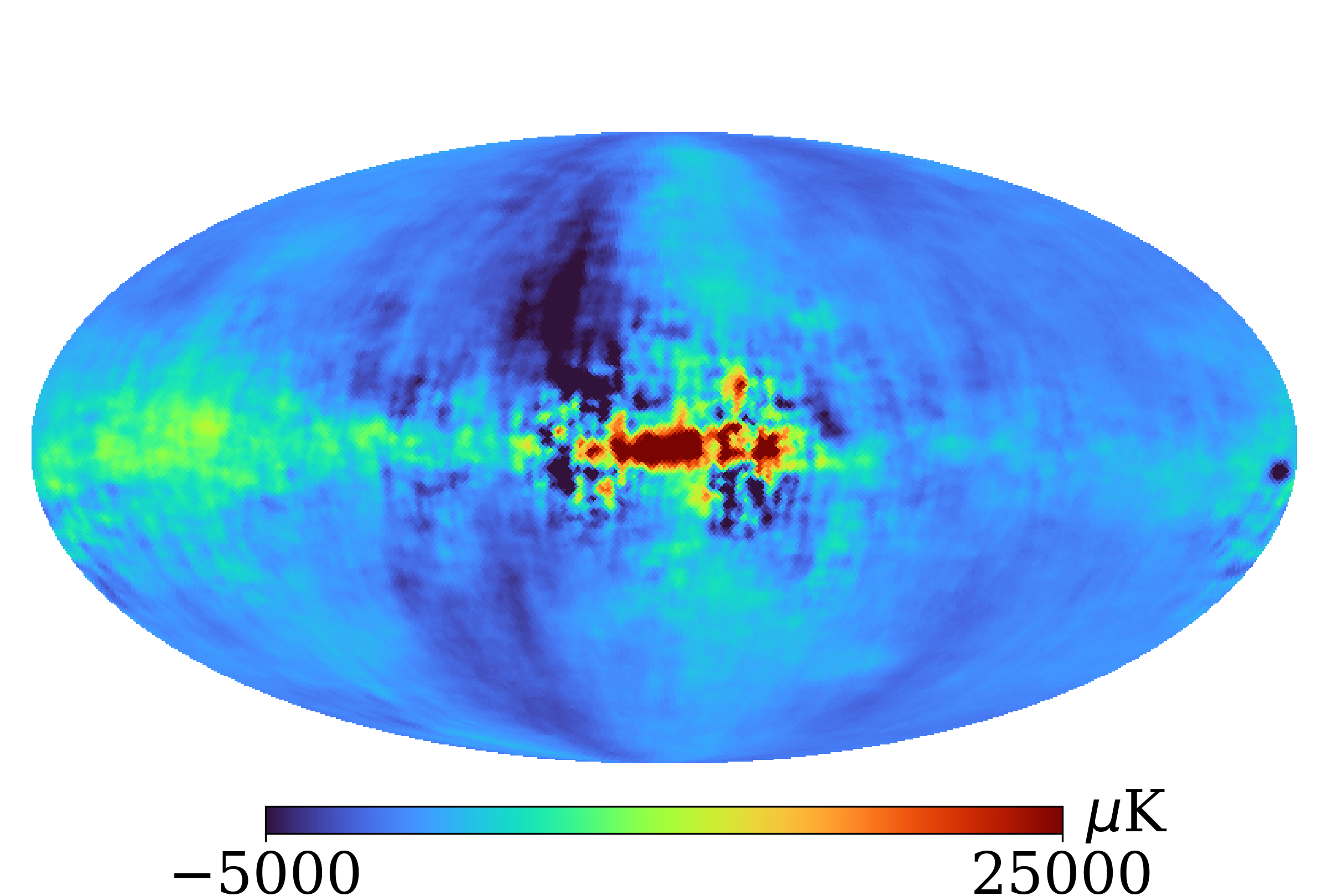} \label{fig: synch Q CPLC}}
\subfigure[H(LC): $\mathbfit{A}_{\mathrm{s}}^Q$ @ $5\,\mathrm{GHz}$]{\includegraphics[width=0.325\textwidth]{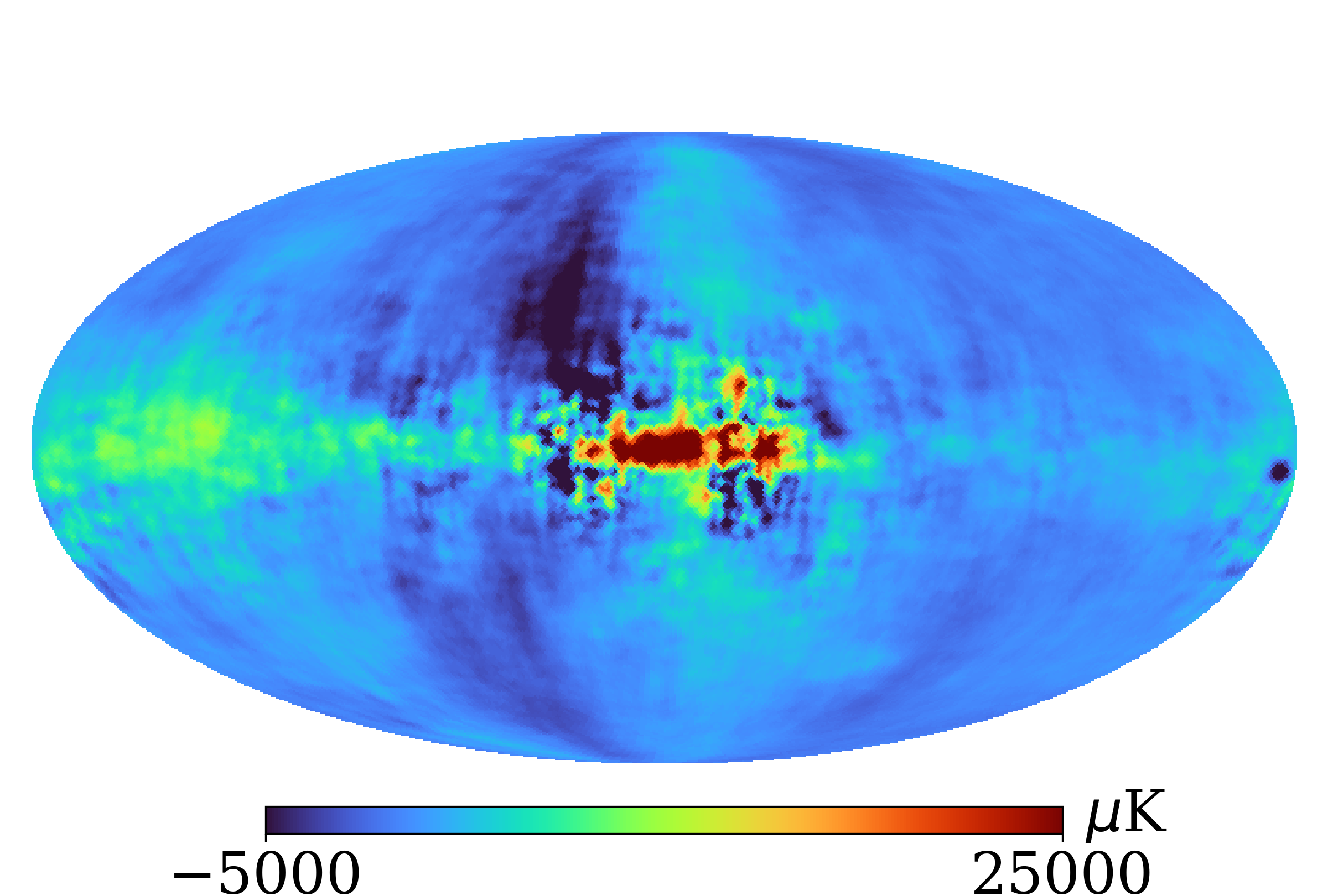} \label{fig: synch Q HLC}}
\subfigure[CP(L): $\mathbfit{A}_{\mathrm{s}}^U$ @ $40\,\mathrm{GHz}$]{\includegraphics[width=0.325\textwidth]{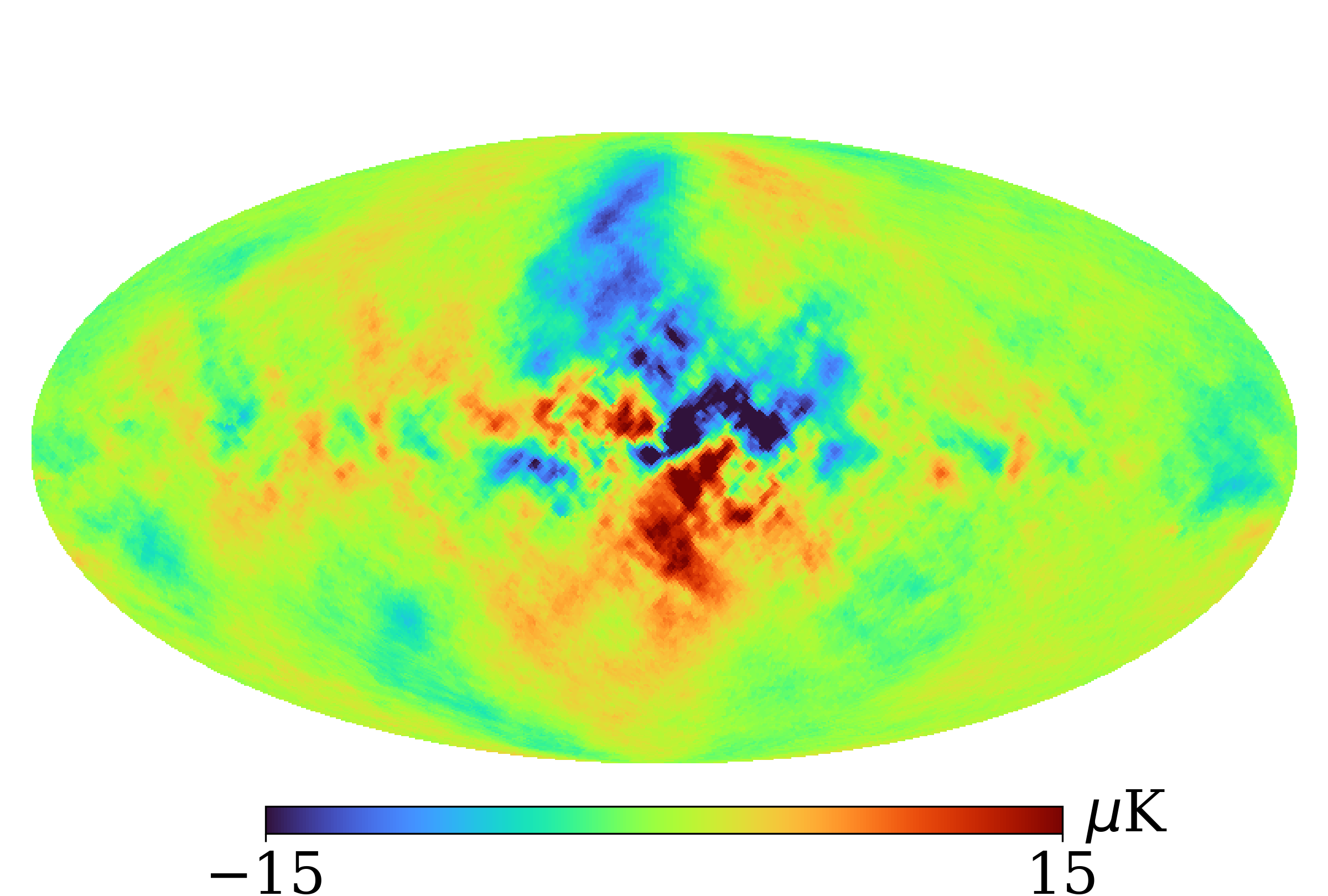} \label{fig: synch U CPL}}
\subfigure[CP(LC): $\mathbfit{A}_{\mathrm{s}}^U$ @ $5\,\mathrm{GHz}$]{\includegraphics[width=0.325\textwidth]{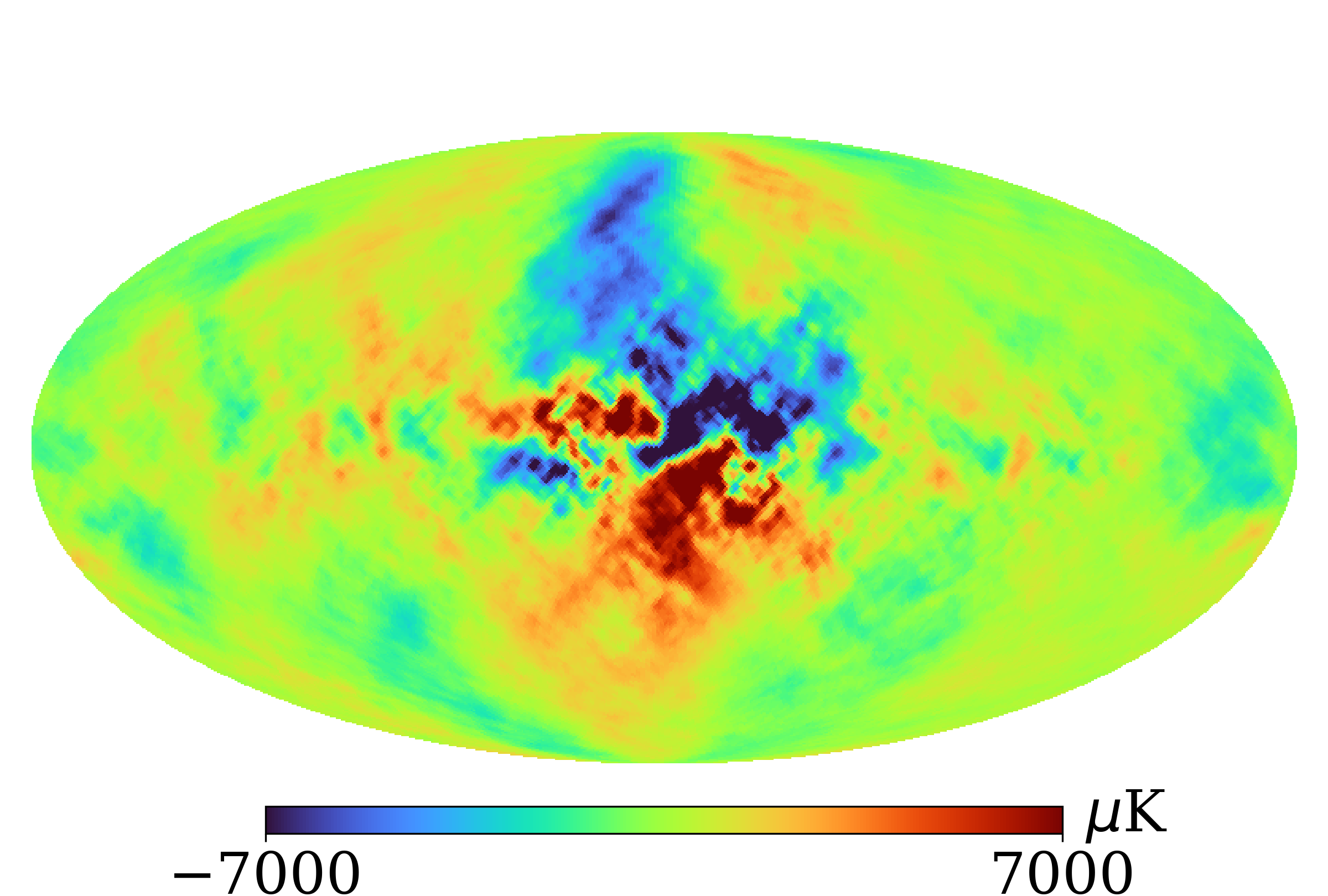} \label{fig: synch U CPLC}}
\subfigure[H(LC): $\mathbfit{A}_{\mathrm{s}}^U$ @ $5\,\mathrm{GHz}$]{\includegraphics[width=0.325\textwidth]{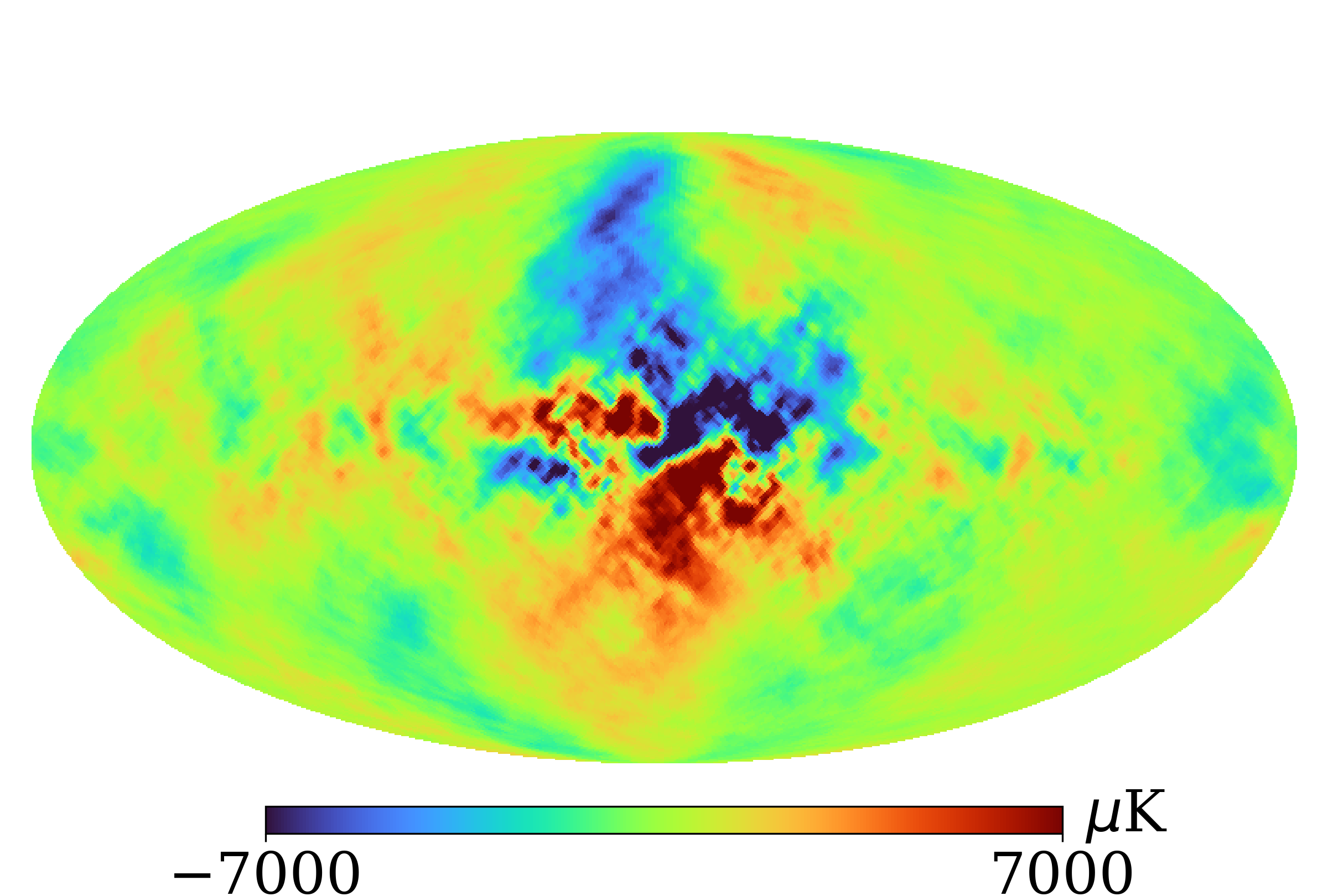} \label{fig: synch U HLC}}
\subfigure[CP(L): $\mathbfit{A}_{\mathrm{d}}^Q$ @ $402\,\mathrm{GHz}$]{\includegraphics[width=0.325\textwidth]{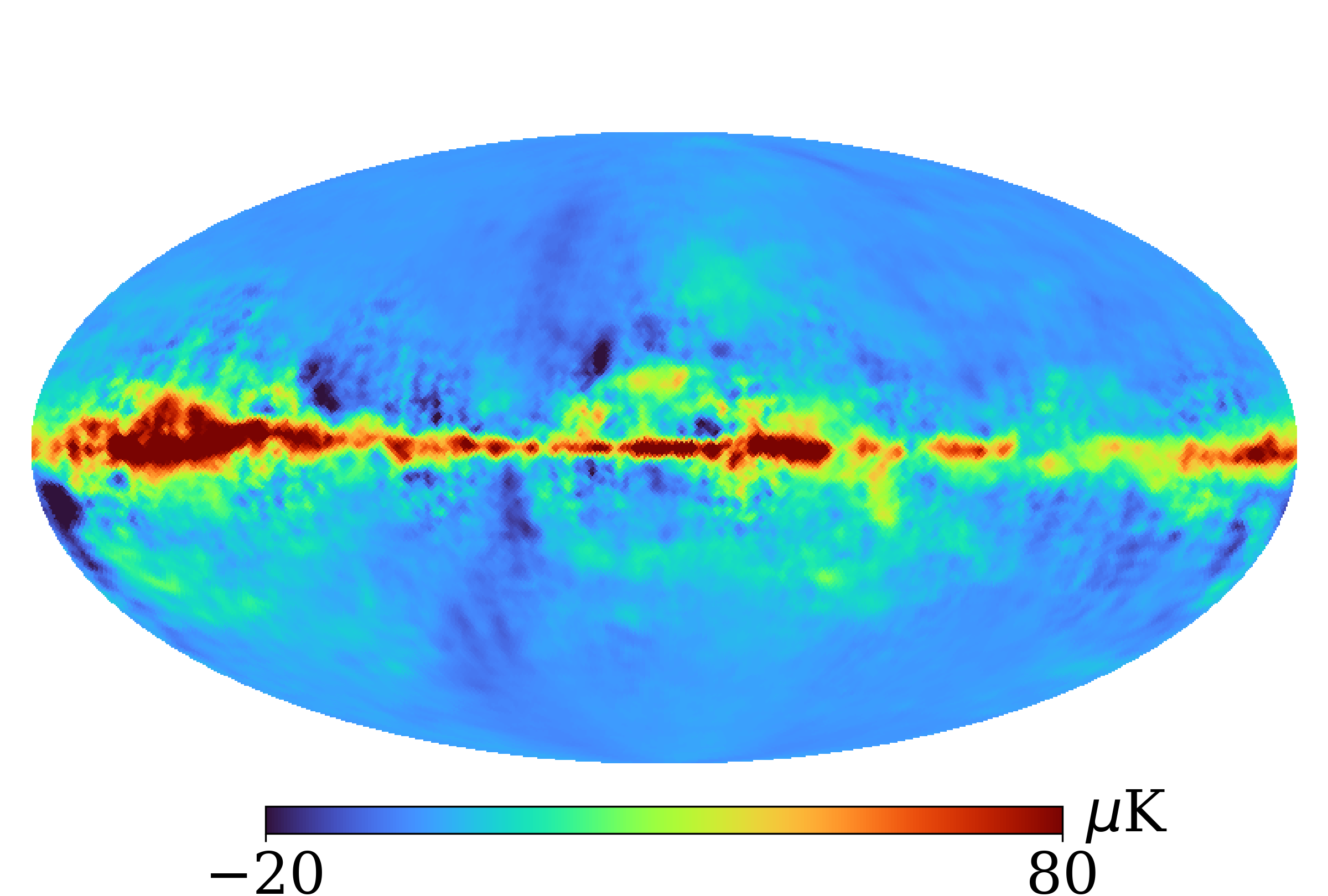} \label{fig: dust Q CPL}}
\subfigure[CP(LC): $\mathbfit{A}_{\mathrm{d}}^Q$ @ $402\,\mathrm{GHz}$]{\includegraphics[width=0.325\textwidth]{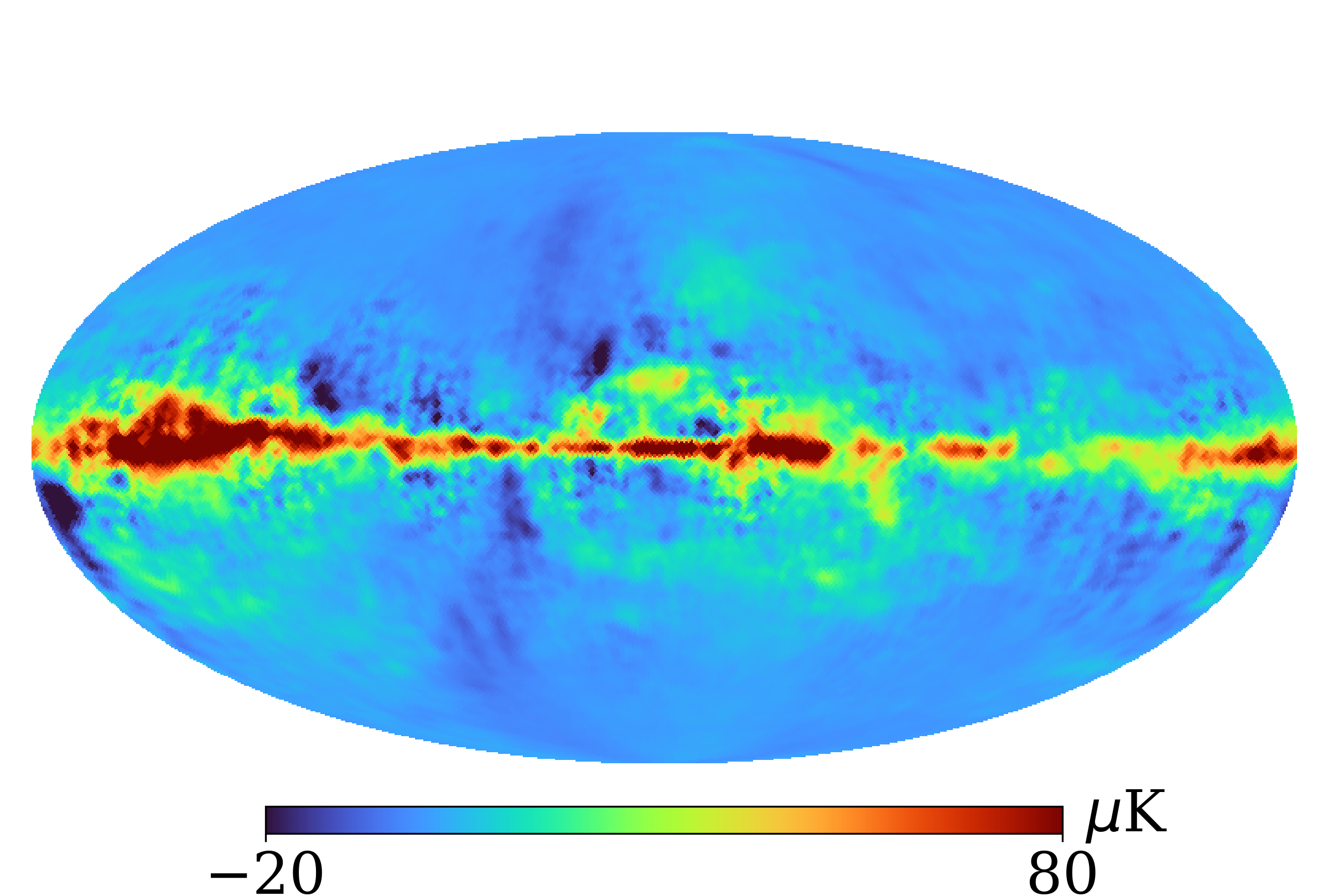} \label{fig: dust Q CPLC}}
\subfigure[H(LC): $\mathbfit{A}_{\mathrm{d}}^Q$ @ $402\,\mathrm{GHz}$]{\includegraphics[width=0.325\textwidth]{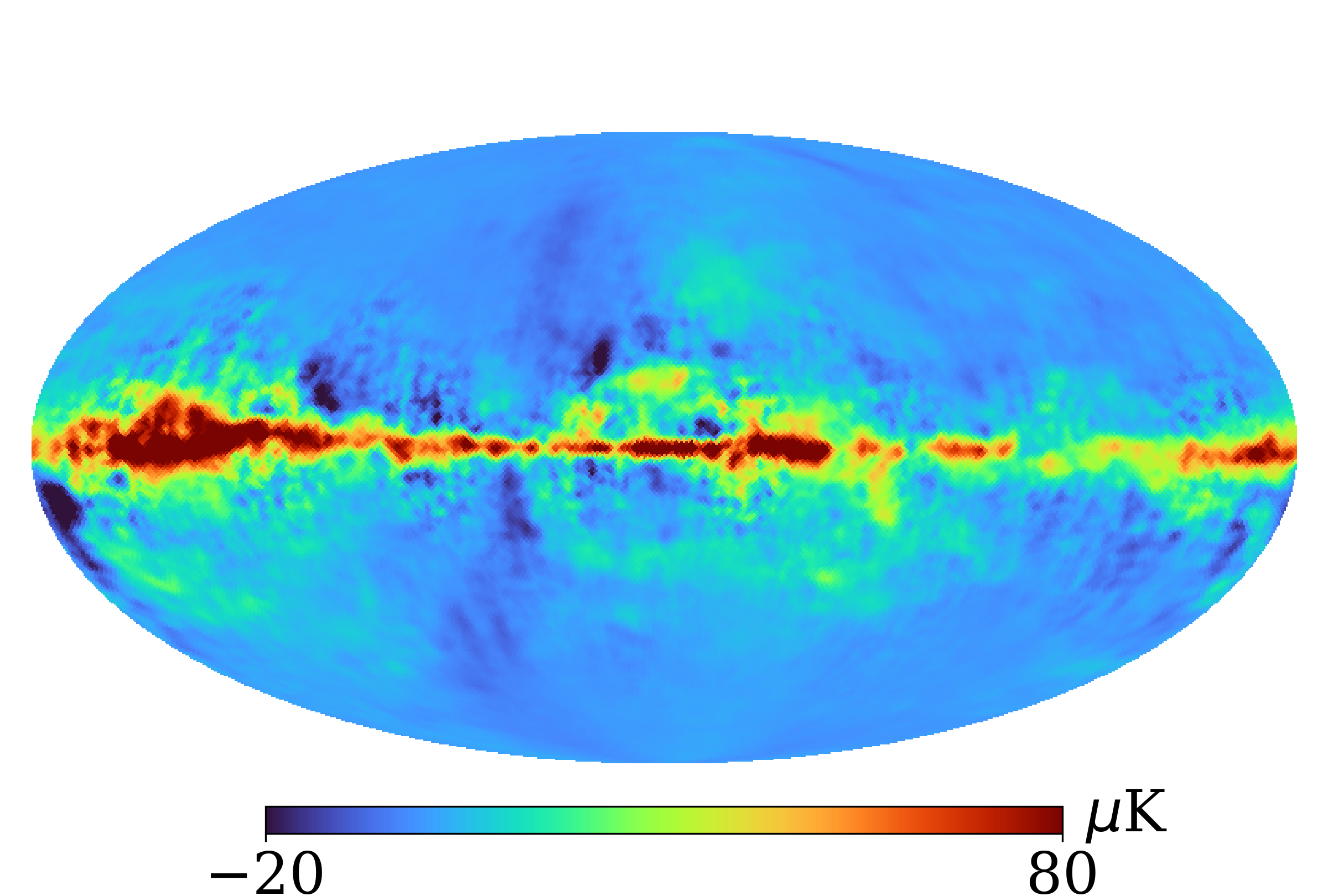} \label{fig: dust Q HLC}}
\subfigure[CP(L): $\mathbfit{A}_{\mathrm{d}}^U$ @ $402\,\mathrm{GHz}$]{\includegraphics[width=0.325\textwidth]{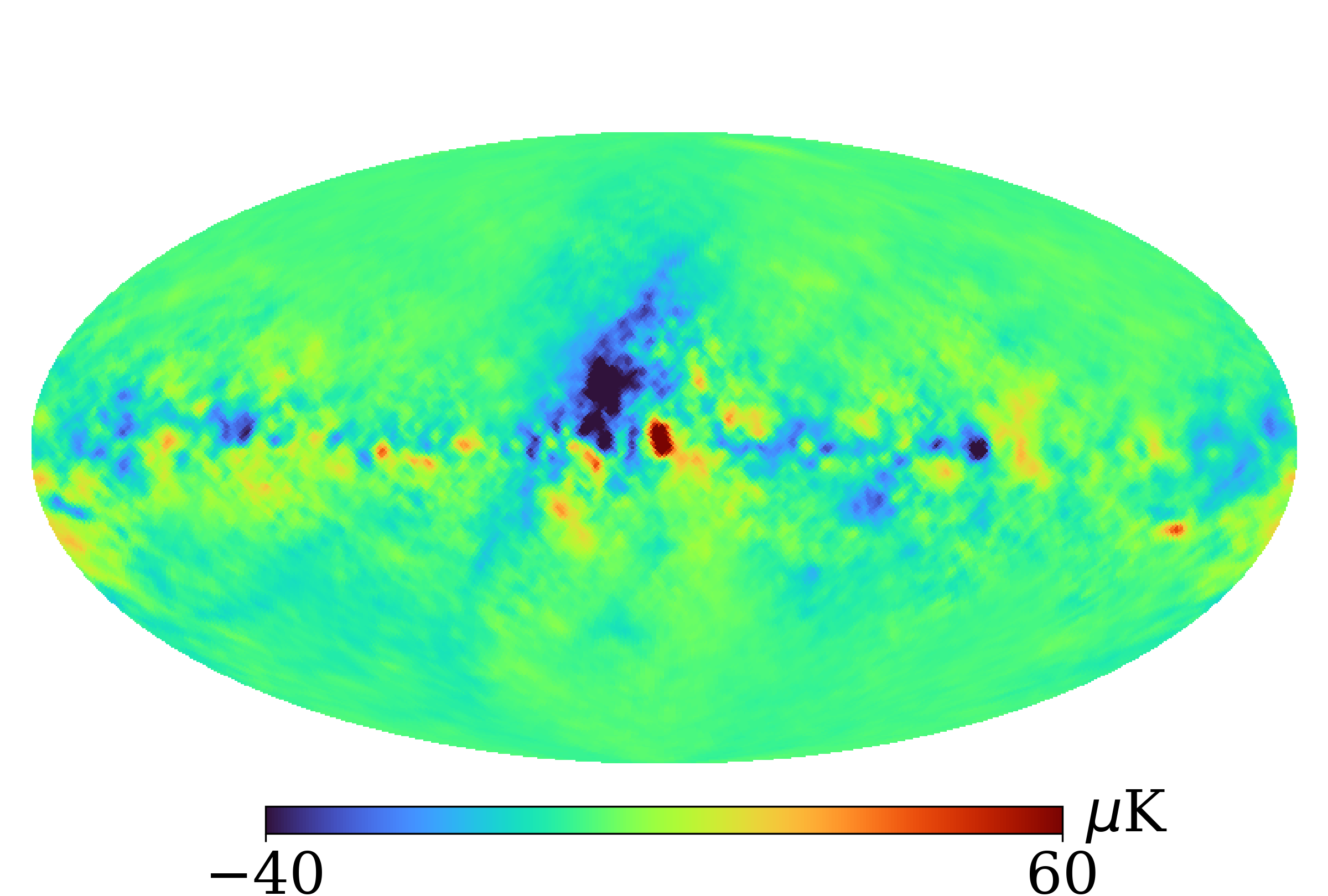} \label{fig: dust U CPL}}
\subfigure[CP(LC): $\mathbfit{A}_{\mathrm{d}}^U$ @ $402\,\mathrm{GHz}$]{\includegraphics[width=0.325\textwidth]{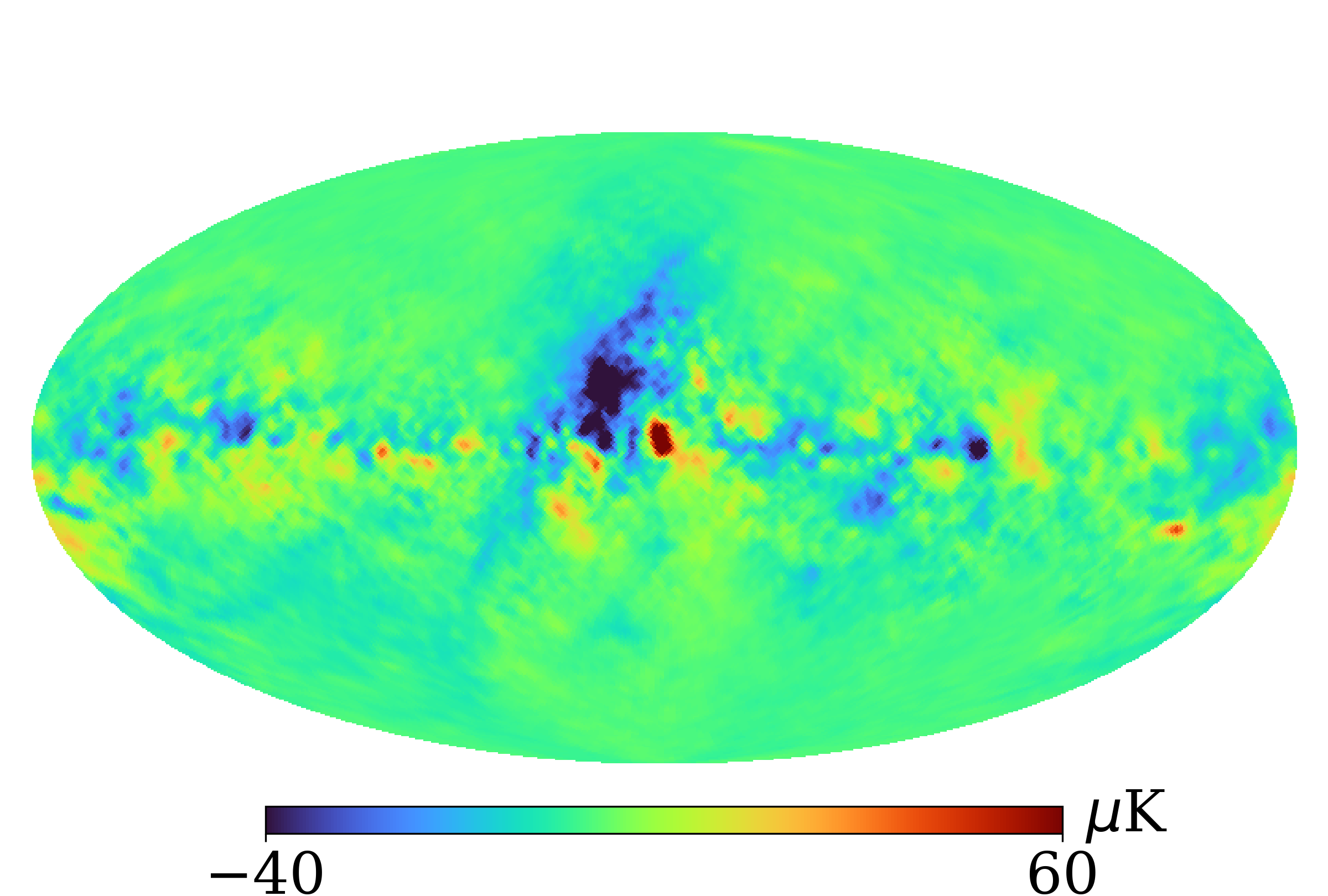} \label{fig: dust U CPLC}}
\subfigure[H(LC): $\mathbfit{A}_{\mathrm{d}}^U$ @ $402\,\mathrm{GHz}$]{\includegraphics[width=0.325\textwidth]{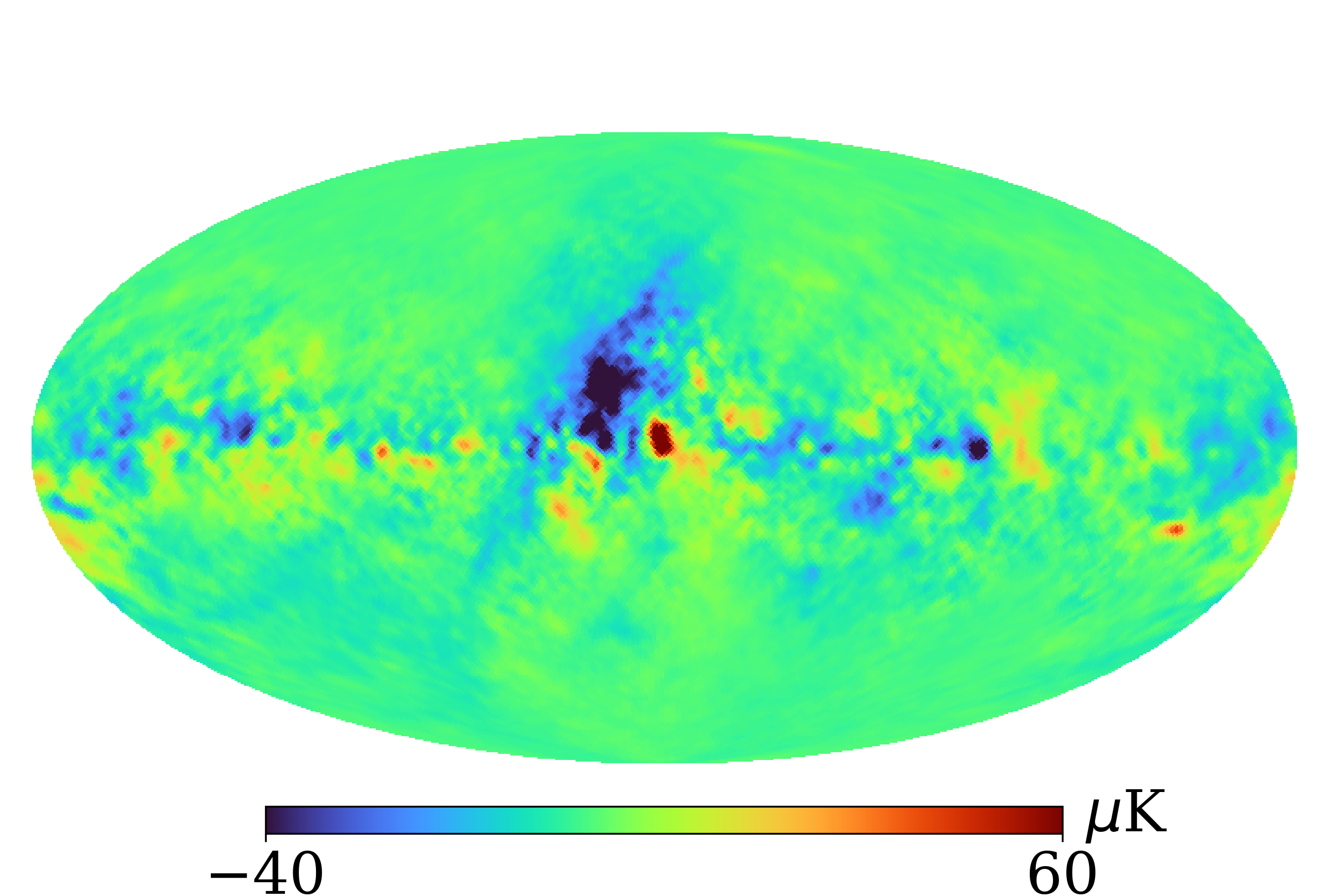} \label{fig: dust U HLC}}
\caption{Synchrotron and dust amplitude maps obtained for the three validation sets. In panels (a) to (f) we show the synchrotron amplitude maps, shown at the reference frequency of $\nu_0=40\,\mathrm{GHz}$ for the CP(L) validation set, and $\nu_0=5\,\mathrm{GHz}$ for the CP(LC) and H(LC) validation sets. In panels (g) to (l) we show the dust amplitude maps, displayed at the reference frequency of $\nu_0=402\,\mathrm{GHz}$. In all three cases the component separation algorithm recovers accurate estimates of the foreground amplitude maps at their reference frequencies. This is to be expected, and is to some extent a result of our choice of synchrotron and dust reference frequencies. The typical residuals in synchrotron amplitude maps are $\sim 5$ per cent lower for the H(LC) set compared to the CP(LC) set. Similarly, the dust amplitude residuals are $\sim 40$ per cent lower for the H(LC) set compared to the CP(LC) set.}
\label{fig: synch dust amp}
\end{figure*}

In Fig. \ref{fig: synch dust amp} we show the dust and synchrotron amplitude maps. Dust amplitude maps are shown at the reference frequency of $\nu_0=402\,\mathrm{GHz}$ for all three validation sets. Synchrotron amplitude maps are shown at a reference frequency of $\nu_0=40\,\mathrm{GHz}$ for the CP(L) validation set, and at $\nu_0=5\,\mathrm{GHz}$ for the CP(LC) and H(LC) validation sets. In all cases we can see the recovered component amplitude maps trace the input component amplitude maps well. However, this is to some extent a result of our choice of reference frequency, with the amplitudes being constrained by the pixel values at those frequencies. The overall level of residuals in the synchrotron amplitude maps is reduced by $\sim 5$ per cent for the H(LC) set compared to the CP(LC) set, and the typical residuals in the dust amplitude maps are reduced by $\sim 40$ per cent. These reductions were estimated using the MAD values of the residuals. Whilst the absolute value of these reductions is small at synchrotron and dust frequencies, they ultimately propagate through to significant biases in the recovered CMB maps, as seen in our tensor-to-scalar ratio estimates. 

In Fig. \ref{fig: synch amp norm dev} we show histograms of the normalized deviations for the synchrotron amplitude, and in Fig. \ref{fig: dust amp norm dev} we show histograms of the normalized deviations for the dust amplitude. For the synchrotron amplitude we obtain similar distributions across all three validation sets. The normalized deviations for the synchrotron amplitudes are wider than the standard Gaussian, indicating the standard deviations of the posterior samples underestimate the uncertainties on the synchrotron amplitudes. Conversely, for the dust amplitudes the normalized deviations are narrower than the standard Gaussian.

\subsection{Synchrotron spectral parameters} \label{subsec: synch spectral val}

\begin{figure*}
\centering
\subfigure[CP(L): $\beta_s$]{\includegraphics[width=0.325\textwidth]{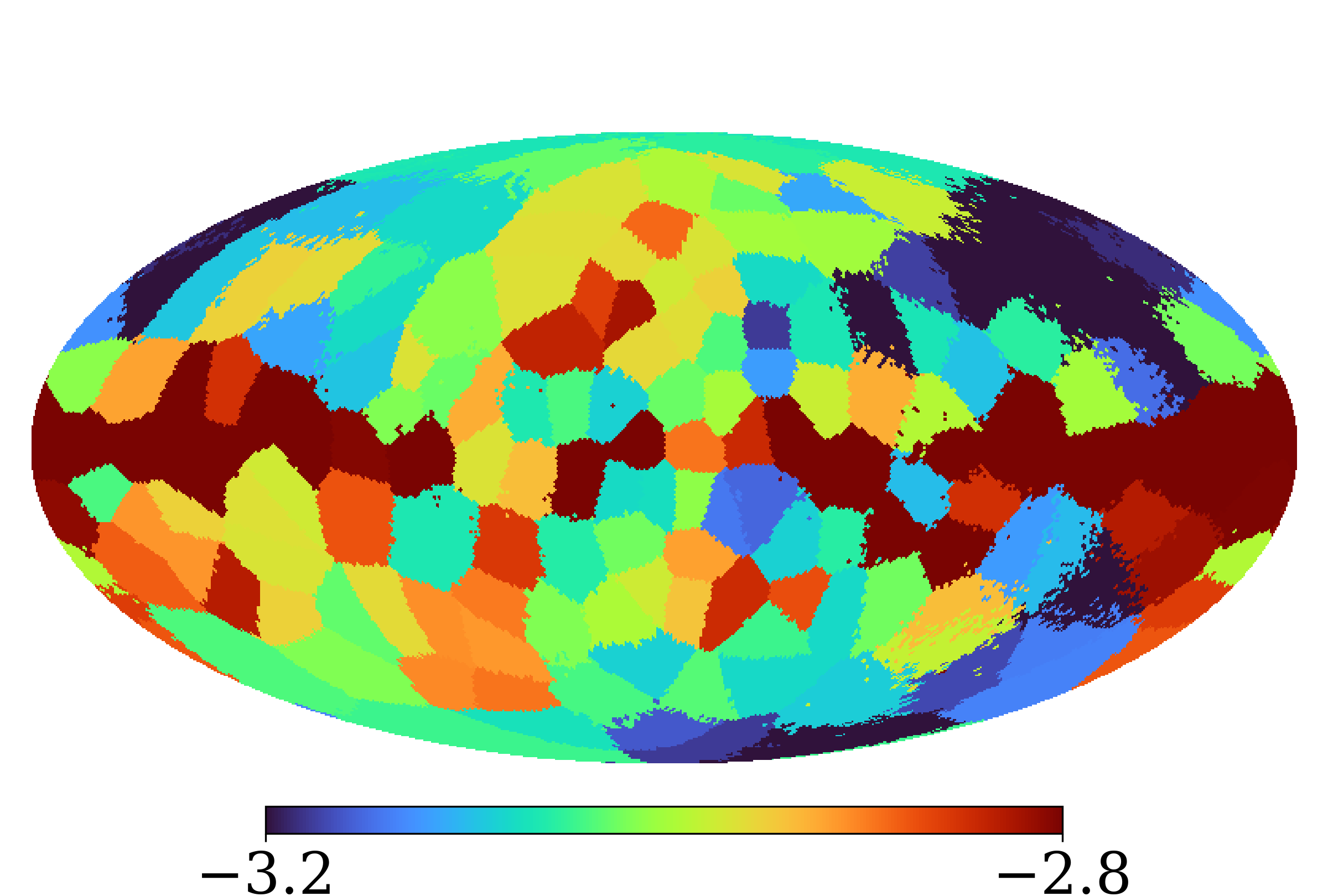} \label{fig_synch_beta_CPL}}
\subfigure[CP(LC): $\beta_s$]{\includegraphics[width=0.325\textwidth]{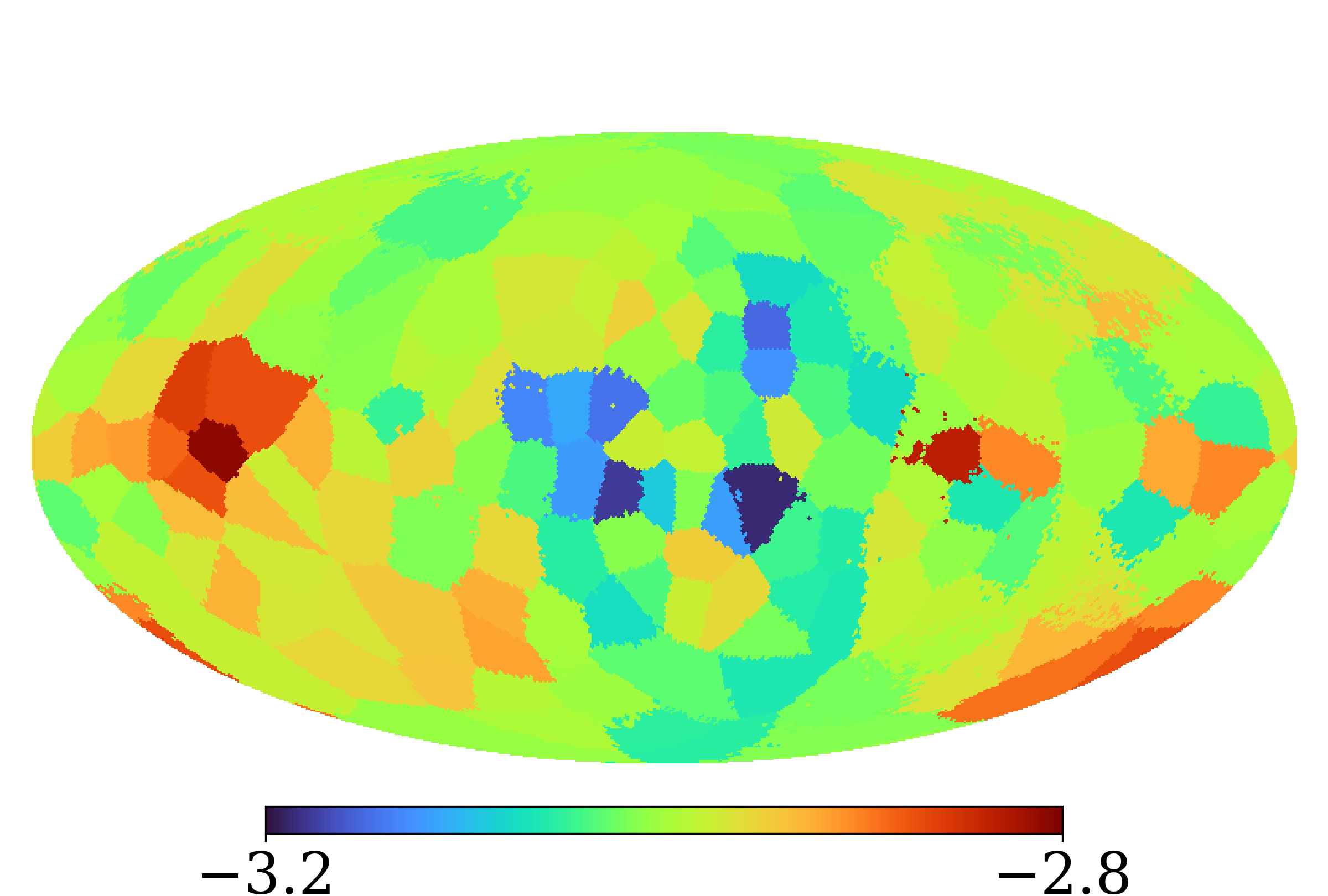}\label{fig_synch_beta_CPLC}}
\subfigure[H(LC): $\beta_s$]{\includegraphics[width=0.325\textwidth]{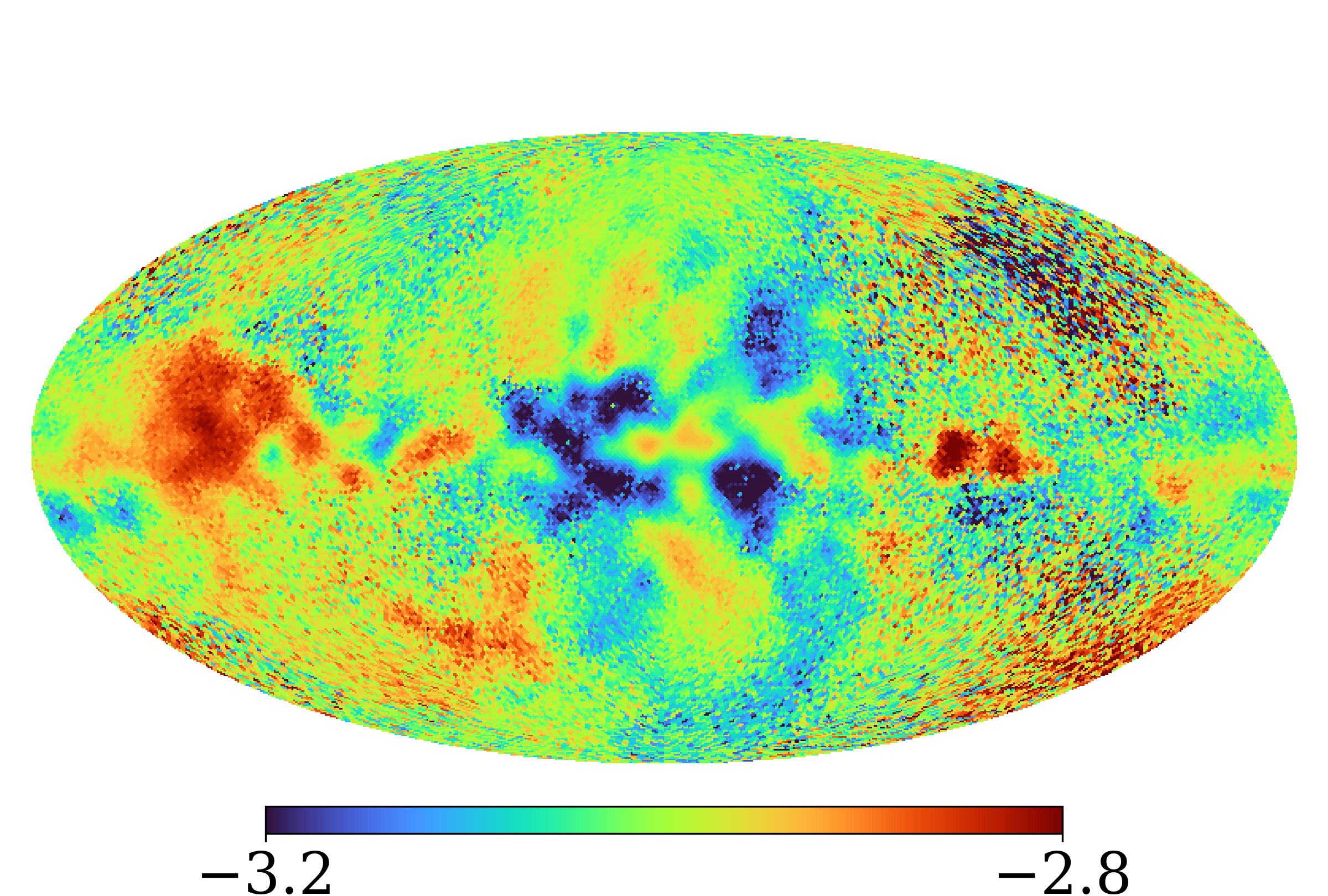} \label{fig_synch_beta_HLC}}
\subfigure[CP(L): $\beta_d$]{\includegraphics[width=0.325\textwidth]{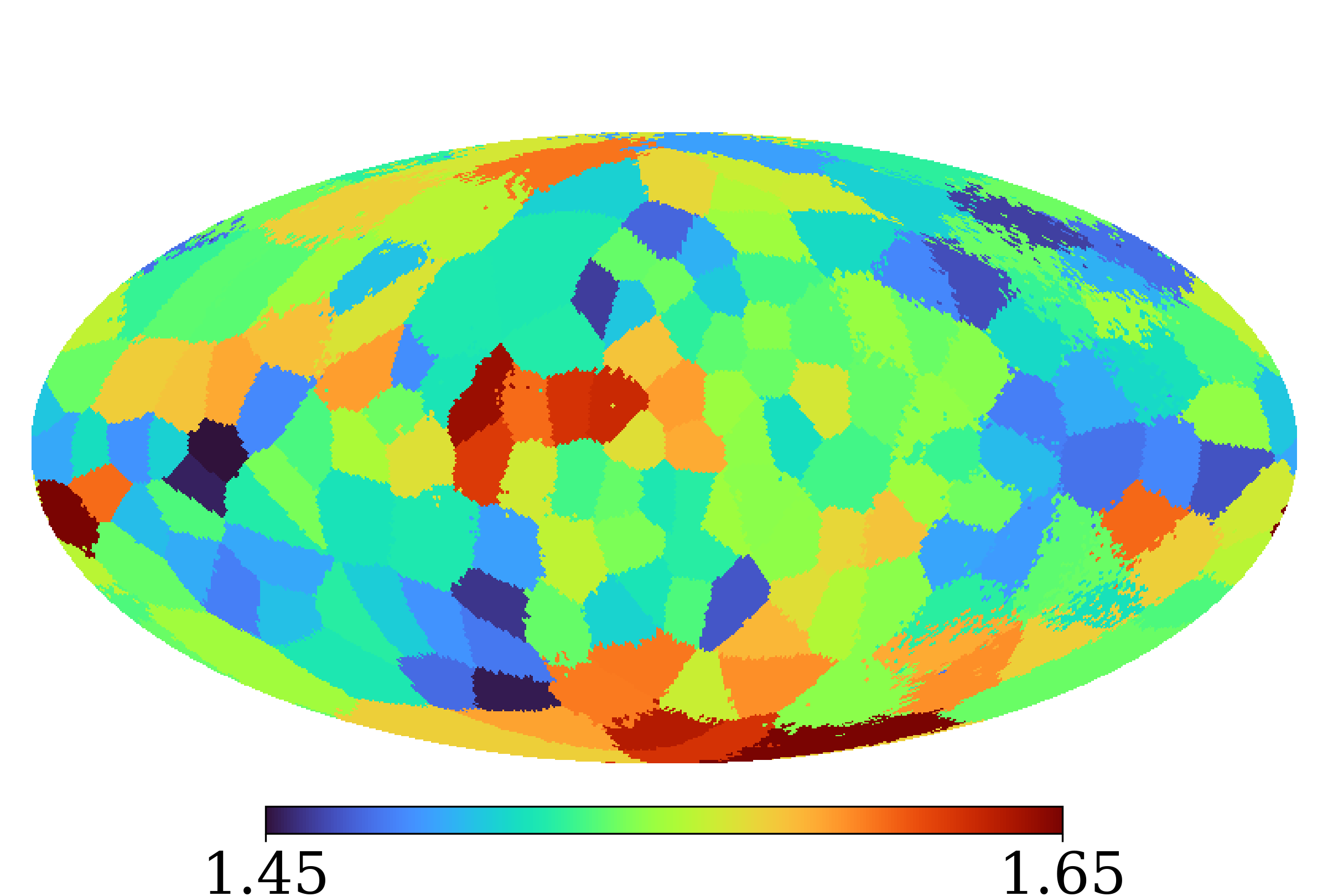} \label{fig_dust_beta_CPL}}
\subfigure[CP(LC): $\beta_d$]{\includegraphics[width=0.325\textwidth]{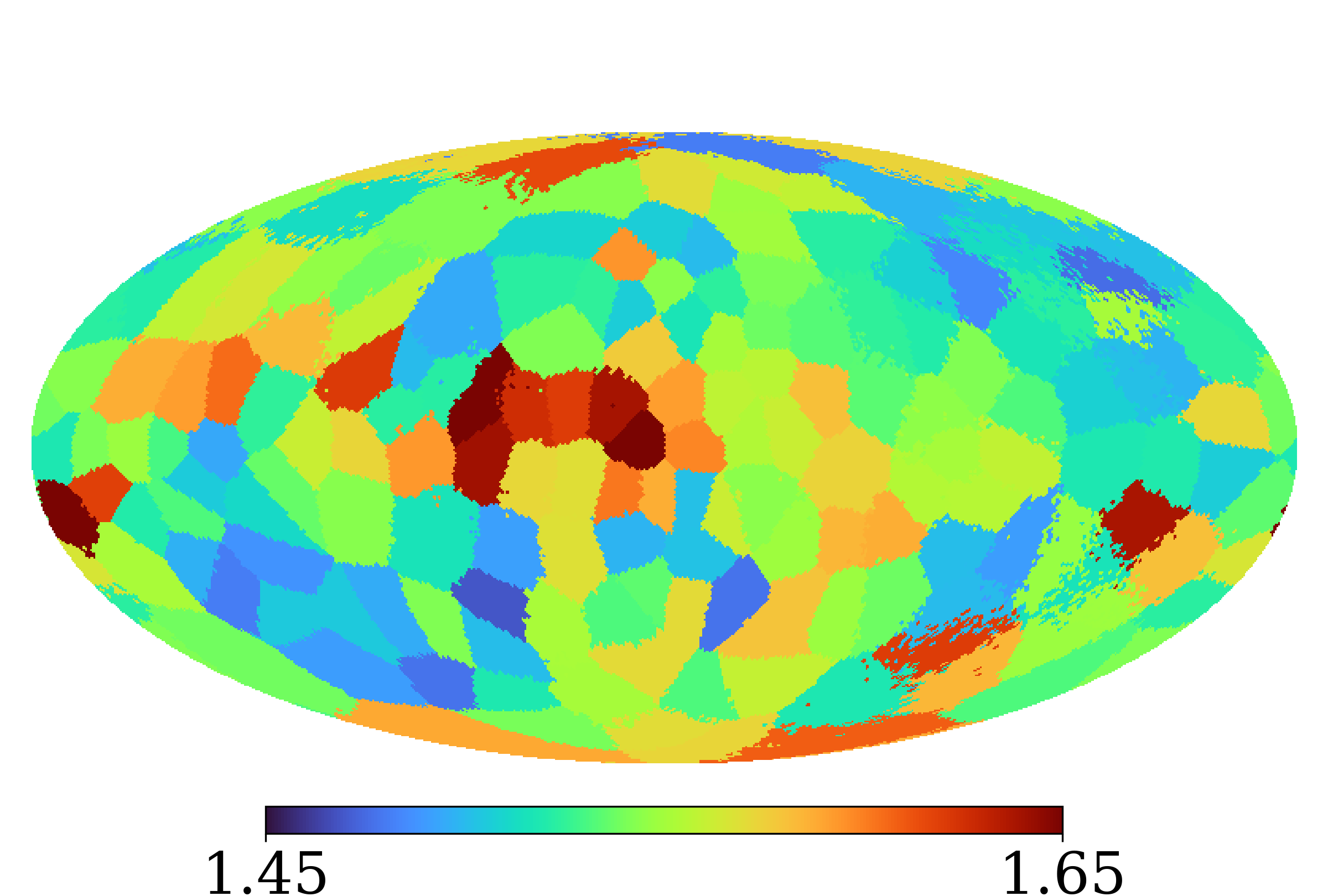} \label{fig_dust_beta_CPLC}}
\subfigure[H(LC): $\beta_d$]{\includegraphics[width=0.325\textwidth]{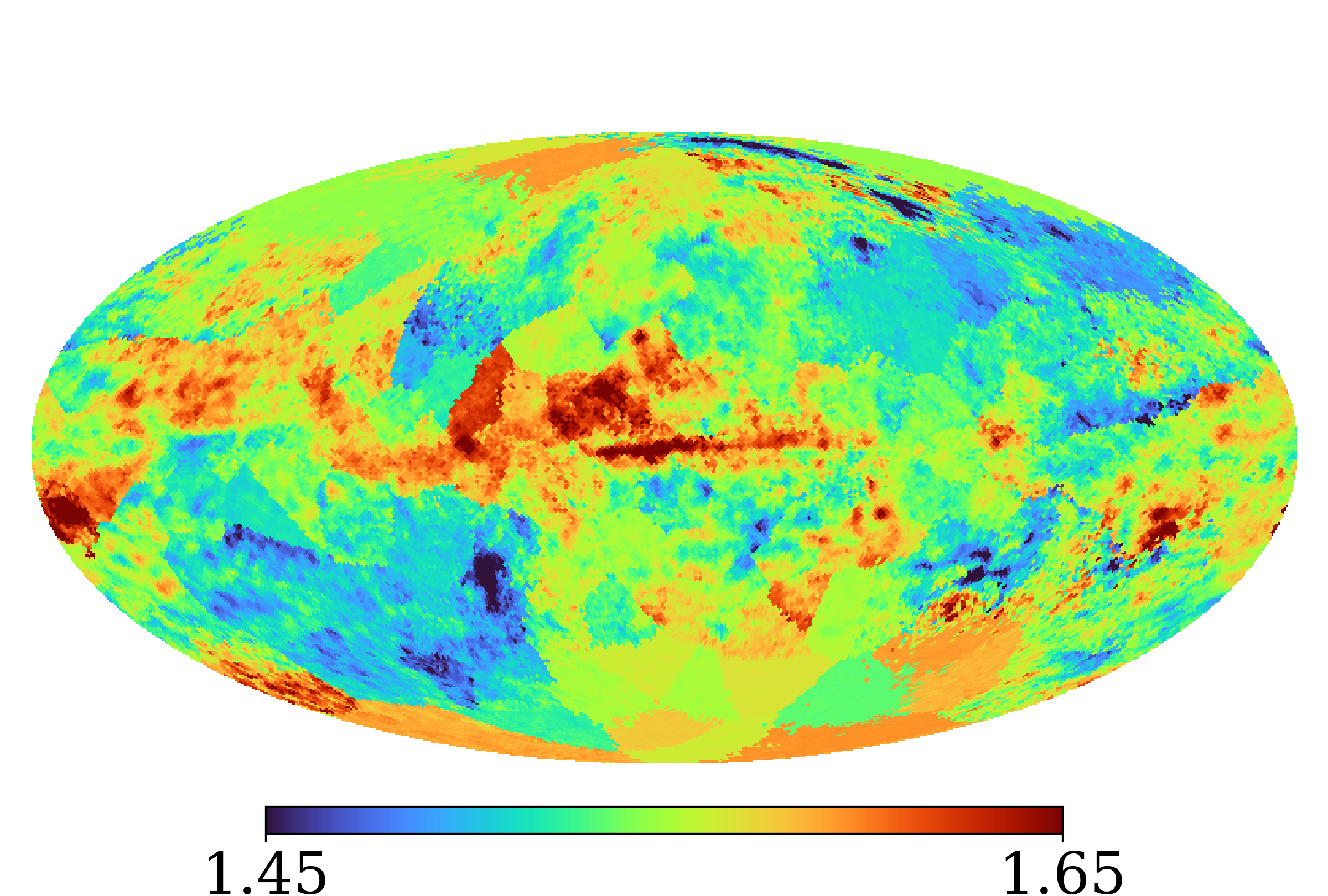} \label{fig_dust_beta_HLC}}
\subfigure[CP(L): $T_d$]{\includegraphics[width=0.325\textwidth]{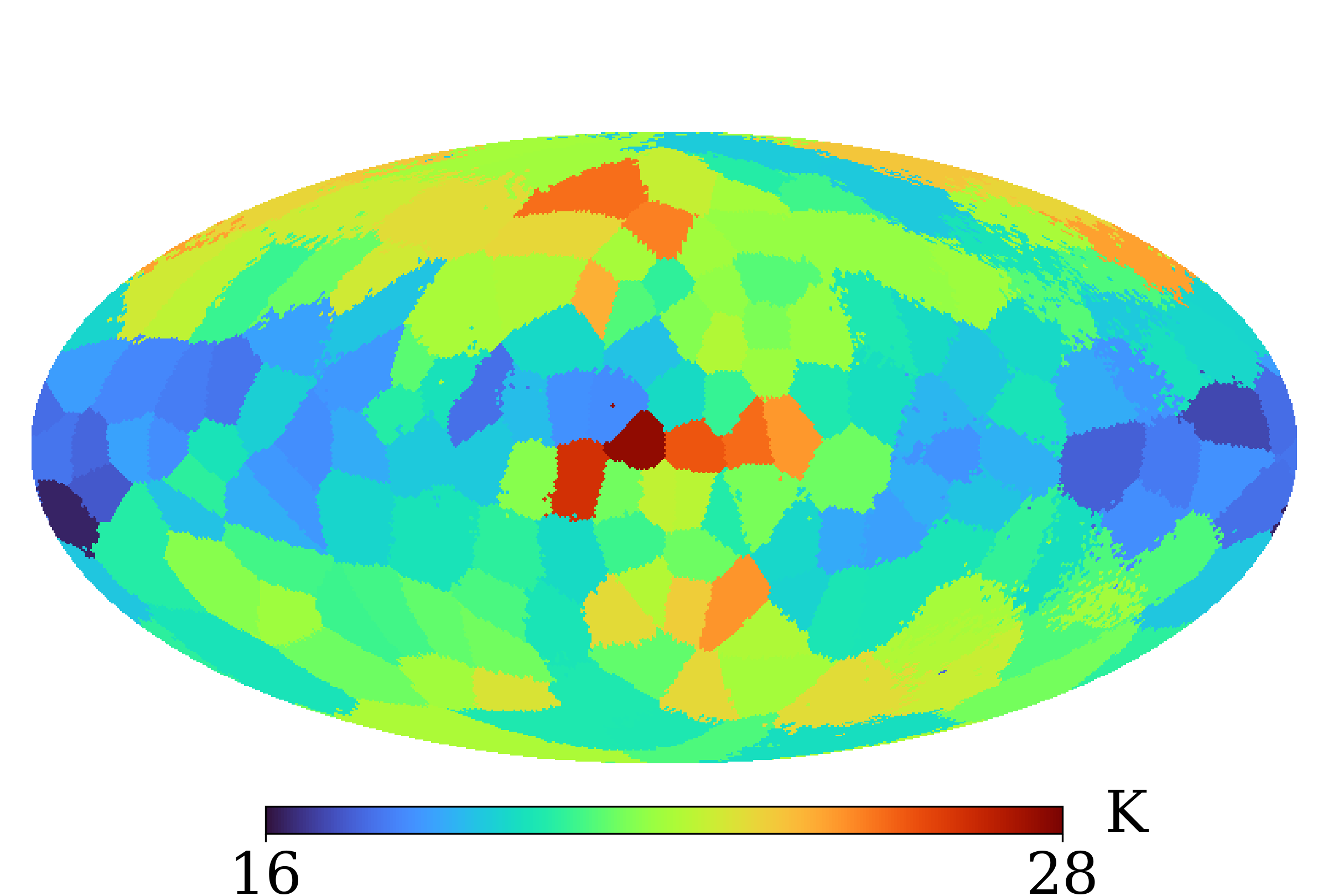} \label{fig_dust_temp_CPL}}
\subfigure[CP(LC): $T_d$]{\includegraphics[width=0.325\textwidth]{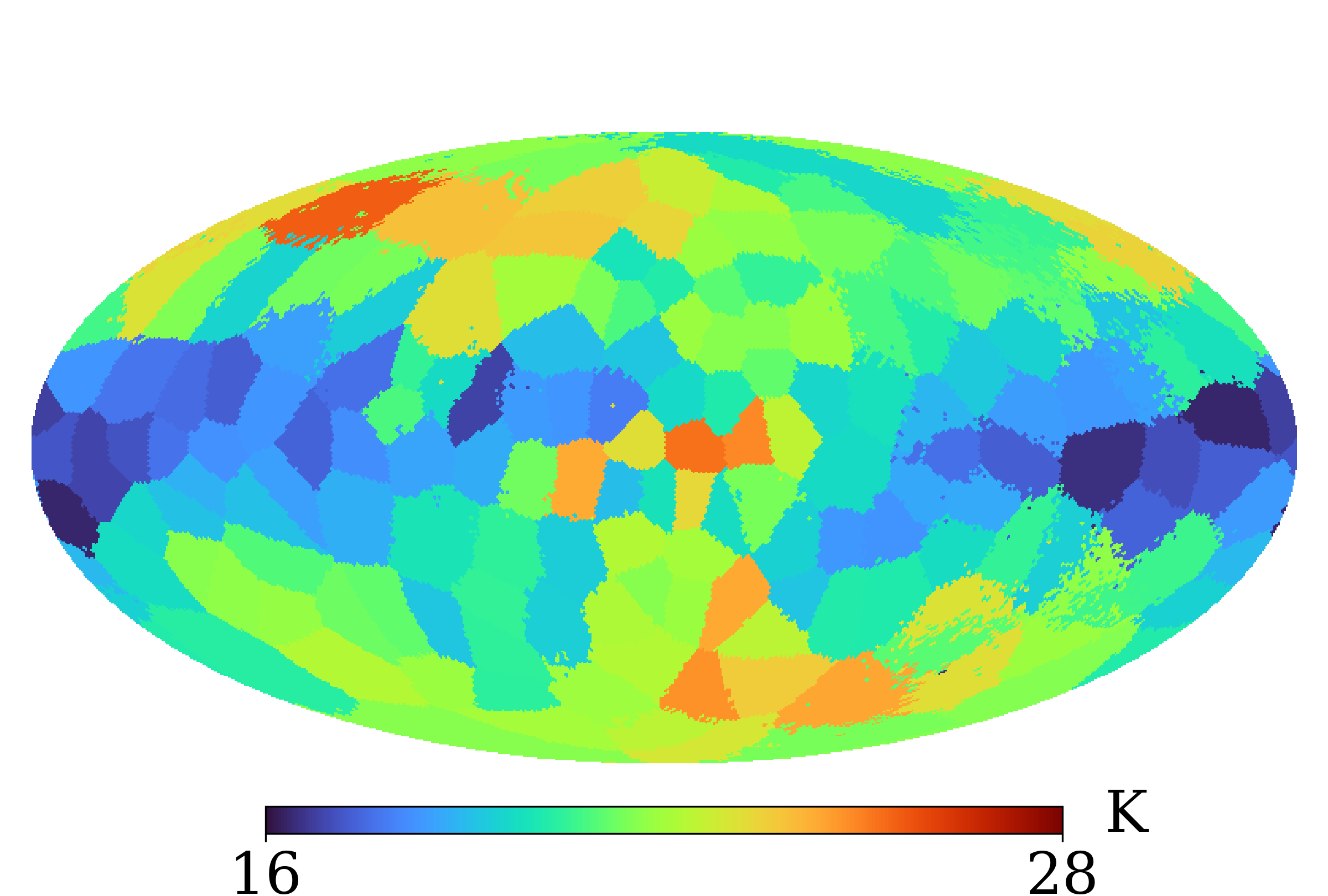} \label{fig_dust_temp_CPLC}}
\subfigure[H(LC): $T_{\mathrm{d}}$]{\includegraphics[width=0.325\textwidth]{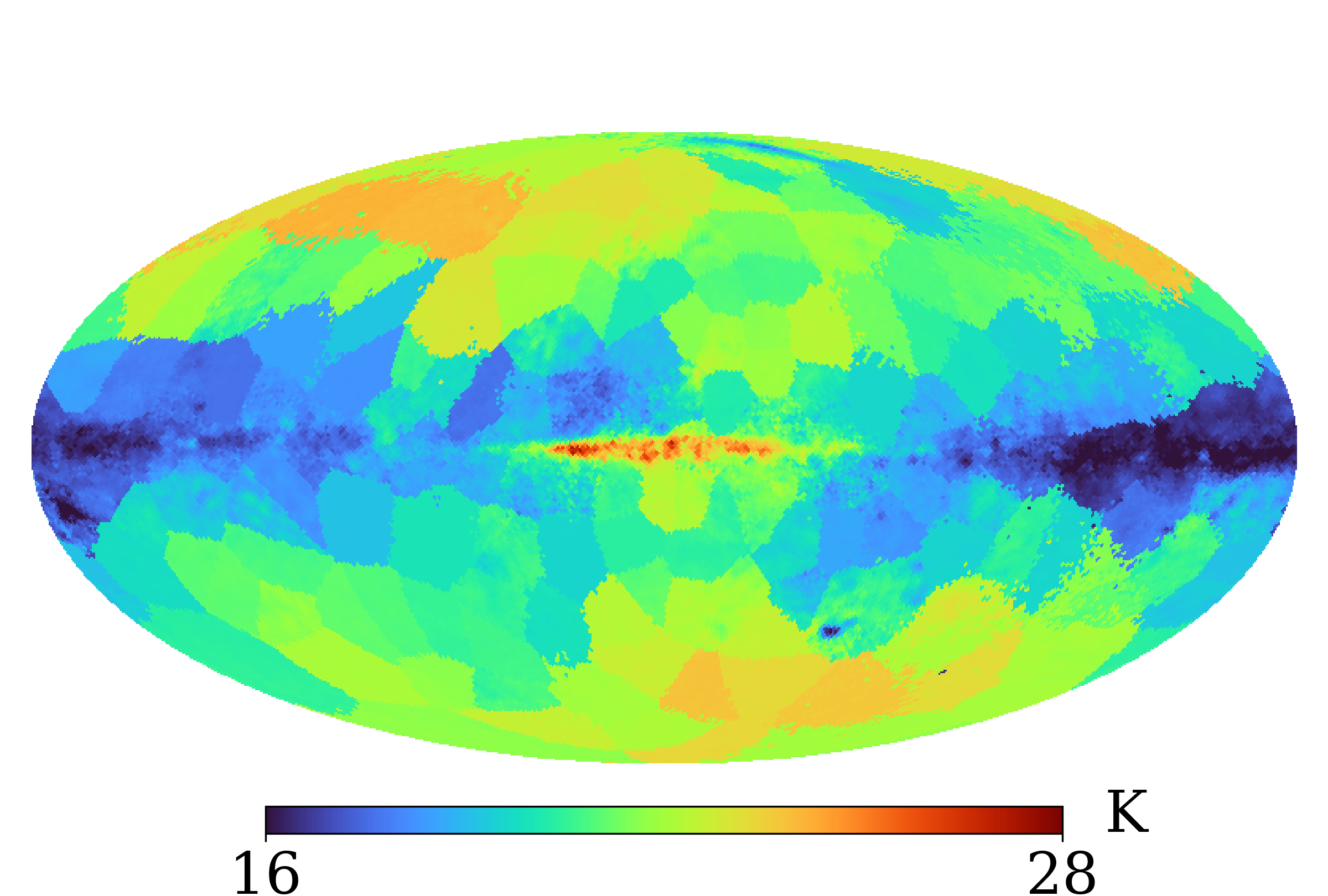} \label{fig_dust_temp_HLC}}
\caption{Spectral parameter maps obtained for the three validation sets. Panels (a) to (c): Recovered synchrotron spectral index maps. Panels (d) to (f): Recovered dust spectral index maps. Panels (g) to (i): Recovered dust temperature maps. For the CP(L) set, we struggle to accurately constrain the synchrotron spectral index, due to the absence of low-frequency channels below $40\,\mathrm{GHz}$. With the addition of a $5\,\mathrm{GHz}$ C-BASS channel for the CP(LC) and H(LC) sets, we are able to place improved constraints on the synchrotron spectral index. Given the frequency coverage of \textit{LiteBIRD} we struggle to constrain dust spectral parameters for all three validation sets, reflected in low effective sample sizes for the dust spectral parameters. The hierarchical model does allow us to capture some of the pixel-level variations in our spectral parameters. Improvements in the point estimates obtained with the hierarchical model can likely be achieved by defining larger regions in areas of low SNR, increasing the smoothing effect of the hyper-distributions. The typical residuals for the synchrotron spectral index maps are comparable between the CP(LC) and H(LC) sets. However, the performance of the CP(LC) set is likely exaggerated by the lack of small-scale features in the input synchrotron spectral index map. The typical residuals for the dust spectral index maps are reduced by $\sim 50$ per cent for the H(LC) set compared to the CP(LC) set. The typical residuals in the recovered dust temperature maps are $\sim 25$ per cent lower for the H(LC) set compared to the CP(LC) set. The simulated dust spectral parameter maps contain additional small-scale features compared to the synchrotron spectral index map.}
\label{fig: spectral params}
\end{figure*}

In Fig. \ref{fig_synch_beta_CPL}, \ref{fig_synch_beta_CPLC} and \ref{fig_synch_beta_HLC}, we show the synchrotron spectral indices obtained for the three validation sets. In the case of the CP(L) set, we struggle to place accurate constraints on the synchrotron spectral index. This is to be expected, given the lack of low-frequency channels below $40\,\mathrm{GHz}$, and is consistent with expectations from the single-pixel component separation analysis presented in \cite{2019MNRAS.tmp.2372J}. For the CP(LC) validation set, we are able to place improved constraints on the synchrotron spectral index, with the variations in the synchrotron spectral index from region-to-region tracing the variations in the input spectral index map shown in Fig. \ref{fig: input maps}. 

When using a hierarchical model we are able to more finely model variations in the synchrotron spectral index across the sky. As can be seen in Fig. \ref{fig_synch_beta_HLC}, the spectral index in regions of high SNR traces the same variations seen in the input synchrotron spectral index map. In regions of low SNR, away from the Galactic plane, we do see additional features not present in the input $\beta_s$ map. This is a result of the noisier estimates of the synchrotron spectral index we obtain in these regions. These noisy variations are constrained by the hyper-distribution, which penalises individual estimates of $\beta_s$ being too far from the population mean, $\mu_{\beta_s}$. Further, even in these noisier regions of the sky we can see that many of the large-scale variations in the spectral index are still traced by the individual spectral indices. This is a well known property of hierarchical models, known as posterior shrinkage, and is one of the main advantages of adopting the hierarchical approach i.e., we obtain improved point estimates of our latent variables \citep{KATAHIRA201637}. 

For the simulations we have considered here, the typical residuals in the synchrotron spectral index maps are similar between the CP(LC) and H(LC) validation sets. This is partly a result of the input synchrotron spectral index map being highly idealized, lacking in small-scale features \citep{2018A&A...618A.166K}. Improvements can also likely obtained for the hierarchical model by refining the region definition, such that we define larger regions in areas of low SNR, increasing the smoothing effect of the hyper-distributions.

In Fig. \ref{fig: synch beta norm dev} we show histograms of the normalized deviations for the synchrotron spectral index. When using a complete pooling model, the distributions of the normalized deviations are significantly wider than the standard Gaussian. This is particularly apparent for the CP(LC) set. With the addition of C-BASS, we have greater constraining power on the average spectral index across each sky region. However, normalized deviations are evaluated at the pixel-level. The uncertainties obtained for the average spectral index in a given region will not be representative of the pixel-level uncertainties. For the range of values shown in Fig. \ref{fig: synch beta norm dev}, the CP(L) set appears to show a positive bias. However, the median is driven to $-0.44$ by a set of large magnitude, negative normalized deviations. Given the lack of low-frequency channels below $\nu=40\,\mathrm{GHz}$, \textit{LiteBIRD} data alone is unable to properly constrain synchrotron spectral parameters.

\subsection{Dust spectral parameters} \label{subsec: dust spectral val}

Most of the constraining power for dust spectral parameters comes from high-frequency channels i.e., $\nu\gtrsim 100\,\mathrm{GHz}$. In our three validation sets these remained identical, being the high-frequency $\textit{LiteBIRD}$ channels. In \cite{2019MNRAS.tmp.2372J} it was found that, given a \textit{LiteBIRD}-like frequency coverage, it is difficult to constrain dust spectral parameters. Indeed, this was the case for our own analysis here, where informative priors were needed on the dust spectral parameters. The difficulty in constraining dust spectral parameters is reflected in their low effective sample size, which was typically $\lesssim 1000$ over much of the sky in all three validation sets.

In Fig. \ref{fig_dust_beta_CPL}, \ref{fig_dust_beta_CPLC} and \ref{fig_dust_beta_HLC} we show the recovered dust spectral index maps, and in Fig. \ref{fig_dust_temp_CPL}, \ref{fig_dust_temp_CPLC} and \ref{fig_dust_temp_HLC} we show the recovered dust temperature maps. In using a hierarchical model for the dust spectral parameters, we are able to fit for variations in the spectral parameters in regions of high SNR close to the Galactic plane. Away from the Galactic plane, the individual variations become much smaller in each region, with the resulting maps of $\beta_d$ and $T_d$ very obviously tracing out the crude structure of the regions used in the component separation. Given the limited frequency coverage, there is simply not enough information to constrain the low-level variations of the dust spectral parameters in each region. In this case the marginal distributions for $\sigma_{\beta_{\mathrm{d}}}$ and $\sigma_{T_{\mathrm{d}}}$ have a large fraction of their probability mass close to zero, constraining individual spectral parameters to be very close to their population means, $\mu_{\beta_d}$ and $\mu_{T_d}$. The typical residuals for the dust spectral index maps are reduced by $\sim 50$ per cent for the H(LC) set compared to the CP(LC) set. The typical residuals in the recovered dust temperature maps are $\sim 25$ per cent lower for the H(LC) set compared to the CP(LC) set. By allowing for some of the low-level variations in the dust spectral parameters we are able to obtain smaller residuals in the recovered parameter maps, propagating through to reduced biases in the CMB amplitude estimates.

In Fig. \ref{fig: dust beta norm dev} we show histograms of the normalized deviations for the dust spectral index, and in Fig. \ref{fig: dust temp norm dev} we show histograms of the normalized deviations for the dust temperature. For both the dust spectral index and the dust temperature we find biases in the medians of the normalized deviations. The MAD values obtained with a complete pooling model are significantly larger than one. This is again a result of the fact that the uncertainty obtained on the average spectral parameters in a region are not representative of the pixel-level uncertainty in those parameters. The distributions of normalized deviations for the H(LC) set are closer to the standard Gaussian than for the CP(LC) and CP(L) sets. Improved point estimates could potentially be obtained by conducting a more detailed exploration of informative priors on the hyper-parameters of dust spectral parameters, or considering model re-parametrizations. However, the hierarchical model is still limited by the lack of additional frequency channels at $\nu>402\,\mathrm{GHz}$.

\section{Conclusions}\label{sec: conclusions}

We have developed a new implementation of Bayesian CMB component separation, using the NUTS algorithm to draw samples from the full posterior distribution. The NUTS algorithm is a self-tuning variant of HMC, that avoids the random walk behaviour that leads to slow convergence when using standard Metropolis-Hastings and Gibbs sampling algorithms. Measured against the rate of effective sample generation, NUTS offers performance improvements of $\sim 10^{3}$ compared to Metropolis-Hastings when fitting the complete pooling model. Geometrical pathologies typical of hierarchical models often make variants of HMC the only reliable option for the diagnosis of divergences and biased inferences \citep{hmc_hierarchical2013}.   

We apply this component separation algorithm to simulations of the \textit{LiteBIRD} and C-BASS experiments to validate the algorithm performance and fidelity. These simulations use a tensor-to-scalar ratio of $r=5\times 10^{-3}$ and a lensing amplitude of $A_L = 1$. Component separation is performed over a set of separate sky regions, defined using the mean-shift algorithm. This clusters sky regions according to the similarity in their synchrotron and dust spectral properties, and their location on the sky. In each region we adopt two different modelling approaches, namely complete pooling and a hierarchical model. In the complete pooling model we assume the spectral parameters in each region are constant. In the hierarchical model we assume spectral parameters are drawn from underlying Gaussian distributions, fitting for the hyper-parameters defining the mean and variance of the Gaussian hyper-distributions, along with the individual pixel-by-pixel spectral parameters constrained by these hyper-distributions. 

When using the complete pooling model we are able to recover accurate estimates of the CMB over much of the sky. However, component separation artefacts are present close to the Galactic plane where the CMB is highly sub-dominant to foregrounds. Using the hierarchical model, these artefacts are removed from the recovered CMB. Estimating the CMB power spectra with these maps, we find the complete pooling model induces large scale foreground residuals in the recovered power spectra. Using multipoles between $30\leq\ell\leq 180$ and fixing the lensing amplitude, we are able to translate these power spectrum estimates into tensor-to-scalar ratio constraints. With only \textit{LiteBIRD} frequency channels, and using the complete pooling model, we find $r=(12.9\pm 1.4)\times 10^{-3}$. Applying the complete pooling model with an additional C-BASS channel at $5\,\mathrm{GHz}$ we find $r=(9.0\pm 1.1)\times 10^{-3}$, and using the hierarchical model with C-BASS and \textit{LiteBIRD} we find $r=(5.2\pm 1.0)\times 10^{-3}$. The addition of C-BASS reduces the bias in the recovered tensor-to-scalar ratio for the complete pooling model. However, the crude assumptions made regarding the behaviour of spectral parameters still leaves a $\sim 3.6\sigma$ bias in the estimate of $r$. We find that the hierarchical model offers an effective generative approach to the modelling of spectral parameters, that helps to mitigate the propensity for fitting outliers when assuming total independence between spectral parameters. The reduced foreground residuals in the recovered CMB maps also means that the hierarchical model does not suffer from the same issues around model misspecification at the power spectrum level, which results in increased uncertainties and biases on recovered cosmological parameters for the complete pooling model.  

For the analysis in this paper we have developed a simple proof-of-concept implementation of our algorithm. Potential future work includes the extension of the algorithm to allow for the joint fitting of dipole and monopole corrections, along with instrumental effects such as bandpass corrections. It would also be prudent to extend the power spectrum estimation, to include the direct joint sampling of CMB amplitudes and power spectra at low multipoles, and expand the code to allow for multi-resolution analyses. From the perspective of modelling, one may also consider more sophisticated approaches to clustering. For the purposes of validation in this study we used naive spectral indices as tracers of the foreground spectral properties, which will be contaminated by noise. In future work, it would be worthwhile examining improvements in region definition by using more sophisticated estimators of foregrounds spectral properties, along with studying the optimal datasets to be used as foreground templates. Finally, significant gain can potentially be made by exploring optimizations to the component separation code e.g., through re-parametrizations and GPU acceleration.

\section{Data Availability}

The simulated data used in this study were generated using the \textsc{PySM} software package, which can be found at \url{https://pysm-public.readthedocs.io/en/latest/}.

\section*{Acknowledgements}

The authors would like to thank David Alonso for useful comments on a draft of this paper. RDPG acknowledges support from an STFC studentship. CD acknowledges support from an ERC Starting (Consolidator) Grant (no.~307209) and an STFC Consolidated Grant (ST/P000649/1).




\bibliographystyle{mnras}
\bibliography{nuts_paper} 






\bsp	
\label{lastpage}
\end{document}